\documentclass[12pt]{article}

\usepackage{mathptmx}
\usepackage[letterpaper,margin=1in]{geometry}
\usepackage{setspace}
\usepackage{etoolbox}
\makeatletter
\renewcommand{\normalsize}{\@setfontsize\normalsize\@xiipt{14.5}}
\makeatother
\renewenvironment{abstract}{\begin{center}\large\scshape Abstract\end{center}\vspace{1em}\singlespacing\setlength{\parindent}{0pt}\noindent\ignorespaces}{\par}
\AtBeginEnvironment{figure}{\setstretch{1}}
\AtBeginEnvironment{table}{\setstretch{1}}
\AtBeginDocument{\normalsize\doublespacing}

\usepackage[utf8]{inputenc} 
\usepackage[T1]{fontenc}    
\usepackage{float}
\usepackage{placeins}
\usepackage{hyperref}       
\usepackage{url}            
\usepackage{xurl}           
\usepackage{booktabs}
\usepackage{multirow}       
\usepackage{array}
\usepackage{arydshln}      
\usepackage{amsfonts}       
\usepackage{amsmath}
\usepackage{amssymb}
\usepackage{amsthm}
\usepackage{nicefrac}       
\usepackage{microtype}      
\usepackage{cleveref}       
\usepackage{graphicx}
\usepackage{tikz}
\usetikzlibrary{arrows.meta}
\tikzset{
  manuscript dag/.style={x=1.5cm,y=1.5cm},
  manuscript dag node/.style={circle,draw=black,fill=white,line width=0.75pt,
    minimum size=7.2mm,inner sep=0pt,outer sep=1.5pt,font=\large\rmfamily},
  manuscript dag edge/.style={-{Stealth[length=2.2mm,width=1.6mm]},
    line width=0.85pt,draw=black}
}
\usepackage{doi}
\usepackage{natbib}
\renewcommand{\harvardurl}[1]{\textbf{URL:} \url{#1}}
\setcitestyle{authoryear,round,aysep={},yysep={;}}
\newtheorem{proposition}{Proposition}
\newtheorem{propositionA}{Proposition}

\newtheorem{corollary}{Corollary}
\newtheorem{definition}{Definition}
\newtheorem{assumption}{Assumption}

\newcommand{\BHorigCountries}{79}
\newcommand{\BHorigRows}{2,061}
\newcommand{\BHorigCCCountries}{78}
\newcommand{\BHorigCCRows}{1,769}
\newcommand{\BHexpCountries}{132}
\newcommand{\BHexpCCCountries}{115}
\newcommand{\BHexpCCRows}{2,198}
\newcommand{\BHorigCoef}{-0.465}
\newcommand{\BHorigSE}{0.143}

\newcommand{\BHexpCoef}{-0.546}

\newcommand{\BHshareCCDem}{67.2}

\newcommand{\BHshareTgtDem}{48.4}
\newcommand{\BHnCCNegReweightedNS}{three}
\newcommand{\BHnGapSig}{four}
\newcommand{\BHnModerators}{six}
\newcommand{\BHminVarLo}{25}
\newcommand{\BHminVarHi}{38}

\newcommand{\KBattrChi}{61.28}
\newcommand{\KBattrDf}{36}
\newcommand{\KBattrP}{0.0054}
\newcommand{\KBresponseControl}{61.28}
\newcommand{\KBresponseTreated}{62.09}
\newcommand{\KBnEligible}{1,900}
\newcommand{\KBnMissing}{728}
\newcommand{\KBpctMissing}{38.3}
\newcommand{\KBnComplete}{1,172}
\newcommand{\KBpctMissNoCol}{45.0}
\newcommand{\KBpctMissCol}{30.9}

\newcommand{\KBipwEst}{0.0932}

\newcommand{\KBeffNoCol}{0.0497}
\newcommand{\KBeffCol}{0.1107}
\newcommand{\KBshareCCCol}{53.2}
\newcommand{\KBshareTargetCol}{47.5}
\newcommand{\KBtargetAvg}{0.0787}
\newcommand{\KBccPooled}{0.0925}

\newcommand{\KBdiffSameSpec}{-0.0035}
\newcommand{\KBdiffTarget}{-0.0138}
\newcommand{\KBdiffTargetLo}{-0.0488}
\newcommand{\KBdiffTargetHi}{0.0174}
\newcommand{\KBhetP}{0.281}

\newcommand{\LRnEligAPSR}{100}
\newcommand{\LRnEligAJPS}{58}

\newcommand{\LRnEligible}{158}
\newcommand{\LRnFillMain}{42}
\newcommand{\LRpctFillMain}{26.6}
\newcommand{\LRnImpMain}{11}
\newcommand{\LRpctImpMain}{7.0}
\newcommand{\LRpctDeletionMain}{50.6}

\newcommand{\LRnDeletion}{80}
\newcommand{\LRnRegYes}{45}
\newcommand{\LRpctRegYes}{56}

\newcommand{\MSreps}{500}
\newcommand{\MSmiA}{1.057}
\newcommand{\MSsatB}{1.002}
\newcommand{\MSipwB}{1.002}
\newcommand{\MSsatC}{1.541}
\newcommand{\MSipwC}{1.580}

\newcommand{\MARreps}{500}

\newcommand{\RAreps}{10,000}
\newcommand{\RAn}{803}
\newcommand{\RAnSource}{925}
\newcommand{\RAmissPct}{11.5}
\newcommand{\RAmedTotal}{-0.560}
\newcommand{\RAmedDirect}{-0.305}
\newcommand{\RAconfTotal}{-0.305}
\newcommand{\RAconfCCXMean}{-0.285}

\hypersetup{
pdftitle={When Listwise Deletion Works: Causal Diagrams and Complete-Case Regression},
pdfauthor={Wenxin Lin, Shiyao Liu},
}
\RequirePackage[normalem]{ulem} 
\RequirePackage{color}\definecolor{RED}{rgb}{1,0,0}\definecolor{BLUE}{rgb}{0,0,1} 
\providecommand{\DIFaddbegin}{} 
\providecommand{\DIFaddend}{} 
\providecommand{\DIFdelbegin}{} 
\providecommand{\DIFdelend}{} 
\providecommand{\DIFaddbeginFL}{} 
\providecommand{\DIFaddendFL}{} 
\providecommand{\DIFdelbeginFL}{} 
\providecommand{\DIFdelendFL}{} 
\newcommand{\DIFscaledelfig}{0.5}
\definecolor{rlnotecolor}{rgb}{0.75,0,0.6}

\RequirePackage{settobox} 
\RequirePackage{letltxmacro} 
\newsavebox{\DIFdelgraphicsbox} 
\newlength{\DIFdelgraphicswidth} 
\newlength{\DIFdelgraphicsheight} 
\LetLtxMacro{\DIFOincludegraphics}{\includegraphics} 
\newcommand{\DIFaddincludegraphics}[2][]{{\color{blue}\fbox{\DIFOincludegraphics[#1]{#2}}}} 
\newcommand{\DIFdelincludegraphics}[2][]{
\sbox{\DIFdelgraphicsbox}{\DIFOincludegraphics[#1]{#2}}
\settoboxwidth{\DIFdelgraphicswidth}{\DIFdelgraphicsbox} 
\settoboxtotalheight{\DIFdelgraphicsheight}{\DIFdelgraphicsbox} 
\scalebox{\DIFscaledelfig}{
\parbox[b]{\DIFdelgraphicswidth}{\usebox{\DIFdelgraphicsbox}\\[-\baselineskip] \rule{\DIFdelgraphicswidth}{0em}}\llap{\resizebox{\DIFdelgraphicswidth}{\DIFdelgraphicsheight}{
\setlength{\unitlength}{\DIFdelgraphicswidth}
\begin{picture}(1,1)
\thicklines\linethickness{2pt} 
{\color[rgb]{1,0,0}\put(0,0){\framebox(1,1){}}}
{\color[rgb]{1,0,0}\put(0,0){\line( 1,1){1}}}
{\color[rgb]{1,0,0}\put(0,1){\line(1,-1){1}}}
\end{picture}
}\hspace*{3pt}}} 
} 
\LetLtxMacro{\DIFOaddbegin}{\DIFaddbegin} 
\LetLtxMacro{\DIFOaddend}{\DIFaddend} 
\LetLtxMacro{\DIFOdelbegin}{\DIFdelbegin} 
\LetLtxMacro{\DIFOdelend}{\DIFdelend} 
\DeclareRobustCommand{\DIFaddbegin}{\DIFOaddbegin \let\includegraphics\DIFaddincludegraphics} 
\DeclareRobustCommand{\DIFaddend}{\DIFOaddend \let\includegraphics\DIFOincludegraphics} 
\DeclareRobustCommand{\DIFdelbegin}{\DIFOdelbegin \let\includegraphics\DIFdelincludegraphics} 
\DeclareRobustCommand{\DIFdelend}{\DIFOaddend \let\includegraphics\DIFOincludegraphics} 
\LetLtxMacro{\DIFOaddbeginFL}{\DIFaddbeginFL} 
\LetLtxMacro{\DIFOaddendFL}{\DIFaddendFL} 
\LetLtxMacro{\DIFOdelbeginFL}{\DIFdelbeginFL} 
\LetLtxMacro{\DIFOdelendFL}{\DIFdelendFL} 
\DeclareRobustCommand{\DIFaddbeginFL}{\DIFOaddbeginFL \let\includegraphics\DIFaddincludegraphics} 
\DeclareRobustCommand{\DIFaddendFL}{\DIFOaddendFL \let\includegraphics\DIFOincludegraphics} 
\DeclareRobustCommand{\DIFdelbeginFL}{\DIFOdelbeginFL \let\includegraphics\DIFdelincludegraphics} 
\DeclareRobustCommand{\DIFdelendFL}{\DIFOaddendFL \let\includegraphics\DIFOincludegraphics} 
\RequirePackage{listings} 
\RequirePackage{color} 
\lstdefinelanguage{DIFcode}{ 
  moredelim=[il][\color{red}\sout]{\%DIF\ <\ }, 
  moredelim=[il][\color{blue}\uwave]{\%DIF\ >\ } 
} 
\lstdefinestyle{DIFverbatimstyle}{ 
	language=DIFcode, 
	basicstyle=\ttfamily, 
	columns=fullflexible, 
	keepspaces=true 
} 
\lstnewenvironment{DIFverbatim}{\lstset{style=DIFverbatimstyle}}{} 
\lstnewenvironment{DIFverbatim*}{\lstset{style=DIFverbatimstyle,showspaces=true}}{} 

\begin{document}
\addtocontents{toc}{\protect\setcounter{tocdepth}{-10}}
\begin{center}
{\Large When Listwise Deletion Works:\\ Causal Diagrams and Complete-Case Regression}\footnote{Authors are listed in alphabetical order. We thank Jongyoon Baik, Adam Berinsky, Frederick Boehmke, Naoki Egami, Macartan Humphreys, Kosuke Imai, Hongyi Jiang, In Song Kim, Dean Knox, Licheng Liu, Naijia Liu, Adeline Lo, Jacob Montgomery, Xun Pang, Yiqing Xu, Teppei Yamamoto, Soichiro Yamauchi, Junjian Yi, Wei Zhong, Meixi Zhuang, and Cai Zuo, as well as participants at the Poster Session of the 2026 Annual Meeting of the Society for Political Methodology, the Second Asia-Pacific Politics and Public Administration Conference, the 2026 Visions in Methodology conference at the Hertie School, the 3rd Forum for Women in Political Science (China) in Nanjing, and seminar participants of the Political Economy workshop at Peking University, for helpful comments. We used Anthropic's Claude (Opus 5 and 5.5 and Sonnet 5 and 5.5), OpenAI's Codex (GPT-5.6, GPT-6, and GPT-6.1), and Alibaba's Qwen 3.8 for language editing, \LaTeX{} formatting, drafting, coding, debugging, pipeline management, reproducibility checks, and conducting adversarial reviews. All AI-assisted passages and code were reviewed, revised, and approved by the authors, who take full responsibility for the manuscript's scientific content.}\par
\vspace{1.5em}
Wenxin Lin\footnote{National School of Development, Peking University.}\quad
Shiyao Liu\footnote{China Center for Economic Research, Institute of South-South Cooperation and Development, National School of Development, Peking University; Governance/Lab, Massachusetts Institute of Technology. Corresponding author: \texttt{shiyaoliu at nsd.pku.edu.cn}.}
\end{center}
\vspace{2em}

\begin{abstract}
Listwise deletion (LWD) appears more common than imputation: at least 50\% of the observational causal studies we review in leading journals state or imply it, while 7\% describe an imputation model. We develop a framework for when complete-case regression consistently estimates the population average treatment effect (ATE). With a constant unit-level treatment effect, two checks suffice: whether covariates and the missingness indicator satisfy the adjustment criterion in an augmented causal diagram (DAG), and whether the estimator pools within-covariate comparisons appropriately, possibly using inverse propensity weighting. A collider--mediator illustration highlights the first check: the same missing-at-random (MAR) mechanism can require different adjustment choices. With heterogeneous effects, conditional effects must be transportable to the population and be reweighted to its population distribution. These formal results provide practical checks for applied researchers to run complete case regressions, recover existing guidance as special cases, and are illustrated with applications across fields.

\end{abstract}
\vspace{\baselineskip}
\noindent\textbf{Word count:} 9,985

\clearpage
\section{Introduction}
\label{sec:introduction}

Missing data are common in social science research. Through a review of published observational causal studies in the \emph{American Journal of Political Science} (AJPS) and \emph{American Political Science Review} (APSR) between 2019 and 2024, \citet{huangLiu2026} found that more than 50\% of the studies mentioned missing values. Our review of a reconstructed sample of theirs finds that \LRnDeletion{} of \LRnEligible{} articles (\LRpctDeletionMain{}\%) state or imply listwise deletion in the main text, compared with \LRnFillMain{} (\LRpctFillMain{}\%) reporting any fill-in method and \LRnImpMain{} (\LRpctImpMain{}\%) describing an imputation model. Further, listwise deletion could also arise from software defaults: in R, \texttt{lm()}, \texttt{glm()}, and \texttt{ivreg()} drop incomplete observations under default \texttt{na.action = na.omit}, and mediation analyses may inherit these sample restrictions from their component models\footnote{\texttt{mediate()} refits mediator and outcome models on their overlapping observations when \texttt{dropobs = TRUE} \citep{tingley2014mediation}.}.

Multiple imputation uses partially observed records and can reduce bias and improve precision \citep{king2001incomplete,honaker2010missing,lall2016multiple}, but under nonignorable missingness it can perform worse than listwise deletion \citep{pepinsky2018note}. Listwise deletion avoids an imputation model and can yield consistent coefficient estimates under some nonrandom missingness \citep{arel2018can}, yet it discards observed information and can introduce selection bias \citep{heckman1979selection,king2001incomplete,huangLiu2026}.

In this paper, we study when regression after listwise deletion can be justified. Among the \LRnDeletion{} articles that state or imply listwise deletion, \LRnRegYes{} (\LRpctRegYes{}\%) report at least one regression estimated after dropping cases with missing values\footnote{These counts are lower bounds. Appendix~\ref{app:literature-review} details the sample construction and coding rules.}. Thus, we consider the following causal-regression workflow for listwise deletion: a researcher chooses an adjustment set intended to identify the population average treatment effect (ATE) and a regression specification intended to estimate it as if all required values were observed, then applies listwise deletion and reports the treatment coefficient from the complete-case regression as an estimate of that ATE. We develop a formal framework for this decision by making the assumed causes of missingness explicit and deriving conditions under which the complete-case regression consistently estimates the researcher's target.

Taking the oracle regression's identification of a variance-weighted average treatment effect as given, we formalize sufficient conditions under which regression after deletion consistently estimates the population ATE. Under a constant unit-level treatment effect, researchers must check two boxes: whether treatment comparisons remain valid after selection into the complete-case sample, and whether the fitted regression pools those comparisons appropriately. The first asks researchers to add the missing indicator and its causes to the DAG used to justify their original adjustment set, then reassess adjustment and treatment overlap. The second asks whether the outcome specification remains correct among complete cases, or whether treatment-propensity weighting can instead recover their average effect. With heterogeneous effects, a third box connects complete-case comparisons to the population target: when conditional effects are transportable to the population, researchers need to average them over the population covariate distribution rather than the complete-case covariate distribution. These checks provide a formal framework for assessing the causal and modeling assumptions behind complete-case regression, which require substantive judgment beyond statistical tests: as Subsection~\ref{sec:covariate-roles} illustrates, the same observed data can imply different conclusions about the validity of listwise deletion under different causal structures.

We also show why each box requires a separate check. In one simulation, selection leaves within-covariate treatment comparisons valid but changes the outcome specification, causing complete-case regression to pool those comparisons incorrectly. In the collider--mediator illustration, the same missing mechanism and MAR classification call for different adjustment choices depending on whether a covariate is a mediator or a confounder. Finally, the reweighting applications show how valid complete-case effects can differ from the population ATE when treatment effects vary and the covariate distributions differ.

Our approach builds on graphical analyses of selection bias and missing data. \citet{bareinboim2012controlling}, \citet{bareinboim2014recovering}, and \citet{bareinboim2016causal} study when statistical and causal quantities can be recovered from selected data. \citet{mohan2021graphical} develop the corresponding missing-data perspective. \citet{daniel2012using} and \citet{saadati2020adjustment} establish graphical conditions for causal analysis using complete records. \mbox{\citet{ross2020complete}} show how effect modification and differences in covariate distributions affect marginal complete-case estimates. \citet{hughes2019accounting}, most closely related, use causal diagrams to illustrate when complete-case regression yields unbiased estimates, taking the validity of the full-data analysis as given, but do not provide formal proofs. Building on their perspective, we derive and prove sufficient conditions that distinguish valid treatment comparisons among complete cases, which graphical criteria address, from the regression's pooling of those comparisons, which they alone do not guarantee. Together, these results bring graphical criteria for listwise deletion into the regression workflow used by applied political scientists, who, to our knowledge, do not currently use causal diagrams to justify deletion. Our framework thus gives applied researchers a practical set of checks for assessing whether their complete-case regression consistently estimates the ATE of interest.

Our framework also reconciles the differing recommendations of \citet{pepinsky2018note} and \citet{arel2018can} by specifying the causal and modeling assumptions under which they apply, complements guidance on choosing imputation models \citep{marbach2022choosing} by clarifying when deletion is justified in the first place, and generalizes the identification conditions underlying \citet{huangLiu2026}: we clarify when their imputation step is needed to recover the population covariate distribution and when complete-case effects can instead be averaged using an observed population distribution.

In Section~\ref{sec:motivating-simulation}, we use a simulation to build intuition for the two sufficient conditions: with a constant unit-level treatment effect, we contrast three missing-not-at-random mechanisms under which complete-case regression respectively recovers the population ATE, requires a pooling remedy, and fails because within-covariate comparisons are invalid.

Section~\ref{sec:causal-conditions} then derives formal sufficient conditions under which complete-case regression consistently estimates the population ATE: when within-strata treatment comparisons among complete cases identify causal effects, when regression pools them appropriately, and how the resulting effect connects to the population ATE under constant or heterogeneous effects.

Section~\ref{sec:specification-propensity} revisits conventional guidance, recovering familiar recommendations as special cases and showing how a covariate's role as mediator or confounder changes the validity of listwise deletion under the same missingness pattern and MAR classification.

Section~\ref{sec:reweighting-hte} develops reweighting for heterogeneous effects, generalizing the identification argument underlying \citeauthor{huangLiu2026}'s (\citeyear{huangLiu2026}) two-stage projection estimator, and then illustrates it in a reanalysis of \citet{bodeaHicks2015}. Appendix~\ref{app:glm-reweighting} shows the potential of extension into generalized linear models.

Section~\ref{sec:comprehensive-illustration} brings all three checks together in a reanalysis of the canvassing experiment \citep{kallaBroockman2023}, illustrating the application of the three box roadmap we propose.

\section{Motivating Simulation: Identification and Weighting}
\label{sec:motivating-simulation}

We begin with a simulation to illustrate conditions under which complete-case regression consistently estimates the population ATE, the causal effect that the oracle adjustment set identifies. The oracle sample contains \(N_{\mathrm{or}}=20{,}000\) observations generated by the following process, with a constant unit-level treatment effect \(\tau=1\):

\[
Y_i=D_i+X_i+U_i+\xi_i,
\qquad \xi_i\sim N(0,0.5^2).
\]

The pretreatment confounder \(X_i\) is uniform on \(\{-1,0,1\}\), while the unobserved cause of the outcome, \(U_i\), is also uniform on \(\{-1,1\}\). \(X_i\), \(U_i\), and \(\xi_i\) are mutually independent. Treatment assignment depends only on \(X_i\): \(D_i\mid X_i=x\sim\operatorname{Bernoulli}((1+x^2)/4+0.30x)\).

In the observed data, \(Y_i\) and \(D_i\) are available for all 20,000 observations, but \(X_i\) is missing for some. Let \(M_i=0\) when \(X_i\) is observed and \(M_i=1\) otherwise. We consider three mechanisms governing whether \(X_i\) is observed:

\[
\begin{aligned}
\text{A:}\quad P(M_i=0\mid X_i=x,U_i=u)&=0.95-0.30x^2,\\
\text{B:}\quad P(M_i=0\mid X_i=x,U_i=u)&=(0.95-0.30x^2)(1+0.50ux^2),\\
\text{C:}\quad P(M_i=0\mid X_i=x,U_i=u,D_i=d)&=(0.95-0.30x^2)[1+0.50u(2d-1)x^2].
\end{aligned}
\]

All three mechanisms are not missing at random (NMAR): whether \(X_i\) is observed depends on its unobserved value. Under B, observation further depends on \(U_i\), while under C, it additionally depends on the interaction between \(U_i\) and \(D_i\). Averaging over \(U_i\), each mechanism gives \(P(M_i=0\mid X_i=x)=0.95-0.30x^2\). Because \(X_i\) is uniform on \(\{-1,0,1\}\), the missingness rate is \(P(M_i=1)=0.25\) in all three cases.

An analyst could use listwise deletion to fit the following linear regression among complete cases:
\[
Y_i=\alpha_{\mathrm{cc}}+\beta_{\mathrm{cc}}D_i+\gamma_{\mathrm{cc}}X_i+\varepsilon_{\mathrm{sim},i},
\qquad M_i=0.
\]
One rationale for this choice is the advice\footnote{An analyst might read \citet{arel2018can} as endorsing listwise deletion here, because the oracle error \(U_i+\xi_i\) is independent of \(X_i\) and \(D_i\) and missingness depends only on \(X_i\). But their advice also requires the complete-case indicator to be independent of the regression error---exactly what mechanisms B and C violate, since \(M_i\) is associated with \(U_i+\xi_i\) given \(X_i\) and \(D_i\). Their advice therefore excludes these cases.} of \citet[p.~243]{arel2018can}:

\begin{quotation}
Under [the pattern of missingness driven by values of the independent variables], LWD can be used without fear that it will introduce bias in regression estimates.
\end{quotation}

Alternatively, an analyst might use multiple imputation to fit the same regression, guided by the general expectation expressed by \citet[p.~51]{king2001incomplete}:

\begin{quotation}
Thus, multiple imputation will normally be better than, and almost always not worse than, listwise deletion.
\end{quotation}

\begin{figure}[H]
\centering
\includegraphics[width=\textwidth]{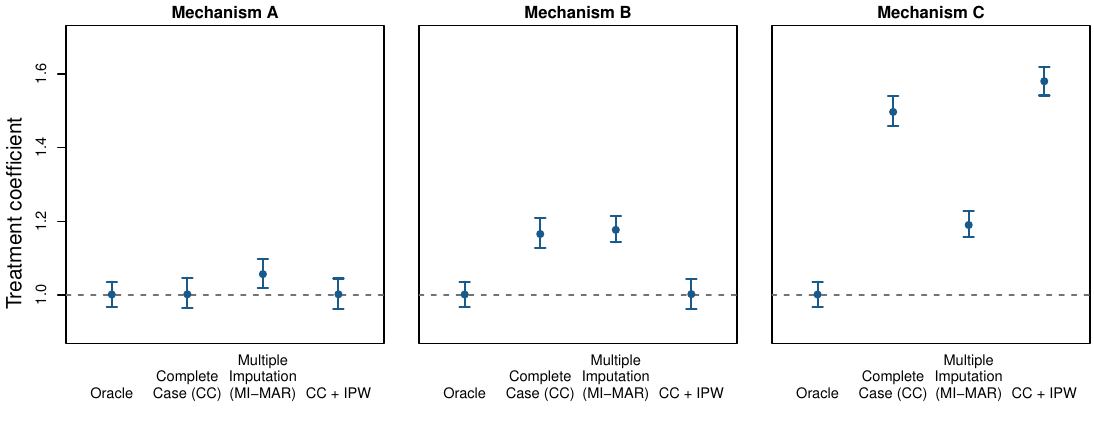}
\caption{Estimated Treatment Coefficients under Different Missing Mechanisms}
\label{fig:motivating-simulation}
\vspace{0.4em}
\begin{minipage}{0.92\linewidth}
\footnotesize\emph{Note:} Points show mean treatment-coefficient estimates across \MSreps{} replications. Bars show the 2.5th and 97.5th percentiles. The dashed line marks the true effect, \(\tau=1\). MI-MAR uses a missing-at-random imputation model, while CC + IPW uses inverse treatment-propensity weights. See Appendix Table~\ref{tab:motivating-simulation} for numerical results.
\end{minipage}
\end{figure}

We examine how both choices perform under each mechanism, benchmarking against the oracle estimator and adding a complete-case estimator that uses inverse treatment-propensity weights (CC + IPW) \citep{choi2019comparison}. Figure~\ref{fig:motivating-simulation} summarizes the treatment-coefficient estimates across \MSreps{} replications:

Under mechanism A, the unweighted complete-case estimates are centered on \(\tau=1\), consistent with \citet{arel2018can}'s result that listwise deletion need not bias regression estimates when missingness depends on regressor values. The multiple-imputation estimates instead average \MSmiA{}: being NMAR in mechanism A, the consistency guarantee for correctly specified imputation under MAR does not apply. Under mechanism B, both the unweighted complete-case and multiple-imputation estimates are centered above \(\tau=1\). For complete-case regression, \(M_i\) depends on both \(X_i\) and \(U_i\) and is therefore associated with the oracle regression error \(U_i+\xi_i\), violating the indicator--error independence condition of \citet{arel2018can}. The CC + IPW estimates remain centered near one. Under mechanism C, none of the three feasible estimators produces estimates centered on \(\tau=1\).

\paragraph{Sources of Error} These patterns reflect two sources of error: treatment comparisons within \(X_i\) strata may fail to identify \(\tau\), or the pooled regression may fail to preserve valid within-stratum comparisons in its treatment coefficient. Table~\ref{tab:motivating-subset-regressions} reports the mean estimated intercepts and treatment coefficients from separate complete-case regressions of \(Y_i\) on \(D_i\) in each \(X_i\) stratum. 

\begin{table}[htbp]
\centering
\caption{Complete-Case Regression Coefficients within Each \(X_i\)}
\label{tab:motivating-subset-regressions}
\footnotesize
\setlength{\tabcolsep}{2.8pt}
\begin{tabular}{l*{9}{r}}
\toprule
 & \multicolumn{3}{c}{Mechanism A ($M=0$)}
 & \multicolumn{3}{c}{Mechanism B ($M=0$)}
 & \multicolumn{3}{c}{Mechanism C ($M=0$)} \\
\cmidrule(lr){2-4}\cmidrule(lr){5-7}\cmidrule(lr){8-10}
Coefficient & $X=-1$ & $X=0$ & $X=1$
            & $X=-1$ & $X=0$ & $X=1$
            & $X=-1$ & $X=0$ & $X=1$ \\
\midrule
Intercept & $-1.000$ & $-0.000$ & $1.000$ & $-0.499$ & $-0.000$ & $1.500$ & $-1.501$ & $-0.000$ & $0.501$ \\
          & $(0.018)$ & $(0.016)$ & $(0.038)$ & $(0.016)$ & $(0.016)$ & $(0.033)$ & $(0.016)$ & $(0.016)$ & $(0.035)$ \\
\addlinespace
$D$ & $0.999$ & $1.004$ & $1.002$ & $1.001$ & $1.004$ & $1.001$ & $2.002$ & $1.004$ & $2.001$ \\
    & $(0.041)$ & $(0.032)$ & $(0.043)$ & $(0.036)$ & $(0.032)$ & $(0.037)$ & $(0.036)$ & $(0.032)$ & $(0.040)$ \\
\bottomrule
\end{tabular}
\vspace{0.4em}
\begin{minipage}{0.92\linewidth}
\footnotesize\emph{Note:} Entries are mean coefficients across 500 replications. Parentheses below give their standard deviations.
\end{minipage}
\end{table}

\subparagraph{Weighting} To examine pooling, we first allow the intercept and treatment coefficient to vary across values of \(X_i\):
\[
Y_i=\alpha_{X_i}+\beta_{X_i}D_i+\varepsilon_{\mathrm{str},i},
\qquad M_i=0,\qquad
E[\varepsilon_{\mathrm{str},i}\mid D_i,X_i,M_i=0]=0.
\]

Among complete cases with \(X_i=x\), untreated units have mean outcome \(\alpha_x\), and treated units have mean outcome \(\alpha_x+\beta_x\). Thus, \(\beta_x\) is their observed difference in means. Whether this difference identifies the causal effect \(\tau\) depends on the missing mechanism. The residual \(\varepsilon_{\mathrm{str},i}\) captures individual deviations from these group means.

Under mechanisms A and B, \(\beta_{-1}=\beta_0=\beta_1=\tau=1\), as reflected in Table~\ref{tab:motivating-subset-regressions}. The baseline means \(\alpha_x\) differ: \((\alpha_{-1},\alpha_0,\alpha_1)=(-1,0,1)\) under A and \((-0.5,0,1.5)\) under B. A linear \(X_i\) term captures \(\alpha_x\) under A but not under B. Thus, the pooled complete-case regression on \(D_i\) and linear \(X_i\) consistently estimates \(\tau\) under A but not under B, due to the misspecification of the control variables $X_i$ under B.

Under mechanism B, selection into complete cases depends on \(U_i\): at \(X_i=\pm1\), units with \(U_i=1\) are more likely to be complete than units with \(U_i=-1\). The oracle regression error therefore has a nonzero conditional mean among complete cases, \(\mathbb E[U_i+\xi_i\mid X_i,D_i,M_i=0]=X_i^2/2\). Since \(Y_i(0)=X_i+U_i+\xi_i\), the mean untreated outcome becomes \(\mathbb E[Y_i(0)\mid X_i,M_i=0]=X_i+X_i^2/2\), the added \(X_i^2/2\) cannot be represented by the intercept and linear \(X_i\) term in the fitted regression. Thus, correct oracle specification, linear $X_i$, does not ensure correct specification among complete cases. 

Alternatively, IPW can estimate causal effects without specifying how \(X_i\) enters the outcome regression \citep{rosenbaum1983central,hirano2003efficient}. Under mechanism B, the IPW estimator in Figure~\ref{fig:motivating-simulation} weights each complete case by the inverse probability of its observed treatment given \(X_i\) and \(M_i=0\). Its estimates are centered near the complete-case ATE, which equals the constant population ATE\footnote{Appendix~\ref{app:correct-treatment-weighting} explains how IPW and saturated regression weight the within-\(X\) comparisons differently under mechanisms B and C.}.

\begin{figure}[htbp]
\centering
\resizebox{0.92\textwidth}{!}{%
\begin{tikzpicture}[x=73.7pt,y=73.7pt,
  bdn/.style={circle,draw=black,fill=white,line width=1.65pt,minimum size=35.6pt,inner sep=0pt,outer sep=0pt,
    font=\fontsize{16.8}{20}\selectfont\rmfamily},
  bde/.style={line width=1.95pt,shorten <=3.2pt,shorten >=3.2pt,
    -{Straight Barb[length=10.8pt,width=9.6pt]}}]
\useasboundingbox (-2.049,-1.452) rectangle (6.146,3.134);
\begin{scope}[xshift=0pt]
\node[font=\fontsize{16.8}{20}\selectfont\rmfamily] at (0,2.91) {Mechanism B};
\node[bdn] (X) at (-1.05,0.75) {X};
\node[bdn,dash pattern=on 6.6pt off 6.6pt] (U) at (1.05,0.75) {U};
\node[bdn] (D) at (-1.05,-0.75) {D};
\node[bdn] (Y) at (1.05,-0.75) {Y};
\node[bdn] (M) at (0,1.95) {M};
\foreach \a/\b in {X/D,X/Y,D/Y,X/M,U/Y,U/M}
  \draw[bde] (\a) -- (\b);
\end{scope}
\begin{scope}[xshift=301.8pt]
\node[font=\fontsize{16.8}{20}\selectfont\rmfamily] at (0,2.91) {Mechanism C};
\node[bdn] (X) at (-1.05,0.75) {X};
\node[bdn,dash pattern=on 6.6pt off 6.6pt] (U) at (1.05,0.75) {U};
\node[bdn] (D) at (-1.05,-0.75) {D};
\node[bdn] (Y) at (1.05,-0.75) {Y};
\node[bdn] (M) at (0,1.95) {M};
\foreach \a/\b in {X/D,X/Y,D/Y,X/M,U/Y,U/M,D/M}
  \draw[bde] (\a) -- (\b);
\end{scope}
\end{tikzpicture}%
}
\caption{Causal Diagrams for Missing Mechanisms B and C}
\label{fig:motivating-dags-bc}
\end{figure}
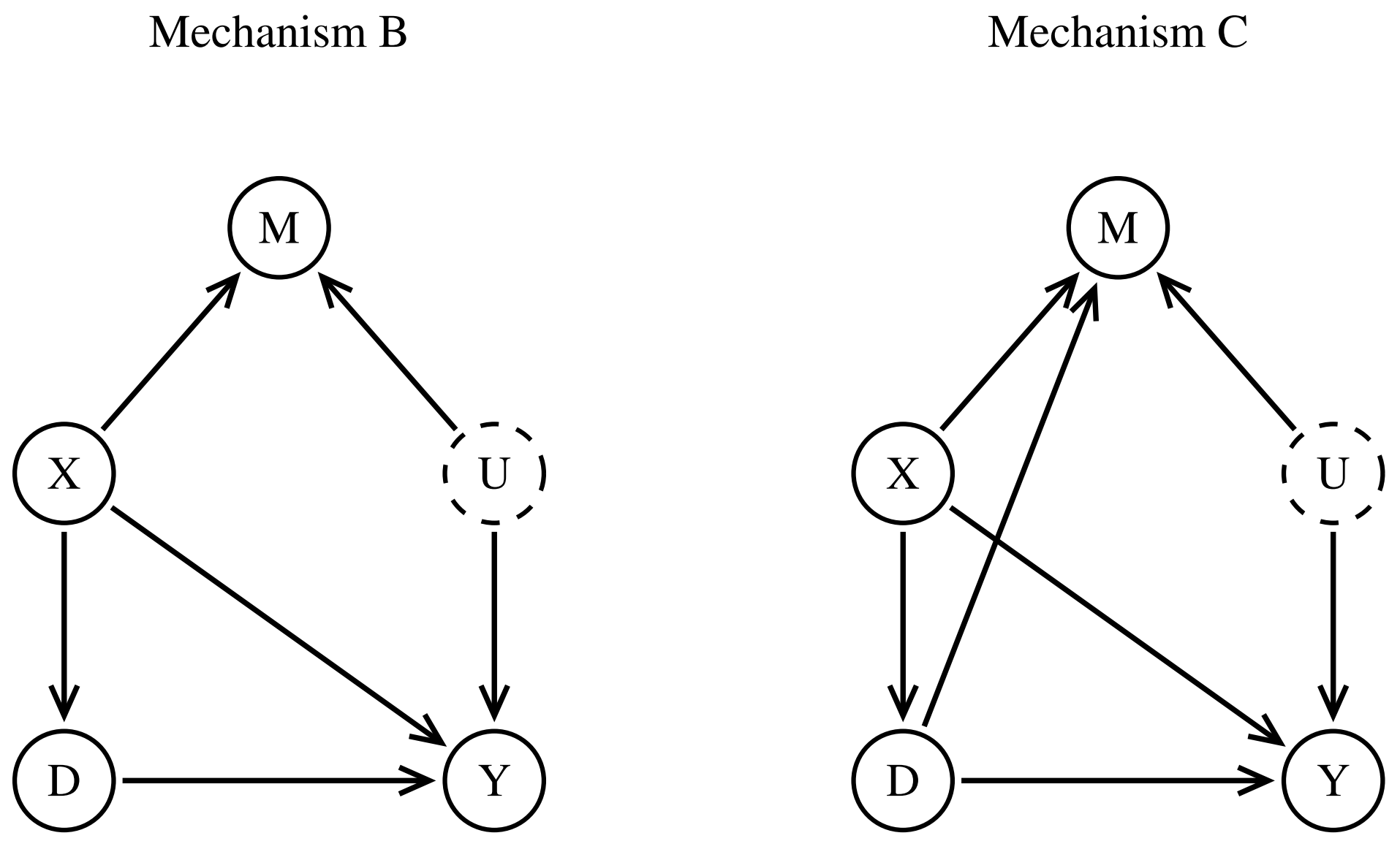

\subparagraph{Back-door adjustment criterion} We next compare mechanisms B and C. In Figure~\ref{fig:motivating-simulation}, the IPW complete-case estimates are centered near \(\tau=1\) under B but not C, although treatment probabilities are correctly specified within each \(X_i\) stratum under both mechanisms. The difference is that \(\beta_x=\tau=1\) in every \(X_i\) stratum under B, whereas \(\beta_{-1}\) and \(\beta_1\) do not identify \(\tau\) under C.

To assess why the within-stratum comparisons differ, we use Directed Acyclic Graphs (DAGs) from \cite{pearl1995causal}, including the missingness indicator \(M_i\) as a node, to trace the paths that conditioning on complete-case status opens. Figure~\ref{fig:motivating-dags-bc} presents the DAGs for mechanisms B and C.

Under B, conditioning on the collider \(M_i\) opens the noncausal path \(D_i\leftarrow X_i\rightarrow M_i\leftarrow U_i\rightarrow Y_i\), but conditioning on \(X_i\) blocks it. Thus, \(\{X_i,M_i\}\) satisfies the adjustment criterion, yielding \(Y_i(d)\perp D_i\mid X_i,M_i\) for \(d\in\{0,1\}\)---the within-complete-case exchangeability condition emphasized by \citet{choi2020reply}. The within-\(X_i\) treatment contrast among complete cases therefore identifies \(\tau=1\).

Under C, conditioning on \(M_i=0\) opens the noncausal path \(D_i\rightarrow M_i\leftarrow U_i\rightarrow Y_i\), which conditioning on \(X_i\) does not block. Thus, \(\{X_i,M_i\}\) fails the adjustment criterion's path-blocking requirement. In this simulation, the within-\(X_i\) treatment contrasts at \(X_i=-1\) and \(X_i=1\) do not identify \(\tau\).

To sum up, all three mechanisms are NMAR and, by construction, share the same probability of missingness conditional on \(X_i\), yet they differ in whether complete-case regression consistently estimates the population ATE, for two reasons. Under A, the within-\(X_i\) comparisons identify \(\tau\) and the pooled linear regression recovers it. Under B, the comparisons still identify \(\tau\), and entering \(X_i\) categorically or using treatment IPW restores consistent estimation. Under C, the comparisons fail to identify \(\tau\) in two strata, and neither approach recovers it. These differences motivate the formal results in Section~\ref{sec:causal-conditions}, which establish sufficient causal and modeling conditions for complete case regressions to consistently estimate the population ATE.

\section{Identification of Population ATE under Complete-Case Regression}
\label{sec:causal-conditions}

In this section, we first prove that two conditions are sufficient for complete-case regression to consistently estimate the complete-case average treatment effect: (a) treatment comparisons within covariate strata identify the stratum-specific causal effects, and (b) the fitted estimator aggregates those comparisons correctly. Under a constant unit-level treatment effect, the complete-case ATE equals the population ATE. With heterogeneous effects, recovering the population ATE requires a third condition: the conditional effects identified among complete cases must also transport to the population, and complete cases must cover the population covariate support. These effects can then be averaged over the population covariate distribution.

Our proof develops in four parts. First, we define the oracle regression benchmark and distinguish missing patterns from missing mechanisms. Second, we use a DAG with a missing indicator to assess within-covariate treatment comparisons among complete cases. Third, we examine when outcome specification or treatment-propensity weighting allows regression to aggregate those comparisons correctly. Finally, we connect the complete-case ATE to the population ATE. Section~\ref{sec:reweighting-hte} develops reweighting under heterogeneous effects.

\subsection{Oracle Regression and Missingness}
\label{subsec:oracle-regression}

We first distinguish the missing mechanism from the missing pattern. A missing pattern describes which variables are missing for a unit. A missing mechanism specifies how the probability of that pattern depends on the full-data values \citep{little2019statistical}. Missing completely at random (MCAR), missing at random (MAR), and not missing at random (NMAR) classify missing mechanisms\footnote{Under MCAR, missingness probabilities are independent of data values. Under MAR, missingness may depend on variables whose values are observed, but not on the values that are missing under that pattern. NMAR permits dependence on those missing values.}. While frequently used, these classifications usually do not specify which variables are missing or, more importantly for our framework, what causes the missingness.

\paragraph{Missing Pattern and Missing Mechanism} In our motivating simulation, mechanisms A, B, and C are all NMAR and share the same missing pattern: \(X_i\) is missing for some units. They differ, however, in how observation of \(X_i\) depends on \(X_i\), \(U_i\), and \(D_i\). These differences determine whether complete-case regression consistently estimates \(\tau\) and, if not, whether saturation or treatment IPW can help. The NMAR label alone therefore does not distinguish the assumptions that matter for these estimators\footnote{The importance of specifying the missing mechanism precisely has also been raised by \citet{seaman2013what}, who distinguish different formal definitions of MAR.}.

Formally, for a unit $i$, let \(X_i=(X_i^1,\ldots,X_i^p)\) collect the
pretreatment covariates in the researcher's chosen adjustment set\footnote{Throughout, we take the chosen adjustment set to include any pretreatment treatment-effect moderators used in the analysis.}, and let \(V_i=(Y_i,D_i,X_i^1,\ldots,X_i^p)\) additionally contain the treatment indicator $D_i$ and the outcome $Y_i$.

\begin{definition}[Missing pattern and omnibus missing indicator]
For unit $i$, let \(M_i^Y\), \(M_i^D\), and \(M_i^{X^j}\), for \(j=1,\ldots,p\), equal one when the corresponding value of \(Y_i\), \(D_i\), or \(X_i^j\) is missing, and zero otherwise. The vector
\[
\mathbf M_i=(M_i^Y,M_i^D,M_i^{X^1},\ldots,M_i^{X^p})
\]
is unit \(i\)'s missing pattern. Complete-case regression uses only units for which every component of \(V_i\) is observed. To represent this restriction, we summarize the missing pattern with an omnibus indicator. Define
\[
M_i=\mathbf 1\!\left\{M_i^Y+M_i^D+
\sum_{v=1}^{p}M_i^{X^v}>0\right\}.
\]
Thus \(M_i=0\) if and only if \(Y_i\), \(D_i\), and every component of \(X_i\) are observed\footnote{The omnibus missing indicator corresponds to \citeauthor{wooldridge2002econometric}'s (\citeyear{wooldridge2002econometric}, p.~553) selection indicator with reversed coding: \(M_i=1-s_i\). Both indicators are defined for all units in the underlying population, including those excluded from estimation. They can therefore represent listwise deletion due to missing values in the outcome, regressors, or both.}.
\end{definition}

If a researcher changes the adjustment set, the missing indicator and complete-case sample may also change, so the researcher must redefine the indicator and reassess the conditions for the revised specification.

Appendix~\ref{app:missing-mechanism} provides the conventional probabilistic definition of a missing mechanism \citep{little2019statistical,seaman2013what}. In the next subsection, we give the mechanism a causal formulation, building on graphical models of missing data \citep{mohan2021graphical}.

Our analysis from now on focuses on the missing mechanism: what causes a unit to enter the complete-case sample and how that selection affects treatment effect identification and estimation. We represent selection into this sample using the omnibus missing indicator \(M_i\), regardless of which variable makes a unit incomplete.

\paragraph{Oracle Samples} To assess how selection into complete cases affects the regression workflow, we first define its benchmark before listwise deletion: the oracle sample, in which all required values are observed and for which the researcher chooses an adjustment set and regression specification. We distinguish two versions of these sampled units to connect the regression benchmark to the causal effect of interest.

The oracle potential-outcome sample contains both potential outcomes and lets us define the causal effect. The oracle sample contains each unit's realized outcome, treatment, and required covariates, including values missing in the observed data. We call the chosen regression applied to the oracle sample the oracle regression and ask whether fitting the same specification to complete cases recovers the population ATE that this adjustment set identifies.

We maintain Assumption~\ref{ass:sutva} (SUTVA) to link potential outcomes to each unit's realized outcome, and denote the population distribution from which the oracle potential-outcome sample is drawn by \(P_{\mathrm{po}}\). The assumption and formal definitions of both oracle samples are in Appendix~\ref{app:adjustment-criterion}.

\paragraph{Oracle Regression} The target causal estimand is the population average treatment effect:
\[
\tau\equiv\mathbb E_{P_{\mathrm{po}}}[Y_i(1)-Y_i(0)].
\]
We take the researcher's chosen oracle regression as our benchmark:
\begin{equation}
\label{eq:oracle-regression}
Y_i=\beta_{\mathrm{or}}D_i+b(X_i)^\top\gamma_{\mathrm{or}}+\varepsilon_{\mathrm{or},i}.
\end{equation}
Here \(b(X_i)\) contains an intercept and the chosen covariate terms. \(\beta_{\mathrm{or}}\) and \(\gamma_{\mathrm{or}}\) are the population treatment and covariate coefficients, and \(\varepsilon_{\mathrm{or},i}\) is the residual. Their sample estimates are \(\widehat{\beta}_{\mathrm{or}}\) and \(\widehat{\gamma}_{\mathrm{or}}\).

We assume that the researcher's causal identification assumptions make \(X_i\) a valid adjustment set, so that the population ATE would be identified if all required values were observed. Even with a valid adjustment set, however, the oracle regression coefficient \(\beta_{\mathrm{or}}\) may not equal the population ATE. It equals \(\tau\) under a constant treatment effect with correctly specified covariate terms, or under a constant treatment propensity, as in a randomized experiment. With heterogeneous effects, it is instead a variance-weighted average of conditional effects \citep{angrist2009mostly,aronow2016regression}. We therefore target \(\tau\) directly without requiring \(\beta_{\mathrm{or}}=\tau\). Under standard OLS regularity conditions, the sample coefficient \(\widehat{\beta}_{\mathrm{or}}\) converges in probability to \(\beta_{\mathrm{or}}\). Although we focus on the population ATE, some results also apply to other causal estimands targeted by the oracle regression, as discussed after Proposition~\ref{prop:graphical-cc-ipw}.

After listwise deletion, the researcher fits the same specification to complete cases:
\begin{equation}
\label{eq:complete-case-regression}
Y_i=\beta_{\mathrm{cc}}D_i+b(X_i)^\top\gamma_{\mathrm{cc}}+\varepsilon_{\mathrm{cc},i},
\qquad M_i=0.
\end{equation}
Here \(\beta_{\mathrm{cc}}\) and \(\gamma_{\mathrm{cc}}\) are the population regression coefficients among complete cases, and \(\varepsilon_{\mathrm{cc},i}\) is the corresponding residual. Under standard OLS regularity conditions, the sample treatment coefficient \(\widehat{\beta}_{\mathrm{cc}}\) converges in probability to \(\beta_{\mathrm{cc}}\).

To summarize the problem, our question is whether complete-case regression consistently estimates the population ATE identified by the oracle regression's adjustment set. When \(\beta_{\mathrm{or}}=\tau\), this requires
\[
\beta_{\mathrm{cc}}=\beta_{\mathrm{or}}.
\]
When \(\beta_{\mathrm{or}}\) differs from \(\tau\), equality between the two regression coefficients would no longer be sufficient to recover the population ATE. In either case, we take the oracle adjustment set's validity as given and focus on the additional causal and modeling conditions needed for regression after listwise deletion to recover \(\tau\).

\subsection{DAG for Within-X Complete-Case Comparisons}

We first use a causal DAG to establish when treatment comparisons within covariate strata among complete cases identify their conditional treatment effects. The next subsection then examines when regression pools these comparisons correctly.

Consider complete cases with the same pre-treatment covariate values \(X_i=x\), and let \(\delta_{\mathrm{cc}}(x)\) denote their average treatment effect. When both treatment arms have positive probability in this stratum, the population coefficient from regressing \(Y_i\) on an intercept and \(D_i\) equals the treated-minus-control difference in mean outcomes. We show when this difference identifies \(\delta_{\mathrm{cc}}(x)\).

We augment the causal DAG for the oracle regression variables \(Y_i,D_i,X_i\) with the omnibus missing indicator \(M_i\) and its causes. This recasts \emph{selection on being observed}, the restriction that defines complete-case regression, as \emph{selection on the observable indicator} \(M_i\), specifically \(M_i=0\). As Figure~\ref{fig:motivating-dags-bc} illustrates, this conditioning can open noncausal paths, where \(\{X_i,M_i\}\) satisfies the adjustment criterion under mechanism B but not mechanism C.

\paragraph{Adjustment criterion} To assess whether \(\{X_i,M_i\}\) is a valid adjustment set, we use the criterion of \citet[Definition~5]{shpitser2010validity}, a generalization of the back-door criterion of \citet{pearl1995causal} (see also \citealt{cinelli2024crash}), stated here for a single treatment \(D\) and outcome \(Y\).

\begin{definition}[Causal missingness mechanism for listwise deletion]
Let \(G\) be a causal DAG containing the oracle variables \(Y_i,D_i,X_i\), the omnibus missing indicator \(M_i\), and any other relevant observed or latent causes. The causal missingness mechanism for listwise deletion consists of the causal relationships leading to \(M_i\) and the process determining it, while \(G\) represents its causal structure.
\end{definition}

\begin{definition}[Adjustment criterion for a single treatment]
\label{def:adjustment-criterion}
Let \(G\) be a causal DAG with treatment \(D\), outcome \(Y\), and a set \(\mathcal Z\) of variables disjoint from \(\{D,Y\}\). The set \(\mathcal Z\) satisfies the adjustment criterion if:
\begin{enumerate}
    \item \(\mathcal Z\) contains no node other than \(D\) on a directed path from \(D\) to \(Y\), nor any descendant of such a node\footnote{A node counts as its own descendant. Unlike the back-door criterion, the adjustment criterion can permit a descendant of \(D\) on a branch that does not lead to \(Y\), provided all noncausal paths are blocked. See Appendix~\ref{app:adjustment-criterion} for the general forms.}.
    \item Conditioning on \(\mathcal Z\) blocks every noncausal path from \(D\) to \(Y\)\footnote{A noncausal path is one that is not directed entirely from \(D\) to \(Y\) \citep[Section~2]{shpitser2010validity}}.
\end{enumerate}
\end{definition}

A path is blocked given \(\mathcal Z\) if it contains a noncollider in \(\mathcal Z\), or a collider with neither itself nor any descendant in \(\mathcal Z\) \citep[Section~2]{shpitser2010validity}. Applying the criterion to \(\mathcal Z=\{X_i,M_i\}\) gives the following result.

\begin{proposition}[Adjustment criterion, conditional ignorability, and within-\(X\) complete-case comparison]
\label{prop:adjustment-ignorability}
Write \(\delta_{\mathrm{cc}}(x)=\mathbb E[Y_i(1)-Y_i(0)\mid X_i=x,M_i=0]\). If \(\{X_i,M_i\}\) satisfies the adjustment criterion relative to \((D_i,Y_i)\) in \(G\), then:
\begin{enumerate}
    \item[(i)] in every causal model inducing \(G\), for each \(d\in\{0,1\}\),
\[
Y_i(d)\perp D_i\mid X_i,M_i;
\]
    \item[(ii)] for any
complete-case stratum \(X_i=x\) with treatment overlap, the population
coefficient on \(D_i\) from regressing \(Y_i\) on an intercept and \(D_i\)
equals
\[
\mathbb E[Y_i\mid D_i=1,X_i=x,M_i=0]
-\mathbb E[Y_i\mid D_i=0,X_i=x,M_i=0]
=\delta_{\mathrm{cc}}(x).
\]
With a constant unit-level treatment effect \(Y_i(1)-Y_i(0)=\tau\) for every unit, \(\delta_{\mathrm{cc}}(x)=\mathbb E[Y_i(1)-Y_i(0)\mid X_i=x,M_i=0]=\tau\).
\end{enumerate}
\end{proposition}

Proof in Appendix~\ref{app:adjustment-criterion}.

The proposition shows that \(\delta_{\mathrm{cc}}(x)\) is identified when the covariates and missing indicator satisfy the adjustment criterion and both treatment arms are represented within each complete-case covariate stratum, which provides the foundation for aggregating the conditional effects later.

Although the criterion is formal, checking it adds little to the diagram work that already supports the oracle regression. A researcher who uses a DAG to choose its adjustment set can add the missing indicator \(M_i\) and its causes, including relevant unobserved ones, to that same diagram and ask whether the original set still works among complete cases.

\subsection{Pooling Within-\(X\) Comparisons into Regression}

As mechanism B in Section~\ref{sec:motivating-simulation} illustrates, valid within-covariate treatment comparisons do not ensure that the pooled regression consistently estimates the complete-case ATE. The following two propositions give sufficient conditions under which unweighted complete-case regression does so.

\begin{proposition}[Correct specification and the complete-case ATE]
\label{prop:cc-ols-preservation}
Suppose \(\{X_i,M_i\}\) satisfies the adjustment criterion and complete-case treatment overlap holds (Assumption~\ref{ass:cc-treatment-overlap}). Consider the complete-case regression
\[
Y_i=g(X_i)^\top\alpha_{\mathrm{cc}}
+D_i h(X_i)^\top\beta_{\mathrm{cc}}+\varepsilon_{\mathrm{cc},i},
\qquad M_i=0,
\]
where \(g(X_i)\) contains an intercept and the chosen baseline covariate terms, and \(h(X_i)\) contains an intercept and the chosen terms interacted with treatment. Suppose this regression is correctly specified in conditional mean:
\[
\mathbb E[\varepsilon_{\mathrm{cc},i}\mid D_i,X_i,M_i=0]=0.
\]
Then, under SUTVA (Assumption~\ref{ass:sutva}),
\[
\delta_{\mathrm{cc}}(x)=h(x)^\top\beta_{\mathrm{cc}},
\]
and averaging over the complete-case covariate distribution gives the complete-case ATE, \(\tau_{\mathrm{cc}}\):
\[
\tau_{\mathrm{cc}}\equiv\mathbb E[h(X_i)^\top\beta_{\mathrm{cc}}\mid M_i=0]
=\mathbb E[Y_i(1)-Y_i(0)\mid M_i=0].
\]
Let \(\widehat\beta_{\mathrm{cc}}\) be the complete-case OLS estimate and \(n_{\mathrm{cc}}\) the number of complete cases. Under standard regularity conditions (Assumptions~\ref{ass:cc-ols-regularity} and~\ref{ass:cc-population-regularity} in Appendix~\ref{app:adjustment-criterion}),
\[
\widehat\tau_{\mathrm{cc}}\equiv\frac{1}{n_{\mathrm{cc}}}
\sum_{i:M_i=0}h(X_i)^\top\widehat\beta_{\mathrm{cc}}
\xrightarrow{p}\tau_{\mathrm{cc}}.
\]
\end{proposition}

Proof in Appendix~\ref{app:proof-cc-ols-preservation}.

When \(h(X_i)=1\), there are no treatment--covariate interactions, \(\beta_{\mathrm{cc}}\) is a scalar, and the conditional treatment effect is the same for every \(X_i\), so the coefficient on \(D_i\) is that common effect. Equation~\eqref{eq:complete-case-regression} is this no-interaction case, and Proposition~\ref{prop:cc-ols-preservation} generalizes it by allowing \(\beta_{\mathrm{cc}}\) to be a vector. When interactions are included, \(h(X_i)\) varies with the covariates, the coefficient on \(D_i\) alone describes only the reference group, and the complete-case ATE must be recovered by averaging the fitted conditional effects over the covariate distribution observed among complete cases.

Proposition~\ref{prop:cc-ols-preservation} identifies two requirements for consistently estimating the complete-case ATE, \(\tau_{\mathrm{cc}}\): valid within-covariate comparisons and their correct aggregation. The adjustment criterion, together with treatment overlap, makes those comparisons causally valid. Correct specification of the complete-case outcome regression makes the aggregation valid\footnote{Proposition~\ref{prop:cc-ols-preservation} generalizes the specification and selection argument of \citet{arel2018can}: their indicator--error independence condition, together with correct oracle specification, implies the complete-case zero conditional mean condition required here. Our graphical condition additionally establishes the causal interpretation as the complete-case ATE, which their coefficient-preservation result does not address. See Appendix~\ref{app:oracle-mean-preservation} for a discussion.}: it allows the fitted conditional treatment effects to be averaged over complete cases, and with the stated regularity conditions that average is consistent for \(\tau_{\mathrm{cc}}\).

However, correct specification after listwise deletion may be harder to establish than researchers expect. A regression correctly specified for the oracle sample may be misspecified after deletion because selection can change how the mean outcome depends on the covariates \citep{heckman1979selection}. Section~\ref{sec:motivating-simulation}'s Mechanism B illustrates this: a linear term for \(X_i\) is sufficient in the oracle regression, but the complete-case regression must capture a separate baseline mean for each \(X_i\) stratum. Two practical responses, a more flexible outcome model and residual diagnostics for the specification, are reserved for future work. The main takeaway from Proposition~\ref{prop:cc-ols-preservation} is instead interpretive: whether the specification remains adequate after deletion may require assumptions about unobserved outcome determinants and how they affect missingness, beyond the determinants the oracle regression already includes.

Treatment-propensity weighting, not IPW for missingness\footnote{IPW for missingness weights complete cases by the inverse probability of being a complete case to address selection due to missingness \citep{seaman2013review,hughes2019accounting}. Here, we weight by the inverse treatment probability among complete cases to pool their within-covariate treatment comparisons.}, offers another way to pool valid within-covariate comparisons without requiring a correctly specified outcome equation \citep{rosenbaum1983central,hirano2003efficient}. Appendix Proposition~\ref{prop:cc-ipw} states the standard IPW result for complete cases \citep{hirano2003efficient,choi2019comparison}. Here, we consider its use in a complete-case regression.

\begin{proposition}[Graphical criterion for complete-case IPW regression]
\label{prop:graphical-cc-ipw}
Suppose \(\{X_i,M_i\}\) satisfies the adjustment criterion and complete-case treatment overlap holds (Assumption~\ref{ass:cc-treatment-overlap}).

Among complete cases, consider the IPW-weighted working regression \[Y_i=\alpha+\beta D_i+X_i^\top\gamma+\varepsilon_{\mathrm{ipw},i}, \qquad
\omega_i=\frac{D_i}{e_0(X_i)} +\frac{1-D_i}{1-e_0(X_i)}, \] where \(e_0(x)=P(D_i=1\mid X_i=x,M_i=0)\). Let \(\beta^*\) denote the population counterpart of the IPW-weighted regression coefficient on \(D_i\) among complete cases. Then, under SUTVA and standard regularity conditions (Assumptions~\ref{ass:sutva}, \ref{ass:cc-population-regularity}, and~\ref{ass:cc-wls-moments} in Appendix~\ref{app:adjustment-criterion}),
\[
\beta^*=\mathbb E[Y_i(1)-Y_i(0)\mid M_i=0].
\]
If the propensity score is estimated consistently, in the sense that the estimated inverse weights converge uniformly (Assumption~\ref{ass:cc-ipw-weight-consistency}), the sample coefficient \(\widehat\beta\), computed using \(\widehat e_0\), converges in probability to \(\beta^*\).
\end{proposition}

The proposition follows from Proposition~\ref{prop:adjustment-ignorability} and Proposition~\ref{prop:cc-ipw-wls}, which applies \citet{gabriel2024inverse} to complete cases.

Among complete cases, the weights make each treatment arm reproduce the complete-case covariate distribution and set the weighted treatment probability to one-half within every supported stratum \(X_i=x\) \citep[Section~2.4]{hernanRobins2020}. The weighted regression therefore pools the valid within-\(X\) comparisons without the correctly specified complete-case outcome mean that Proposition~\ref{prop:cc-ols-preservation} requires, relying instead on the estimated propensity weights.

In practice, weighting adds one input, a propensity model for treatment among complete cases. Its inverse weights are informative only where the data have treatment overlap, and the within-\(X\) comparisons still need the adjustment criterion in the augmented DAG. As the motivating simulation shows, weighting addresses the pooling problem under mechanism B but cannot repair the invalid comparisons under mechanism C.

\begin{corollary}[Constant complete-case propensity: unweighted OLS equals IPW]
\label{cor:cc-ols-ipw-equivalence}
Suppose the conditions of Proposition~\ref{prop:graphical-cc-ipw} hold, and additionally there exists \(\bar e_0\in(0,1)\) such that \(e_0(x)=\bar e_0\) for every \(x\) in the complete-case support. Then, for any chosen \(g(X_i)\) containing an intercept, the unweighted additive complete-case regression
\[
Y_i=g(X_i)^\top\alpha_{\mathrm{cc}}+\beta_{\mathrm{cc}}D_i+\varepsilon_{\mathrm{cc},i},\qquad M_i=0,
\]
has \(\beta_{\mathrm{cc}}=\tau_{\mathrm{cc}}\) regardless of whether \(g(X_i)^\top\alpha_{\mathrm{cc}}\) correctly specifies the complete-case baseline mean, and, under Assumption~\ref{ass:cc-ols-regularity}, \(\widehat\beta_{\mathrm{cc}}\xrightarrow{p}\tau_{\mathrm{cc}}\): unweighted OLS coincides with IPW whenever every complete-case unit shares the same treatment propensity.
\end{corollary}

Proof in Appendix~\ref{app:proof-cc-ols-ipw-equivalence}.

A constant complete-case propensity makes \(D_i\) exactly mean-independent of any function of \(X_i\) among complete cases, so adding, omitting, or misspecifying baseline controls does not affect the consistency of the treatment coefficient \citep{lin2013agnostic}. Unlike Proposition~\ref{prop:cc-ols-preservation}, this route does not require correctly specifying the shape of treatment-effect heterogeneity: the additive specification \(h(X_i)=1\) recovers the complete-case ATE regardless of how \(\delta_{\mathrm{cc}}(x)\) varies with \(x\).

Finally, we show in Appendix~\ref{app:proof-cc-ols-preservation} that Propositions~\ref{prop:adjustment-ignorability}--\ref{prop:graphical-cc-ipw} still hold when treatment itself affects missingness, e.g. when treatment induces attrition.

\paragraph{Connection to the Oracle Target} Both regression approaches consistently estimate the complete-case ATE under their respective assumptions\footnote{The researcher's target may be another causal estimand, such as the average treatment effect on the treated (ATT). Proposition~\ref{prop:cc-ols-preservation} extends to the complete-case ATT by averaging the fitted conditional effects over treated complete cases. Under Corollary~\ref{cor:cc-ols-ipw-equivalence}, the complete-case ATE, ATT, and ATU coincide because treatment is independent of covariates among complete cases. Proposition~\ref{prop:graphical-cc-ipw} targets the complete-case ATE, and the ATT requires different weights. Connecting any complete-case estimand to its population target requires additional assumptions.}. To recover the population ATE, we also need to establish that the complete-case ATE equals the population ATE. The next subsection gives conditions for this connection under constant and heterogeneous treatment effects.

\subsection{When Complete-Case Regression Recovers the Population ATE}
\label{subsec:population-ate}

Under a constant unit-level treatment effect, the connection is immediate: \(\tau_{\mathrm{cc}}=\tau\). Either regression approach therefore consistently estimates the population ATE under its respective assumptions.

\begin{corollary}[Constant unit-level treatment effect]
\label{cor:graphical-cc-ipw-constant-effect}
Suppose there is a constant treatment effect for every unit, \(Y_i(1)-Y_i(0)=\tau\), and \(P(M_i=0)>0\). Then
\[\mathbb E[Y_i(1)-Y_i(0)\mid M_i=0]=\mathbb E[Y_i(1)-Y_i(0)]=\tau.\]

Under the conditions of Proposition~\ref{prop:cc-ols-preservation}, \(\tau_{\mathrm{cc}}=\tau\), and \(\widehat{\tau}_{\mathrm{cc}}\) converges in probability to \(\tau\).

Under the conditions of Proposition~\ref{prop:graphical-cc-ipw}, \(\beta^*=\tau\). Under Assumption~\ref{ass:cc-ipw-weight-consistency} (estimated inverse-weight consistency), \(\widehat\beta\) also converges in probability to \(\tau\).
\end{corollary}

Under the conditions of Corollary~\ref{cor:cc-ols-ipw-equivalence}, \(\beta_{\mathrm{cc}}=\tau\) as well, and \(\widehat\beta_{\mathrm{cc}}\) converges in probability to \(\tau\).

With heterogeneous effects, however, \(\tau_{\mathrm{cc}}\) may not equal the population ATE, \(\tau\). Recovering the latter requires connecting the effects among complete cases to those in the population.

To make that connection, we assume \emph{mean exchangeability of selection and treatment effect heterogeneity}: at every value of the covariates used for transport, the average treatment effect is the same among complete cases and in the population. Effects may still vary across those covariates. Within one value, selection need only be unrelated to the effect. This condition also relaxes the outcome-response ignorability assumption of \citet{huangLiu2026}.\footnote{For full-covariate transport, given valid within-\(X_i\) treatment comparisons in both the population and complete cases, outcome-response ignorability \citep{yangWangDing2019,huangLiu2026} is sufficient for mean exchangeability but not necessary. For example, selection on an unobserved characteristic that raises both potential outcomes equally changes their conditional means while preserving the conditional average treatment effect.} This is more plausible when the observed covariates capture characteristics that influence both completeness and responses to treatment, and it may fail when an unobserved characteristic does \citep{egamiHartman2023,chenHuang2026}.

\begin{corollary}[Heterogeneous treatment effects]
\label{cor:cc-effect-transport}
Suppose the conditions of Proposition~\ref{prop:adjustment-ignorability} hold with adjustment covariates \(X_i\). Let \(W_i\) be a subvector of \(X_i\), possibly \(X_i\) itself, and let \(F_{\mathrm{po}}^W\) denote its population distribution. Suppose \(\mathbb E[|Y_i(d)|]<\infty\) for \(d\in\{0,1\}\). Further, suppose \emph{mean exchangeability of selection and treatment effect heterogeneity given \(W_i\)} holds: for almost every \(w\) with respect to \(F_{\mathrm{po}}^W\),
\[
\mathbb E[Y_i(1)-Y_i(0)\mid W_i=w,M_i=0]
=\mathbb E[Y_i(1)-Y_i(0)\mid W_i=w].
\]
Suppose also that \(P(M_i=0\mid W_i=w)>0\) for almost every such \(w\). Define
\[
\tau_{\mathrm{cc}}(w)
=\mathbb E[Y_i(1)-Y_i(0)\mid W_i=w,M_i=0].
\]
Then
\[
\tau=\int\tau_{\mathrm{cc}}(w)\,dF_{\mathrm{po}}^W(w).
\]
Under Assumptions~\ref{ass:sutva} (SUTVA) and~\ref{ass:cc-treatment-overlap} (complete-case treatment overlap),
\[
\tau_{\mathrm{cc}}(w)
=\int\delta_{\mathrm{cc}}(x)\,dF_{\mathrm{cc}}(x\mid W_i=w),
\]
where \(F_{\mathrm{cc}}(x\mid W_i=w)\) is the distribution of \(X_i\) among complete cases with \(W_i=w\), and
\[
\delta_{\mathrm{cc}}(x)
=\mathbb E[Y_i\mid D_i=1,X_i=x,M_i=0]
-\mathbb E[Y_i\mid D_i=0,X_i=x,M_i=0].
\]
\end{corollary}

Proof in Appendix~\ref{app:proof-cc-effect-transport}.

Applying Corollary~\ref{cor:cc-effect-transport} asks researchers for one new judgment. The first requirement, valid within-\(X_i\) comparisons, is the adjustment-criterion check already made for Proposition~\ref{prop:adjustment-ignorability}, and positive complete-case probability across the support of \(W_i\) is an overlap question the data can answer. The new judgment is mean exchangeability given \(W_i\), which data alone cannot establish. With all three in hand, researchers recover \(\tau\) by averaging \(\tau_{\mathrm{cc}}(w)\) over \(F_{\mathrm{po}}^W\), as Section~\ref{sec:reweighting-hte} discusses. Further, the covariates used for adjustment $X_i$ can be different from the covariate used for transport $W_i$. If transport covariates contain missingness, imputation as in \citet{huangLiu2026} may recover their target distribution under additional assumptions.

\begin{table}[htbp]
\centering
\begin{minipage}{0.90\textwidth}
\centering
\caption{From Oracle Regression to Complete-Case Estimation: Conditions and Checks}
\label{tab:complete-case-framework}
\begingroup
\small
\renewcommand{\arraystretch}{1.2}
\begin{tabular}{@{}p{0.20\linewidth}p{0.58\linewidth}p{\dimexpr0.22\linewidth-4\tabcolsep\relax}@{}}
\toprule
\textbf{Step} & \textbf{What researchers should do} & \textbf{Result} \\
\midrule
\raggedright Starting point: the oracle regression &
\raggedright Choose a regression specification whose adjustment set identifies the causal target if all required values were observed. Apply that specification among complete cases\textsuperscript{1}. &
\raggedright Subsection~\ref{subsec:oracle-regression} \tabularnewline
\addlinespace
\raggedright Box 1: Valid comparisons &
\raggedright Augment the original DAG with the omnibus missing indicator and its causes. Check the adjustment criterion and treatment overlap among complete cases. These checks establish valid within-stratum treatment comparisons. &
\raggedright Proposition~\ref{prop:adjustment-ignorability} \tabularnewline
\addlinespace
\raggedright Box 2: OLS pooling (correct specification) &
\raggedright Correctly specify the complete-case conditional outcome mean. Average fitted treatment effects over all, treated, or untreated complete cases to estimate the complete-case ATE, ATT, or ATU, respectively. &
\raggedright Proposition~\ref{prop:cc-ols-preservation} \tabularnewline
\addlinespace
\raggedright Alternative Box 2: IPW pooling &
\raggedright Estimate treatment propensity scores among complete cases and construct consistent inverse weights. IPW regression consistently estimates the complete-case ATE. &
\raggedright Proposition~\ref{prop:graphical-cc-ipw} \tabularnewline
\addlinespace
\raggedright Alternative Box 2: Constant propensity &
\raggedright If the complete-case treatment propensity is constant (for example, a randomized experiment with constant assignment probability and response independent of treatment within covariate strata), regress the outcome on treatment and any chosen baseline covariates. This regression consistently estimates the complete-case ATE. &
\raggedright Corollary~\ref{cor:cc-ols-ipw-equivalence} \tabularnewline
\addlinespace
\raggedright Box 3 under a constant unit-level treatment effect &
\raggedright A constant unit-level treatment effect makes the complete-case effect equal the population ATE. &
\raggedright Corollary~\ref{cor:graphical-cc-ipw-constant-effect} \tabularnewline
\addlinespace
\raggedright Box 3 under heterogeneous effects &
\raggedright Assess mean exchangeability of selection and treatment effect heterogeneity and complete-case support. Section~\ref{sec:reweighting-hte} develops how to average conditional effects over the population covariate distribution. &
\raggedright Corollary~\ref{cor:cc-effect-transport} \tabularnewline
\bottomrule
\end{tabular}
\par\smallskip
\begin{minipage}{\linewidth}
\footnotesize
\textit{Notes:} Boxes 1 and 2 apply under both constant and heterogeneous effects. Regularity conditions are presumed throughout.

\smallskip
\textsuperscript{1} The framework may extend to other oracle estimands, such as the LATE identified by an instrumental-variable regression. Such an extension requires assumptions ensuring instrument validity after selection and connecting the complete-case LATE to the oracle LATE.
\end{minipage}
\endgroup
\end{minipage}
\end{table}

\paragraph{Practical Advice} To sum up, empirical researchers can assess whether their complete-case regression recovers the population ATE by checking three boxes, as in Table~\ref{tab:complete-case-framework}. Box 1 asks whether within-covariate treatment comparisons remain valid after listwise deletion. Researchers who already use a DAG to justify their adjustment set need to add the missing indicator and its causes, then reassess the comparisons. Box 2 asks whether the regression appropriately aggregates those comparisons, either through a correctly specified complete-case outcome model or treatment-propensity weighting. Box 3 asks whether the complete-case effects connect to the population target. Under a constant unit-level treatment effect, the complete-case ATE equals the population ATE. With heterogeneous effects, researchers also need to assess the transport and support conditions and average the conditional effects over the population covariate distribution (Section~\ref{sec:reweighting-hte}). The next section shows how this framework incorporates and extends conventional guidance on complete-case regression.

\section{Revisiting Conventional Guidance on Complete-Case Regression}
\label{sec:specification-propensity}
In this Section, we first show how our framework recovers conventional guidance for missingness driven by regression covariates or the outcome. We follow \citet{arel2018can} in assuming correct oracle specification and their indicator--error independence condition, which together satisfy the specification requirement of Proposition~\ref{prop:cc-ols-preservation}. Our framework thus recovers their recommendations while allowing weaker specification and independence restrictions. We then use the collider--mediator dilemma to show why the justification for complete-case regression depends on the causal graph and the effect of interest.

\subsection{Recovering Familiar Recommendations}
\label{subsec:conventional-guidance}

To begin, we maintain treatment overlap, standard regularity conditions, and a constant unit-level treatment effect, so that valid complete-case treatment comparisons recover the population ATE through the correctly specified regression. The remaining question is whether the adjustment criterion holds after adding the missing indicator and its causes to the oracle DAG. The same graphical assessment still applies under heterogeneous effects (Subsection~\ref{subsec:population-ate} and Section~\ref{sec:reweighting-hte}), and when researchers use the IPW-regression route of Proposition~\ref{prop:graphical-cc-ipw}.

Table~\ref{tab:conventional-missing-guidance} reviews current recommendations about complete-case analysis (CCA), also known as listwise deletion, and multiple imputation, adapted from \citet{pepinsky2018note}. The pattern column identifies which variables contain missing values, while the mechanism column identifies what drives missingness. For example, the NMAR\((X)\) row describes missing values in \(X\) whose probability of being missing depends only on \(X\). This row records that complete-case regression remains consistent when missingness depends on the value of the missing regressor \citep{arel2018can}. We now use our augmented DAG to justify the CCA column and show that our theory reproduces the same recommendations.

\begin{table}[htbp]
\centering
\caption{Conventional Guidance on Complete-Case Analysis and Multiple Imputation}
\label{tab:conventional-missing-guidance}
\begin{tabular}{@{}cccc@{}}
\toprule
\textbf{Pattern} & \textbf{Mechanism} & \textbf{CCA} & \textbf{Imputation} \\
\midrule
\(X/Y\) & MCAR & \(\checkmark\) & \(\checkmark\) \\
\(X/Y\) & MAR\((X)\) & \(\checkmark\) & \(\checkmark\) \\
\(X\) & MAR\((X,Y)\) & \(\times\) & \(\checkmark\) \\
\(X\) & NMAR\((X)\) & \(\checkmark\) & \(\times\) \\
\(X/Y\) & NMAR\((X,Y)\) & \(\times\) & \(\times\) \\
\bottomrule
\end{tabular}
\par\smallskip
\begin{minipage}{\linewidth}
\footnotesize
\textit{Notes:} Adapted from \citet{pepinsky2018note}. \(X\) includes all regressors, including treatment \(D\). Pattern indicates which variables contain missing values; parentheses indicate the drivers of missingness. A check indicates consistency under the stated model assumptions; a cross indicates no such guarantee. For CCA, the checked rows rely on the specification and indicator--error independence conditions of \citet{arel2018can}.
\end{minipage}
\end{table}

Under MCAR, the missing indicator \(M_i\) is independent of the oracle outcome, treatment, and covariates. We represent it as an isolated node in the DAG, so adding \(M_i\) preserves identification by the original adjustment set.

\begin{figure}[htbp]
\centering
\begin{tikzpicture}[manuscript dag]
\node[manuscript dag node] (X) at (-0.302,1.116) {X};
\node[manuscript dag node] (D) at (-1.698,-1.116) {D};
\node[manuscript dag node] (Y) at (1.698,-1.116) {Y};
\node[manuscript dag node] (M) at (1.302,0.447) {M};
\path[manuscript dag edge]
  (X) edge (D) (X) edge (Y) (D) edge (Y)
  (X) edge (M) (D) edge (M);
\end{tikzpicture}
\caption{Selection Driven by Treatment and Covariates}
\label{fig:CCA-Unbiasedness}
\end{figure}
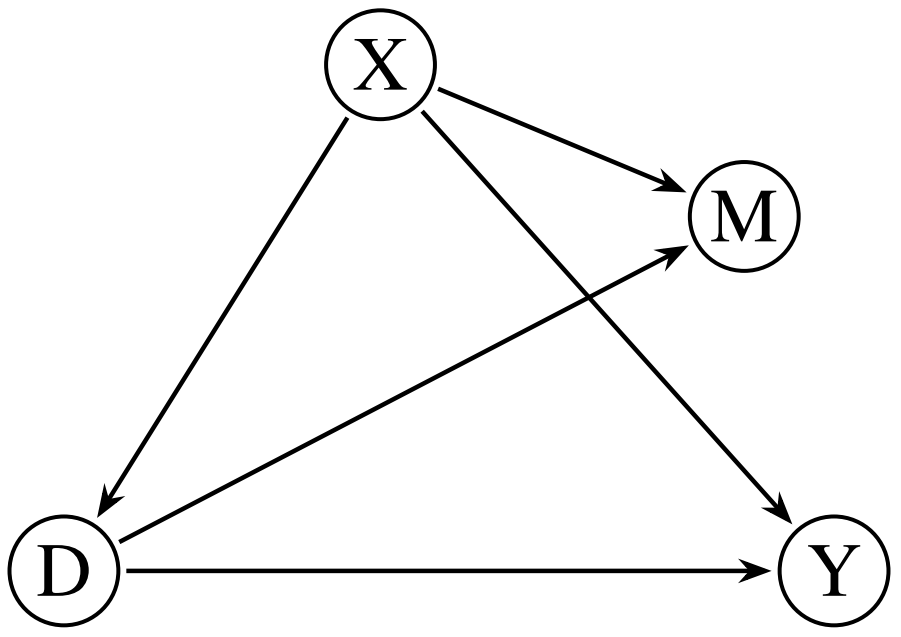

\paragraph{Missingness driven by treatment and covariates}
Figure~\ref{fig:CCA-Unbiasedness} shows a causal structure in which missingness depends only on treatment \(D_i\) and the confounder vector \(X_i\). The same structure represents three cases, each with a different variable missing (pattern) and the others observed, and with different MAR/NMAR classifications:

\begin{enumerate}
\item \textbf{Missing outcome:} \(Y_i\) contains missing values, but its missingness depends only on the observed \(D_i,X_i\). This is MAR and corresponds to the second row of Table~\ref{tab:conventional-missing-guidance}.
\item \textbf{Missing treatment:} \(D_i\) contains missing values, and its own value affects missingness. This is NMAR and corresponds to the fourth row, where the table's \(X\) includes treatment.
\item \textbf{Missing covariates:} Some covariates in \(X_i\) contain missing values, and missingness depends on those covariates' own values. This is also NMAR and corresponds to the fourth row.
\end{enumerate}

In the original DAG, adjusting for \(X_i\) blocks the noncausal path \(D_i\leftarrow X_i\rightarrow Y_i\), identifying the population ATE from the oracle sample. Restricting to complete cases additionally conditions on the collider \(D_i\rightarrow M_i\leftarrow X_i\), opening the path \(D_i\rightarrow M_i\leftarrow X_i\rightarrow Y_i\), which remains blocked by adjustment for \(X_i\). Moreover, \(M_i\) is on a separate branch from the causal path \(D_i\rightarrow Y_i\). Thus, \(\{X_i,M_i\}\) satisfies the adjustment criterion. Under the maintained assumptions, complete-case regression thus consistently estimates the population ATE in all three cases, recovering the recommendation that missingness driven by regressors need not invalidate regression estimates, whether the mechanism is MAR or NMAR \citep{arel2018can}.

\begin{figure}[htbp]
\centering
\begin{tikzpicture}[manuscript dag]
\node[manuscript dag node] (X) at (-1.244,1.116) {X};
\node[manuscript dag node] (D) at (-1.698,-1.116) {D};
\node[manuscript dag node] (Y) at (0.570,0) {Y};
\node[manuscript dag node] (M) at (1.698,0) {M};
\path[manuscript dag edge]
  (X) edge (D) (X) edge (Y) (D) edge (Y) (Y) edge (M);
\end{tikzpicture}
\caption{Selection Driven by the Outcome}
\label{fig:CCA-biasedness}
\end{figure}
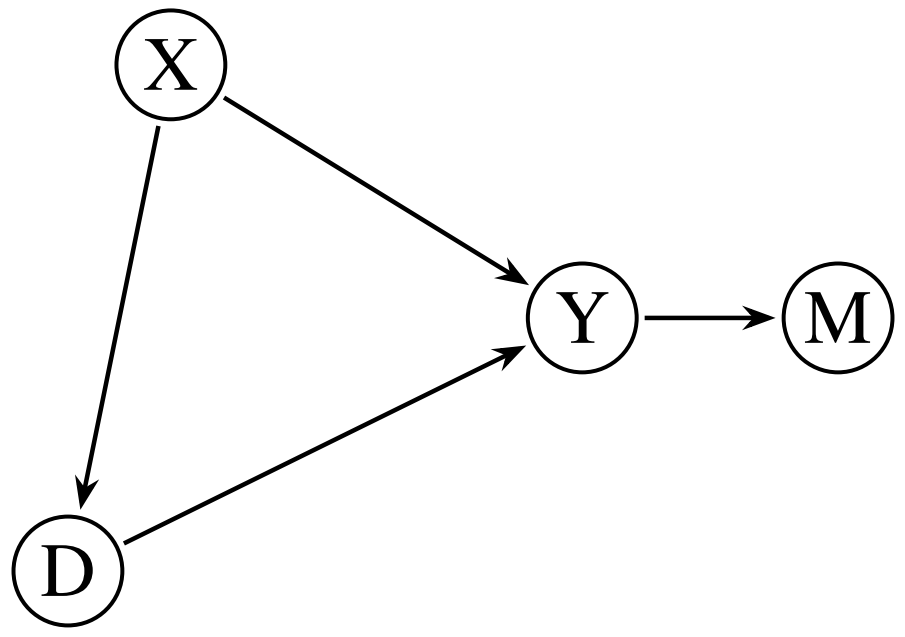

\paragraph{Missingness driven by the outcome}
Figure~\ref{fig:CCA-biasedness} retains the same treatment--outcome relationship and confounder, but missingness now instead depends only on \(Y_i\). This DAG likewise represents three cases with different classifications:

\begin{enumerate}
\item \textbf{Missing outcome:} \(Y_i\) contains missing values, and missingness depends on its own value. This is NMAR and illustrates the outcome-driven selection covered by the fifth row of Table~\ref{tab:conventional-missing-guidance}.
\item \textbf{Missing treatment:} \(D_i\) contains missing values, but missingness depends only on the observed \(Y_i\). This is MAR and corresponds to the third row.
\item \textbf{Missing covariates:} \(X_i\) contains missing values, but missingness depends only on the observed \(Y_i\). This is also MAR and corresponds to the third row.
\end{enumerate}

Restricting to complete cases conditions on \(M_i\), a descendant of \(Y_i\) on the causal path \(D_i\rightarrow Y_i\). Thus, \(\{X_i,M_i\}\) violates the first condition of the adjustment criterion, and complete-case regression is no longer guaranteed to consistently estimate the population ATE. This recovers the warning that outcome-driven missingness can invalidate complete-case regression \citep{heckman1979selection}, even under MAR \citep{arel2018can}.\footnote{A notable exception is a logistic outcome model: if selection depends only on the outcome, the treatment log odds ratio is preserved \citep{prentice1979logistic}, but risk differences, and thus the ATE, are not.}

Together, the two DAGs show that whether complete-case regression is justified depends on the missingness mechanism in the augmented DAG, rather than the missingness pattern or its MAR/NMAR classification.

\subsection{The Collider--Mediator Dilemma}
\label{sec:covariate-roles}

The preceding examples show how our framework recovers existing recommendations for complete-case regression. We now use it to examine a setting for which those recommendations provide no clear guidance: a variable that explains missingness and also mediates the treatment effect.

\citet{rivlin2025personality} study the relationship between the Big Five personality traits and political attitudes using a survey of Jewish Israelis. Their original dataset contains \RAnSource{} respondents. They use listwise deletion for all regressions, leaving 816 complete cases for specifications with sociodemographic controls\footnote{The OLS regressions include all five personality traits, with additional specifications controlling for gender, age group, education, and religious self-identification. }. Our illustration draws on their analysis of conscientiousness and support for a two-state solution, specifically the mediating role of authoritarian tendency.

The authors do not discuss the missingness mechanism underlying their data, so we retain the \RAn{} respondents with complete observations on all variables used in this illustration and impose additional missingness. We assume that respondents with below-average conscientiousness and above-average authoritarian tendencies have a 50\% probability of having their outcome value unreported\footnote{Research on social desirability shows that respondents may conceal politically sensitive preferences, while research on survey effort links conscientiousness to careless responding \citep{carminesNassar2021,zamarroEtAl2018}.}. The missingness mechanism is therefore known in our synthetic data. In practice, researchers must enter that mechanism themselves into their augmented DAG. The left panel of Figure~\ref{fig:example-dag-1} augments the paper's original mediation structure with the missing indicator and its assumed causes\footnote{The other personality traits and sociodemographic controls are omitted from the diagram.}.

\begin{figure}[htbp]
\centering
\begin{tabular}{c}
Left Panel: \(A\) as Mediator\\[4pt]
\begin{tikzpicture}[manuscript dag]
\node[manuscript dag node] (D) at (0,0) {D};
\node[manuscript dag node] (Y) at (2,0) {Y};
\node[manuscript dag node] (A) at (1,-1) {A};
\node[manuscript dag node] (M) at (3,-1) {M};
\path[manuscript dag edge]
  (D) edge (Y) (D) edge (A) (A) edge (Y)
  (A) edge (M) (D) edge (M);
\end{tikzpicture}
\end{tabular}\hspace{3em}%
\begin{tabular}{c}
Right Panel: \(A\) as Confounder\\[4pt]
\begin{tikzpicture}[manuscript dag]
\node[manuscript dag node] (D) at (0,0) {D};
\node[manuscript dag node] (Y) at (2,0) {Y};
\node[manuscript dag node] (A) at (1,-1) {A};
\node[manuscript dag node] (M) at (3,-1) {M};
\path[manuscript dag edge]
  (D) edge (Y) (A) edge (D) (A) edge (Y)
  (A) edge (M) (D) edge (M);
\end{tikzpicture}
\end{tabular}
\par\medskip
\begin{minipage}{0.9\linewidth}
\normalsize
\(A\): authoritarian tendency\\
\(D\): conscientiousness\\
\(Y\): support for a two-state solution\\
\(M\): missing indicator for \(Y\), with \(M=1\) when the response is missing
\end{minipage}
\caption{Authoritarian Tendency as a Mediator or Confounder}
\label{fig:example-dag-1}
\end{figure}

\paragraph{Why the mediator creates a dilemma}
While the original study distinguishes average direct and causal mediation effects, our illustration focuses on the total effect of conscientiousness on support for a two-state solution\footnote{The authors describe their analysis as ``causal mediation analysis,'' while acknowledging its suggestive nature \citep[p.~404]{rivlin2025personality}, so we use it here as an illustration.}. Under the assumed causal structure and correct regression specification, regressing \(Y_i\) on \(D_i\), with the baseline controls but without the mediator \(A_i\), consistently estimates this total effect.

Analyzing the dataset after listwise deletion conditions on \(M_i=0\), opening the noncausal path \(D_i\rightarrow M_i\leftarrow A_i\rightarrow Y_i\), on which \(M_i\) is a collider. Moreover, \(M_i\) is a descendant of the mediator \(A_i\) on the causal path \(D_i\rightarrow A_i\rightarrow Y_i\), so conditioning on it violates the first condition of the adjustment criterion. Thus, in contrast to the case of the oracle sample, a regression of $Y_i$ on $D_i$ can fail to consistently estimate the total effect.

The guidance that listwise deletion preserves the oracle regression coefficients when missingness is driven by included regressors \citep{arel2018can} may tempt researchers to add \(A_i\) to the regression. Indeed, adjusting for \(A_i\) blocks the opened noncausal path, but it also blocks the mediated causal path. Thus, adding the mediator to the complete-case regression does not restore identification of the total effect. The treatment coefficient, instead, identifies the controlled direct effect, which equals the average direct effect when the specification has no treatment--mediator interaction, under sequential ignorability \citep{imai2010general}.

\paragraph{Illustration with Synthetic Data} To demonstrate this, we draw \RAreps{} bootstrap samples of \RAn{} respondents each, sampling with replacement from the complete-case dataset. In each sample, we impose the specified missing mechanism using the sample means of conscientiousness and authoritarian tendency as the thresholds. Specifically,
\[
\Pr(M_i=1\mid D_i,A_i)=
\begin{cases}
0.5, & D_i<\overline D\ \text{and}\ A_i>\overline A,\\
0, & \text{otherwise}.
\end{cases}
\]
On average, \RAmissPct{} percent of outcomes are missing. In each bootstrap sample, we estimate the following models, all of which include the authors' sociodemographic controls.
\begin{itemize}
    \item Oracle estimator: regress \(Y_i\) on \(D_i\), without the mediator \(A_i\), on the oracle dataset before imposing missingness.
    \item Oracle with \(A_i\) (Oracle + A): regress \(Y_i\) on \(D_i\) and \(A_i\) on the oracle dataset.
    \item Complete-case estimator (CCA): regress \(Y_i\) on \(D_i\), without \(A_i\), on the complete-case dataset where \(M_i=0\).
    \item Complete-case estimator with \(A_i\) (CCA + A): regress \(Y_i\) on \(D_i\) and \(A_i\) after listwise deletion.
    \item Multiple-imputation estimator (MI): impute missing \(Y_i\) using \(D_i\), \(A_i\), and the baseline controls, then regress \(Y_i\) on \(D_i\), without \(A_i\), on the imputed datasets and pool the estimates \mbox{\citep{little2019statistical}}.
\end{itemize}

\begin{figure}[htbp]
\centering
\includegraphics[width=0.8\textwidth]{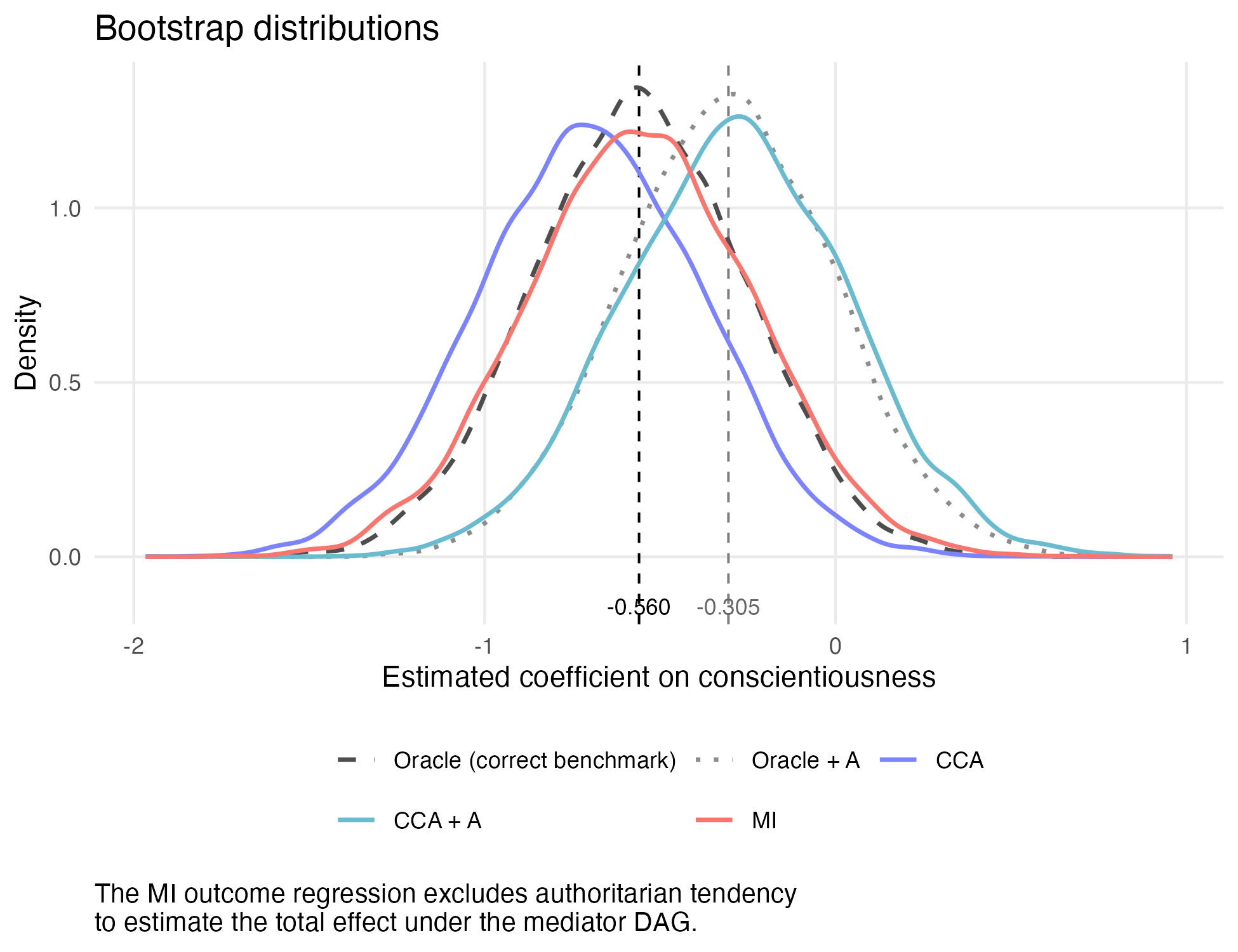}
\caption{Bootstrap Illustration with Authoritarianism as a Mediator}
\label{fig:results-autho-mediator}
\end{figure}

Figure~\ref{fig:results-autho-mediator} summarizes the empirical distributions of these estimators. Neither CCA nor CCA + A is centered around the oracle total-effect benchmark of $\RAmedTotal$\footnote{Our oracle regressions yield total- and controlled-direct-effect estimates of $\RAmedTotal$ and $\RAmedDirect$, compared with $-0.564$ and $-0.310$ in \citet[Table~4]{rivlin2025personality}. The small differences arise because they report averages over 1,000 quasi-Bayesian draws, while we report OLS point estimates.}. Instead, CCA + A is concentrated near the oracle controlled direct-effect benchmark of $\RAmedDirect$. These results are consistent with the predictions from the augmented DAG.

We also observe that the multiple-imputation estimates are centered near the oracle total-effect benchmark of $\RAmedTotal$. This is consistent with the imposed MAR mechanism: missingness depends only on observed conscientiousness and authoritarian tendency. Multiple imputation enables researchers to include both variables and the baseline controls in the imputation model while keeping the mediator out of the total-effect regression\footnote{We analyze the canonical listwise-deletion workflow. G-computation could instead recover the total effect under suitable assumptions by combining a complete-case outcome model with the mediator distribution conditional on treatment and baseline controls, estimated from the original sample.}.

\paragraph{When authoritarian tendency is a confounder}

We now consider an alternative substantive theory in which authoritarian tendency is a confounder of the relationship between conscientiousness and support for a two-state solution. Political science research sometimes treats authoritarianism as a relatively enduring predisposition that shapes political attitudes \citep{stenner2005authoritarian,engelhardt2023advancing}. The right panel of Figure~\ref{fig:example-dag-1} represents this alternative causal structure, retaining the same missing mechanism.

\begin{figure}[htbp]
\centering
\includegraphics[width=0.8\textwidth]{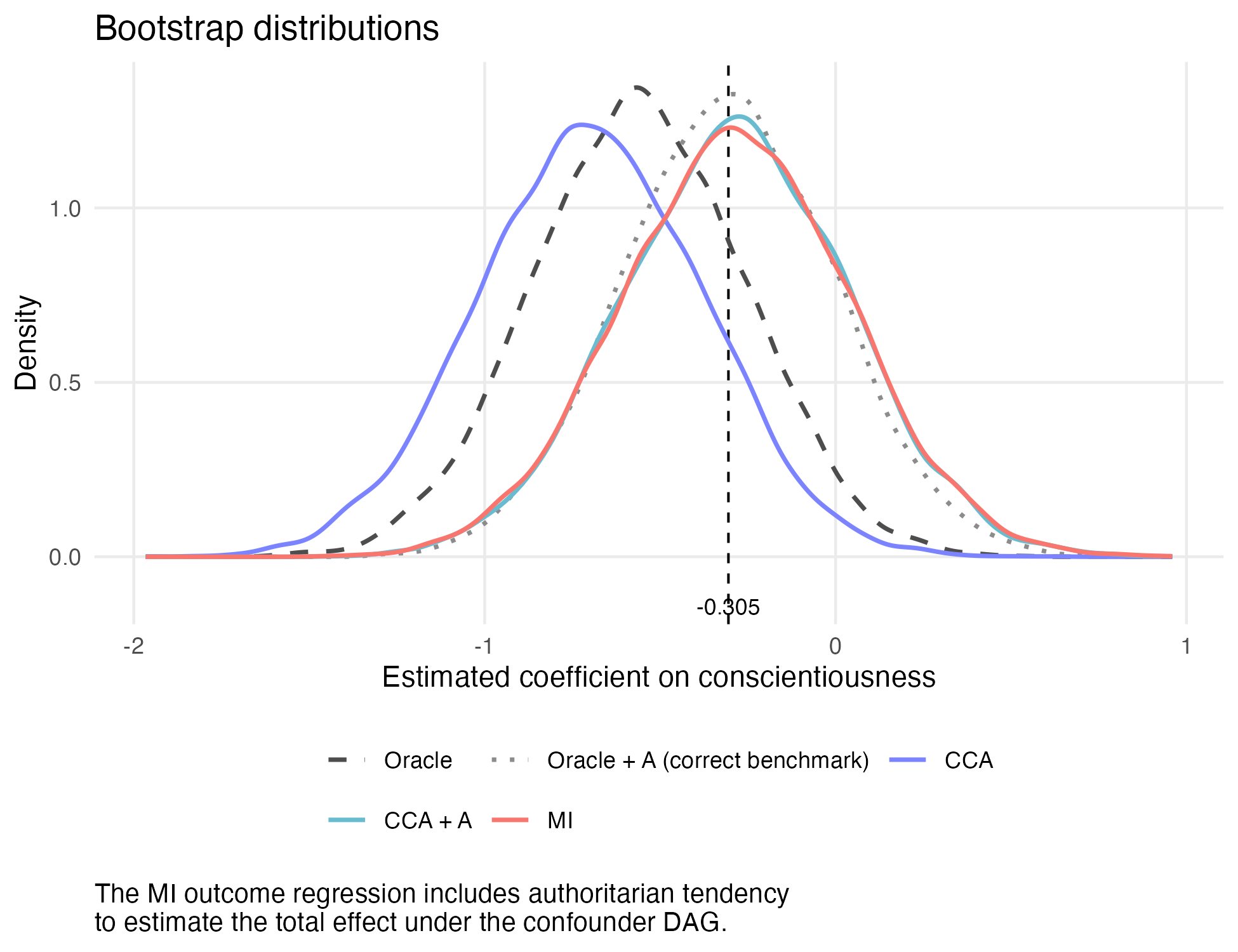}
\caption{Bootstrap Illustration with Authoritarianism as a Confounder}
\label{fig:results-autho-confounder}
\end{figure}

The missingness mechanism is MAR in both left and right DAGs, but the causal role of \(A_i\) differs, as shown by the reversed arrow between \(D_i\) and \(A_i\). When \(A_i\) is a confounder, conditioning on \(M_i\) opens the noncausal path \(D_i\rightarrow M_i\leftarrow A_i\rightarrow Y_i\). Adjusting for \(A_i\) blocks both this path and the original back-door path \(D_i\leftarrow A_i\rightarrow Y_i\). Under the maintained identification and regression-specification assumptions, complete-case regression adjusting for \(A_i\) therefore consistently estimates the total effect. Consistent with the augmented DAG, Figure~\ref{fig:results-autho-confounder} shows that CCA + A is centered near the total-effect benchmark of $\RAconfTotal$\footnote{The oracle regression adjusting for authoritarian tendency yields the same coefficient, \(\RAmedDirect\), as in the mediation illustration, but it identifies the total effect under the confounder DAG. CCA + A averages \(\RAconfCCXMean\) because, in the finite sample used for the bootstrap, respondents subject to missingness happen to lie slightly above the fitted line -- deleting some flattens the slope. This offset would vanish as the original sample grows if the population conditional mean of the outcome is linear in authoritarian tendency.}.

In this illustration, we show that the same dataset, missingness pattern, and MAR classification may call for different strategies under different causal structures. Adjusting for authoritarian tendency identifies the total effect when it is a confounder, but blocks part of that effect when it is a mediator. Different causal structures can generate observationally equivalent data, so researchers must specify their causal assumptions to assess whether complete-case analysis remains justified after conditioning on the missing indicator. Appendix~\ref{app:collider-mediator-simulation} further supplies a separate simulation of the dilemma, including missing-outcome and missing-treatment cases.

\section{Reweighting Complete-Case Comparisons: Heterogeneous Effects}
\label{sec:reweighting-hte}

In this section, we extend the framework to heterogeneous treatment effects by reweighting complete-case comparisons to recover population averages. Appendix~\ref{app:glm-reweighting} further applies the same approach to generalized linear models.

\subsection{Reweighting for Heterogeneous Treatment Effects}
\label{subsec:reweighting-hte}

Let \(W_i\) denote the covariates used for transport in Corollary~\ref{cor:cc-effect-transport}. Let \(\delta_{\mathrm{cc}}(w)\) and \(\delta_{\mathrm{po}}(w)\) denote the conditional average treatment effects among complete cases and in the population, respectively. Let \(F_{\mathrm{cc}}\) and \(F_{\mathrm{po}}\) denote the distributions of \(W_i\) among complete cases and in the population. Then
\[
\tau_{\mathrm{cc}}=\int\delta_{\mathrm{cc}}(w)\,dF_{\mathrm{cc}}(w),
\qquad
\tau=\int\delta_{\mathrm{po}}(w)\,dF_{\mathrm{po}}(w).
\]
Under mean exchangeability of selection and treatment effect heterogeneity, \(\delta_{\mathrm{cc}}(w)=\delta_{\mathrm{po}}(w)\) for almost every \(w\) with respect to \(F_{\mathrm{po}}\). With complete-case support, we therefore have
\[
\tau_{\mathrm{cc}}-\tau
=\int\delta_{\mathrm{cc}}(w)\,dF_{\mathrm{cc}}(w)
-\int\delta_{\mathrm{cc}}(w)\,dF_{\mathrm{po}}(w).
\]
Under these conditions, the difference arises from averaging the same conditional treatment effects over different covariate distributions. This decomposition also situates the two-stage projection estimator of \citet{huangLiu2026}. In the full-covariate case \(W_i=X_i\), their outcome-response ignorability and treatment-identification conditions imply \(\delta_{\mathrm{cc}}(w)=\delta_{\mathrm{po}}(w)\). As some confounders are missing, they estimate outcome models among complete cases, impute missing covariates such that they can project the fitted models onto the completed covariate data to average predicted conditional effects over the full sample.

Our framework generalizes this identification argument by specifying graphical conditions under which complete-case comparisons identify \(\delta_{\mathrm{cc}}(x)\) (Proposition~\ref{prop:adjustment-ignorability}). It also generalizes the transport step in two ways: mean exchangeability allows selection to change conditional outcome means as long as the conditional treatment effect is preserved, and the transport covariates \(W_i\) may be a subvector of the adjustment covariates \(X_i\). Rather than first pooling these comparisons into the complete-case ATE, \(\tau_{\mathrm{cc}}\), through the regression approaches in Propositions~\ref{prop:cc-ols-preservation} and~\ref{prop:graphical-cc-ipw}, Corollary~\ref{cor:cc-effect-transport} provides a direct route to the population ATE: given identified within-\(X_i\) effects, average \(\delta_{\mathrm{cc}}(x)\) over the complete-case distribution of \(X_i\) within each \(W_i=w\) group, then average the resulting effects over the population distribution of \(W_i\), assuming mean exchangeability and complete-case support given \(W_i\).

The identification results also allow continuous moderators, but implementing the population-averaging result in that setting requires additional assumptions about how conditional effects are estimated, which we have not developed here. The reanalysis in Subsection~\ref{subsec:reanalysis-bh} instead uses categorical moderators, for which the averaging step is explicit. 

\subsection{Reanalysis of \citet{bodeaHicks2015}}
\label{subsec:reanalysis-bh}

\citet{bodeaHicks2015} study the conditional effects of central bank independence (CBI) using data on 78 countries from 1973 to 2008. They find that CBI lowers inflation in democracies but not in nondemocracies. Their country fixed-effects regressions interact the CBI index with six measures of democracy and democratic institutions: Polity \citep{marshallGurr2020polity5}, Freedom House \mbox{\citep{freedomHouseWorld}}, executive constraints \citep{marshallGurr2020polity5}, checks and balances \citep{cruzKeeferScartascini2021dpi}, political constraints \citep{henisz2000institutional,henisz2012polcon}, and press freedom \citep{freedomHousePress}.

\begin{figure}[htbp]
\centering
\includegraphics[width=\textwidth]{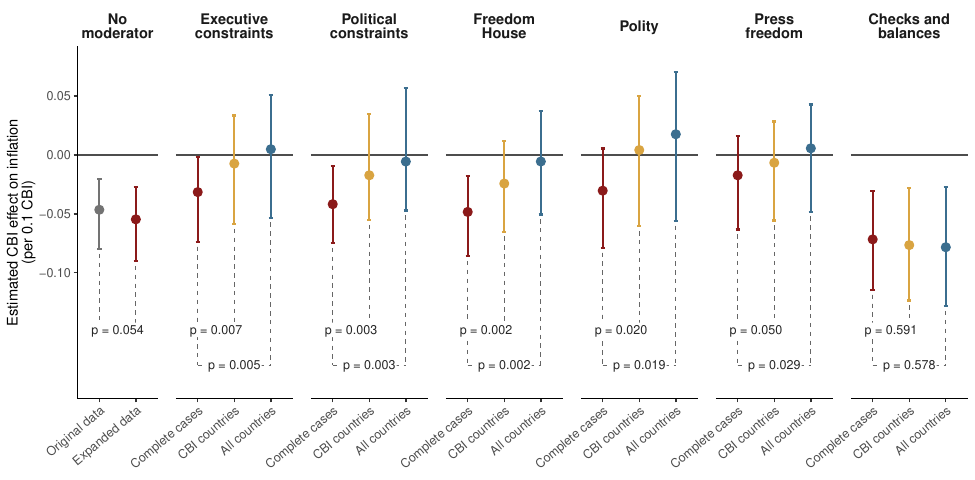}
\caption{Effect of Central Bank Independence on Inflation, Complete Cases and Reweighted Averages}
\label{fig:bodea-hicks}
\par\medskip
\begin{minipage}{0.95\linewidth}
\footnotesize
\textit{Notes:} Points are average effects of a 0.1 increase in the CBI index on signed log inflation. Intervals are 95\% country-cluster bootstrap intervals. The first panel compares the original and expanded data without a moderator. In the other panels, Complete cases uses complete-case category shares, CBI countries uses shares among countries covered by the updated CBI data, and All countries uses shares among countries covered by Polity5. Dashed brackets give two-sided bootstrap $p$-values for differences between linked estimates. Holm-adjusted $p$ values reported in Table~\ref{tab:bodea-hicks-appendix}. Across the \BHnModerators{} moderators, the lowest proportion of countries with within-country CBI variation in any category ranges from \BHminVarLo{}\% to \BHminVarHi{}\%.
\end{minipage}
\end{figure}

We reanalyze their study using the authors' updated and expanded CBI data, which cover 143 countries and the European Central Bank from 1970 to 2020 \citep{bodeaHicksData}, along with updated data from the original sources of the moderators. For 1973--2008, the expanded data contain \BHexpCountries{} countries with CBI and Polity observed, of which \BHexpCCCountries{} (\BHexpCCRows{} country-years) have every control observed. The outcome is signed log inflation, and effects are reported per 0.1 increase in the CBI index. Without a moderator, we reproduce the authors' original CBI coefficient of \(\BHorigCoef\) (SE \BHorigSE) per unit of CBI. The expanded data yield \(\BHexpCoef\) (Figure~\ref{fig:bodea-hicks}, leftmost panel).

In the original data, \BHorigRows{} country-years in \BHorigCountries{} countries have CBI and Polity observed during the analysis period. After listwise deletion, \BHorigCCRows{} country-years in \BHorigCCCountries{} countries enter the authors' baseline regression. Our main comparison concerns heterogeneous effects in the expanded data, retaining the economic controls and country fixed effects. To account for possible nonlinearity in the moderating effects \citep{hainmueller2019much}, we divide each measure into three categories and fit a country fixed-effects regression within each. Holding these category-specific estimates fixed, we compare their averages using complete-case and target-population shares (Table~\ref{tab:bodea-hicks-appendix}, Panel B). Although CBI is continuous, the population-averaging result in Corollary~\ref{cor:cc-effect-transport} can be extended to category-specific CBI slopes, taking their causal interpretation and consistent estimation as given\footnote{The arguments in Propositions~\ref{prop:adjustment-ignorability}--\ref{prop:cc-ols-preservation} and Corollary~\ref{cor:cc-effect-transport} can be adapted to continuous treatments. A sufficient DGP is \(Y_{it}(d)=a_{i,w}+\mu_w(H_{it})+s_wd+u_{it}\), where \(w\) is a pretreatment moderator category and \(H_{it}\) contains pretreatment controls and outcome history. Under valid adjustment, treatment support, consistency, no interference, and a correctly specified complete-case conditional mean, \(s_w\) is the category-specific causal slope. If these slopes transport to the target categories, averaging \(0.1s_w\) using target shares gives the target average effect per 0.1 increase in CBI.}.

\citet{bodeaHicks2015} do not specify a missingness mechanism for their main regression variables. To focus the illustration for Box 3, we assume their complete-case regression uses a valid adjustment set and is correctly specified. In the compatible causal structure in Figure~\ref{fig:CCA-Unbiasedness}, missingness is driven by treatment and covariates, and treatment comparisons within covariate strata remain valid among complete cases. By granting these conditions, we show that even valid complete-case comparisons may not recover the population ATE when treatment effects vary and listwise deletion changes the distribution of moderators.

Following Corollary~\ref{cor:cc-effect-transport}, with the moderator as \(W_i\), we assume mean exchangeability of selection and treatment effect heterogeneity\footnote{Here, we assume that each category's fixed-effects slope equals its complete-case average effect, which in turn equals its target-population average. For example, under Polity, only 11 of the 38 complete-case autocracies have within-country CBI variation. The slope identified from these 11 countries must therefore represent the average effect across all complete-case autocracies and, under mean exchangeability, all target-population autocracies.}: the average effect within each moderator category is the same among complete cases and in the target population, in addition to complete-case support. We then hold the estimated category-specific effects fixed and average them using their population shares to recover the population average.

Whereas \citet{bodeaHicks2015} focus on the conditional effects themselves, we consider a case in which researchers are concerned with the population ATE. Suppose the target population consists of country-years in the countries covered by the expanded CBI data. As Figure~\ref{fig:bodea-hicks} shows, for \BHnCCNegReweightedNS{} of the \BHnModerators{} moderators, the complete-case average is negative and statistically significant, whereas the average reweighted to the target population's moderator distribution is not statistically distinguishable from zero. For \BHnGapSig{} of the \BHnModerators{} moderators, the difference between the complete-case and reweighted averages is statistically significant at the 5\% level using unadjusted \(p\)-values. 

This is because listwise deletion changes the moderator distribution: democracies and countries with stronger democratic institutions are overrepresented among complete cases\footnote{As Panel~C of Table~\ref{tab:bodea-hicks-appendix} shows, in the expanded data, democracies account for \BHshareCCDem{}\% of complete-case country-years, while the corresponding share among country-years in the target population is \BHshareTgtDem{}\%.}, so the complete-case average places more weight on effects in these categories than the population average. Interpreting the complete-case average as a population average can therefore be misleading. The same pattern also holds when the target population is all country-years covered by Polity5, a larger population than the country-years in countries covered by the expanded CBI data. Table~\ref{tab:bodea-hicks-appendix} reports the numerical results.

\section{Checking All Three Boxes: Complete-Case Analysis in a Canvassing Experiment}
\label{sec:comprehensive-illustration}

The preceding two sections illustrate the use of DAGs and reweighting to assess complete-case regression. We now reanalyze \citet{kallaBroockman2023} to show how researchers can justify listwise deletion through all three checks: the validity of treatment comparisons using a DAG, the pooling of those comparisons through treatment weighting, and reweighting to recover the population ATE.

Their Study 1 examines whether canvassing conversations reduce prejudice toward undocumented immigrants. We compare their full intervention, which used personal stories and perspective-taking, with a placebo conversation on an unrelated topic, and use their first-follow-up outcome index, measured approximately one week after canvassing. Among the \KBnEligible{} participants in these two arms, \KBnMissing{} (\KBpctMissing{} percent) have no observed first-follow-up outcome, leaving \KBnComplete{} complete cases. Treatment assignment, baseline education, and the regression controls are fully observed. The missing pattern is purely on the follow-up outcomes: \KBpctMissNoCol{} percent of participants without a college degree and \KBpctMissCol{} percent of those with one lack the outcome.

\paragraph{Box 1: Do Treatment Comparisons Remain Valid after Listwise Deletion?}

Let \(D\) denote assignment to the full intervention rather than the placebo, \(W\) baseline college education, \(X\) the remaining baseline controls, and \(Y\) the first-follow-up outcome. Together, \((W,X)\) form the pretreatment covariate vector, the \(X\) in the general framework. Let \(M=1\) indicate that \(Y\) is missing. We consider two possible missing mechanisms, represented by two DAGs in Figure~\ref{fig:kb-missing-mechanisms}.

\begin{figure}[htbp]
\centering
\begin{tabular}{c}
Left: Treatment and Baseline\\[4pt]
\begin{tikzpicture}[manuscript dag]
\node[manuscript dag node] (D) at (0,0) {D};
\node[manuscript dag node] (Y) at (2.6,0) {Y};
\node[manuscript dag node,dashed] (U) at (3.8,0) {U};
\node[manuscript dag node] (W) at (0,1.6) {W};
\node[manuscript dag node] (X) at (2.6,1.6) {X};
\node[manuscript dag node] (M) at (1.3,2.6) {M};
\path[manuscript dag edge]
 (D) edge (Y) (W) edge (Y) (X) edge (Y) (U) edge (Y)
 (W) edge (M) (X) edge (M) (D) edge (M);
\end{tikzpicture}
\end{tabular}\hspace{3em}%
\begin{tabular}{c}
Right: Outcome\\[4pt]
\begin{tikzpicture}[manuscript dag]
\node[manuscript dag node] (D) at (0,0) {D};
\node[manuscript dag node] (Y) at (2.6,0) {Y};
\node[manuscript dag node,dashed] (U) at (3.8,0) {U};
\node[manuscript dag node] (W) at (0,1.6) {W};
\node[manuscript dag node] (X) at (2.6,1.6) {X};
\node[manuscript dag node] (M) at (1.3,2.6) {M};
\path[manuscript dag edge]
 (D) edge (Y) (W) edge (Y) (X) edge (Y) (U) edge (Y)
 (W) edge (M) (X) edge (M) (D) edge (M);
\path[manuscript dag edge] (Y) edge (M);
\end{tikzpicture}
\end{tabular}
\par\medskip
\begin{minipage}{0.9\linewidth}
\normalsize
\end{minipage}
\begin{minipage}{0.95\linewidth}
\footnotesize
\begin{tabular}{@{}l@{\hspace{2em}}l@{}}
\(D\): assignment to the full intervention rather than the placebo & \(W\): baseline college education\\
\(X\): baseline outcome items and other baseline controls, excluding education & \(Y\): first-follow-up outcome\\
\(U\): unobserved causes of the follow-up outcome & \(M\): missing indicator for \(Y\)
\end{tabular}
\end{minipage}
\caption{Two Possible Missing Mechanisms in the Canvassing Experiment}
\label{fig:kb-missing-mechanisms}
\end{figure}

In the left panel of Figure~\ref{fig:kb-missing-mechanisms}, missingness has three causes: treatment \(D\), baseline college education \(W\), and the remaining baseline controls \(X\), which include prior immigration attitudes. College-educated participants may be more likely to answer an emailed follow-up survey, and the canvassing itself may change willingness to respond\footnote{The authors find no evidence of differential attrition: follow-up response is \KBresponseControl{}\% in the placebo arm and \KBresponseTreated{}\% in the treatment arm. We further test whether response depends on treatment conditional on education and the other baseline controls. In a linear probability model with treatment--covariate interactions, a household-clustered Wald test jointly rejects the treatment and interaction terms (\(\chi^2(\KBattrDf)=\KBattrChi\), \(p=\KBattrP\)). This evidence favors allowing \(D\rightarrow M\).}.

Alternatively, we may believe that follow-up response further depends on post-treatment attitudes. For example, among participants with the same baseline characteristics, those who would express less favorable attitudes toward undocumented immigrants might be less willing to respond. This adds the arrow \(Y\rightarrow M\), shown in the right panel of Figure~\ref{fig:kb-missing-mechanisms}.

In the left panel, conditioning on the collider \(M\) opens the noncausal paths \(D\rightarrow M\leftarrow W\rightarrow Y\) and \(D\rightarrow M\leftarrow X\rightarrow Y\). Adjusting for \(W\) and \(X\) blocks both paths. Moreover, \(M\) is not a descendant of a node on the causal path, so the adjustment criterion in Proposition~\ref{prop:adjustment-ignorability} is satisfied. In the right panel, \(M\) is a descendant of \(Y\) on the causal path \(D\rightarrow Y\). Conditioning on \(M\) therefore violates the first requirement of the adjustment criterion, so complete-case comparisons are no longer justified by Proposition~\ref{prop:adjustment-ignorability}.

Thus, under the left DAG, Box 1 is checked. Under the right DAG, there is no guarantee for valid treatment comparisons under listwise deletion, and researchers need to consider alternative strategies. In the rest of this illustration, we focus on the left DAG.

\paragraph{Box 2: Does the Regression Appropriately Pool the Treatment Comparisons?}

We next assess whether the regression appropriately pools treatment comparisons within baseline-covariate strata into the complete-case ATE. The authors randomly assigned households within blocks, with equal probabilities of assignment to each arm\footnote{For this application, we assume i.i.d. sampling of households rather than individuals, in which consistency still holds under regularity conditions.}. Under the left panel of the DAG, treatment and baseline characteristics jointly determine whether the follow-up outcome is missing, so the complete-case treatment propensity may vary across baseline-covariate strata. We therefore use treatment IPW under Proposition~\ref{prop:graphical-cc-ipw} to pool the comparisons, estimating treatment probabilities among complete cases with a probit model including education and other baseline controls.

The IPW estimate is \KBipwEst{} (Table~\ref{tab:kb-probit-ipw}). We also estimate complete-case effects separately within each education group, fitting a separate treatment propensity model and treatment-IPW regression in each group. The estimates are \KBeffNoCol{} for participants without a college degree and \KBeffCol{} for those with one. We use these education-specific estimates for reweighting below.

\paragraph{Box 3: Does the Complete-Case Regression Recover the Population ATE?}

Having checked Boxes 1 and 2, we ask whether a complete-case regression that omits the education moderator recovers the population ATE. When effects vary by education, the complete-case ATE averages conditional effects over the education distribution among complete cases, whereas the population ATE averages population conditional effects over the population distribution (Corollary~\ref{cor:cc-effect-transport}).

Suppose mean exchangeability of selection and treatment effect heterogeneity holds at the level of education, so that within each education group the average treatment effect is the same among complete cases and in the population. If complete cases are represented in every population education group, we can estimate the population ATE by averaging the complete-case education-specific effects using the population shares. Here, population refers to the popuation represented by the 1,900 participants under original authors' recruitment and sampling procedure.

In the data, college-educated participants account for \KBshareCCCol{}\% of complete cases, but \KBshareTargetCol{}\% of the target sample. Averaging the education-specific effects using the target shares yields \KBtargetAvg{}, compared\footnote{This comparison captures both treatment IPW and reweighting by education. Using the same education-specific estimates, changing from complete-case to target shares gives a difference of \(\KBdiffSameSpec\).} with \KBccPooled{} from the pooled complete-case regression. Their difference is \(\KBdiffTarget\), with a 95\% paired household-bootstrap interval of \([\KBdiffTargetLo, \KBdiffTargetHi]\). 

Thus, reweighting to the population share does not produce a statistically significant change in the estimated effect. This is partly due to the fact that the treatment effect does not statistically differ by education (\(p=\KBhetP\), two-sided household bootstrap). If treatment effects do differ, however, the statistically insignificant difference between the complete-case and reweighted estimates in this finite sample does not establish that the complete-case estimator converges to the population ATE. The difference between the education distributions of complete cases and the population also would not disappear as the sample size grows, so does the resulting difference in averaging weights. 

\section{Concluding Remarks}

This article asks when complete-case regression consistently estimates the population average treatment effect. It does so when complete-case treatment comparisons identify the relevant conditional effects (under the adjustment criterion and treatment overlap) and the regression estimator pools those comparisons appropriately (through correct specification or IPW). With heterogeneous effects, recovering the population ATE also requires averaging the conditional effects over the population covariate distribution. Most importantly, MAR or NMAR labels alone do not establish these conditions.

Our framework brings existing guidance on complete-case regression into an augmented causal diagram. Recommendations for missingness driven by covariates or the outcome follow as special cases, while differences between \citet{arel2018can} and \citet{pepinsky2018note} reflect different assumptions about the missingness mechanism and regression specification. The collider--mediator illustration shows why MAR labels alone cannot determine whether listwise deletion is justified: with the same data, missing pattern, and MAR classification, adjusting for authoritarian tendency recovers the total effect when it is a confounder, but not when it is a mediator. Because these causal structures can be observationally equivalent, researchers must state the assumptions they use to justify listwise deletion, just as they do when choosing an adjustment set for the oracle regression.

For empirical researchers, Table~\ref{tab:complete-case-framework} summarizes the checks: start with the oracle regression's adjustment set, which would identify the population ATE, assess treatment comparisons in its augmented DAG, verify that the estimator pools them appropriately, and, with heterogeneous effects, average conditional effects over the target population's covariate distribution. Our Bodea--Hicks reanalysis illustrates that a negative, statistically significant complete-case average can become indistinguishable from zero after reweighting. In the canvassing reanalysis, the complete-case and reweighted estimates are not statistically distinguishable, but this finite-sample result does not establish consistency: any population gap caused by averaging over the wrong distribution persists as the sample grows. Our framework gives researchers a formal basis for assessing listwise deletion and its consequences. By making the assumed causes of missingness explicit, researchers can explain when deletion is appropriate for their causal question.

When the conditions cannot be established, researchers have two broad options. They may turn to methods that address the missing part of the data, such as imputation or IPW for missingness, and assess the assumptions those methods require. Alternatively, they may retain listwise deletion and use bounds or sensitivity analysis to assess what can be learned when the conditions for complete-case regression are uncertain.

Thus, several limitations and possible extensions remain. First, future work could develop bounds and sensitivity analyses for settings where the conditions for complete-case regression are uncertain, as well as diagnostics for the outcome specification. Second, our framework could be extended to other causal estimands, such as the ATT and the complier average treatment effect, and to estimation and inference with continuous treatment-effect modifiers. Software could also help researchers construct augmented causal diagrams and assess whether they satisfy the graphical conditions for valid complete-case comparisons. Finally, we note that even when complete-case regression is consistent, the number of complete observations can become very small as the number of variables with missing values grows \citep{wangAronow2023}. In such cases, consistency provides limited assurance for the available sample, and imputation may be needed to use information from incomplete observations.

\clearpage
\appendix
\counterwithin*{figure}{section}
\setdisplayskipstretch{1}
\AtBeginEnvironment{align}{\setstretch{1}}
\AtBeginEnvironment{align*}{\setstretch{1}}
\AtBeginEnvironment{aligned}{\setstretch{1}}
\AtBeginEnvironment{gather}{\setstretch{1}}
\AtBeginEnvironment{gather*}{\setstretch{1}}
\addtocontents{toc}{\protect\setcounter{tocdepth}{2}}
\thispagestyle{empty}
\begin{center}
{\Large When Listwise Deletion Works:\\ Causal Diagrams and Complete-Case Regression\par}
\vspace{1em}
{\large Supporting Information\par}
\end{center}
\vspace{1em}
\begingroup
\singlespacing
\renewcommand{\contentsname}{Table of Contents}
\tableofcontents
\thispagestyle{empty}
\endgroup
\clearpage
\pagenumbering{arabic}
\section{Review of Missing Data in \emph{APSR} and \emph{AJPS}, 2019--2024}
\label{app:literature-review}
\setcounter{table}{0}
\renewcommand{\thetable}{A\arabic{table}}
\renewcommand{\theHtable}{A\arabic{table}}

To approximate \citet{huangLiu2026}, we screened 2019--2024 Crossref records for causal terms, such as \emph{causal}, \emph{effect of}, and \emph{confound}, and excluded records mentioning experimental designs. We chose this rule partly to approximate their sample sizes before full-text eligibility review. Full-text eligibility review retained \LRnEligAPSR{} \emph{APSR} and \LRnEligAJPS{} \emph{AJPS} observational causal studies.

Claude Opus 5.5 coded ten pilot articles, and Claude Sonnet 5 coded the remaining articles' main texts and available appendices. We reviewed the AI-generated coding using the same standards we apply to work produced by a human research assistant and found the results reliable. We asked the agent to code deletion as stated when named and implied when missing values were given as the reason for dropping observations, while not to count unexplained sample-size differences. Table~\ref{tab:review-results} summarizes the results.

\begin{table}[H]
\centering
\caption{Missing data in eligible observational causal studies}
\label{tab:review-results}
\begin{tabular}{lcc}
\toprule
 & \emph{APSR} & \emph{AJPS} \\
 & ($n=100$) & ($n=58$) \\
\midrule
\multicolumn{3}{l}{\emph{Mention of missing values}} \\
\quad Main text & 65 (65\%) & 42 (72\%) \\
\quad Main text, including restricted-availability analyses & 88 (88\%) & 52 (90\%) \\
\quad Main text or appendix & 95 (95\%) & 55 (95\%) \\
\quad \citet{huangLiu2026}, main text & 70/130 (54\%) & 41/90 (46\%) \\
\addlinespace
\multicolumn{3}{l}{\emph{Case deletion}} \\
\quad Stated or implied, main text & 47 (47\%) & 33 (57\%) \\
\quad Stated, main text & 3 (3\%) & 1 (2\%) \\
\quad Stated or implied, main text or appendix & 68 (68\%) & 39 (67\%) \\
\quad Baseline column, complete or available cases & 50 (50\%) & 29 (50\%) \\
\quad Regression on a deleted sample, of deletion articles & 27/47 (57\%) & 18/33 (55\%) \\
\addlinespace
\multicolumn{3}{l}{\emph{Imputation}} \\
\quad Imputation model, main text & 7 (7\%) & 4 (7\%) \\
\quad Imputation model, main text or appendix & 10 (10\%) & 8 (14\%) \\
\quad Any fill-in, main text & 27 (27\%) & 15 (26\%) \\
\quad Any fill-in, appendix only & 15 (15\%) & 7 (12\%) \\
\quad \citet{huangLiu2026}, imputation model & fewer than 7 & fewer than 5 \\
\bottomrule
\end{tabular}

\medskip
\begin{minipage}{0.9\textwidth}
\footnotesize \emph{Note:} A paper can fall in several rows. Restricted-availability analyses are those limited to the periods or units for which a variable exists. The baseline-column row comes from the separate baseline-specification codebook. The regression row uses the articles with deletion stated or implied in the main text as its denominator and is a lower bound.
\end{minipage}
\end{table}

\FloatBarrier
\section{Numerical Results for the Motivating Simulation}
\label{app:motivating-simulation}
\setcounter{table}{0}
\renewcommand{\thetable}{\thesection\arabic{table}}
\renewcommand{\theHtable}{\thesection\arabic{table}}
\setcounter{figure}{0}
\renewcommand{\thefigure}{\thesection\arabic{figure}}
\renewcommand{\theHfigure}{\thesection\arabic{figure}}

\begin{table}[H]
\centering
\caption{Estimated Treatment Coefficients under Different Missing Mechanisms}
\label{tab:motivating-simulation}
\begin{tabular}{llrrrr}
\toprule
Mechanism & Regression & Mean & SD & 2.5th pct. & 97.5th pct. \\
\midrule
A & Oracle & 1.001 & 0.018 & 0.968 & 1.037 \\
A & Complete case & 1.002 & 0.021 & 0.965 & 1.047 \\
A & MI (MAR model) & 1.057 & 0.020 & 1.020 & 1.097 \\
A & Complete case with IPW & 1.002 & 0.022 & 0.961 & 1.045 \\
\hdashline
B & Oracle & 1.001 & 0.018 & 0.968 & 1.037 \\
B & Complete case & 1.165 & 0.020 & 1.128 & 1.209 \\
B & Complete case, saturated $X$ & 1.002 & 0.021 & 0.964 & 1.044 \\
B & MI (MAR model) & 1.177 & 0.019 & 1.143 & 1.214 \\
B & Complete case with IPW & 1.002 & 0.021 & 0.963 & 1.043 \\
\hdashline
C & Oracle & 1.001 & 0.018 & 0.968 & 1.037 \\
C & Complete case & 1.497 & 0.021 & 1.460 & 1.539 \\
C & Complete case, saturated $X$ & 1.541 & 0.022 & 1.503 & 1.584 \\
C & MI (MAR model) & 1.190 & 0.018 & 1.157 & 1.228 \\
C & Complete case with IPW & 1.580 & 0.021 & 1.542 & 1.620 \\
\bottomrule
\end{tabular}
\vspace{0.4em}
\begin{minipage}{0.92\linewidth}
\footnotesize\emph{Note:} MI imputes missing $X_i$ from observed $D_i$ and $Y_i$ using a multinomial MAR working model with 20 imputations. The treatment coefficient is pooled across completed datasets. SD and percentiles summarize variation across the 500 replications.
\end{minipage}
\end{table}

\subsection{Saturated Regression and IPW under Mechanisms B and C}
\label{app:correct-treatment-weighting}

Recall the within-\(X_i\) complete-case model from the main text, \(Y_i=\alpha_{X_i}+\beta_{X_i}D_i+\varepsilon_{\mathrm{str},i}\). Here \(\beta_x\) is the observed treated-minus-untreated contrast at \(X_i=x\). For both mechanisms, the \emph{Complete case, saturated \(X\)} row in Table~\ref{tab:motivating-simulation} estimates, among \(M_i=0\) cases, the unweighted regression
\[
Y_i=\alpha^{\mathrm{sat}}_{X_i}+\beta_{\mathrm{sat}}D_i
+r_i^{\mathrm{sat}},\qquad M_i=0.
\]
The fitted intercept \(\alpha^{\mathrm{sat}}_{X_i}\) varies across the three values of \(X_i\), while \(\beta_{\mathrm{sat}}\) is common to them. \(r_i^{\mathrm{sat}}\) is the regression residual. The simulation implements these intercepts with \(X_i\), \(\mathbf{1}\{X_i=-1\}\), and \(\mathbf{1}\{X_i=1\}\). In our DGP, one of these terms is dropped because of perfect collinearity.

The \emph{Complete case with IPW} row estimates the weighted regression
\[
Y_i=\alpha_{\mathrm{IPW}}+\beta_{\mathrm{IPW}}D_i
+\gamma_{\mathrm{IPW}}X_i+r_i^{\mathrm{IPW}},\qquad M_i=0,
\]
with \(\omega_i=1/\hat e_0(X_i)\) for treated units and \(\omega_i=1/[1-\hat e_0(X_i)]\) for untreated units. \(\hat e_0(x)\) is the treated proportion among complete cases with \(X_i=x\), truncated to \([0.01,0.99]\) before weighting. \(r_i^{\mathrm{IPW}}\) is the regression residual.

The two mean treatment coefficients in Table~\ref{tab:motivating-simulation} agree  under B: \(\MSsatB\) for saturated OLS and \(\MSipwB\) for IPW. Under C, they differ: \(\MSsatC\) and \(\MSipwC\), respectively.

To explain the difference, let \(e_0(x)=P(D_i=1\mid X_i=x,M_i=0)\). In the population, the two coefficients pool the \(\beta_x\) contrasts as
\[
\begin{aligned}
\text{IPW:}&\quad \sum_x P(X=x\mid M=0)\,\beta_x,\\
\text{Saturated-}X\text{ OLS:}&\quad
\frac{\sum_x P(X=x\mid M=0)e_0(x)[1-e_0(x)]\,\beta_x}
{\sum_x P(X=x\mid M=0)e_0(x)[1-e_0(x)]}.
\end{aligned}
\]
Thus, saturated-\(X\) OLS weights each stratum by the within-stratum variance of treatment as well as its complete-case frequency \citep{angrist2009mostly}. Under B, \(\beta_x=1\) for every \(x\), so the different weights produce the same population coefficient. Under C, \((\beta_{-1},\beta_0,\beta_1)=(2,1,2)\), so the weights produce different coefficients. 
\FloatBarrier
\section{MAR Companion to the Motivating Simulation}
\label{app:motivating-mar}
\setcounter{table}{0}
\renewcommand{\thetable}{\thesection\arabic{table}}
\renewcommand{\theHtable}{\thesection\arabic{table}}
\setcounter{figure}{0}
\renewcommand{\thefigure}{\thesection\arabic{figure}}
\renewcommand{\theHfigure}{\thesection\arabic{figure}}

We retain the outcome process, treatment assignment, and regression from Section~\ref{sec:motivating-simulation}. The oracle sample has 20,000 observations, \(Y_i=D_i+X_i+U_i+\xi_i\), and a constant unit-level treatment effect \(\tau=1\). We add an always-observed pretreatment noisy proxy \(\widetilde X_i=X_i+U_i+\nu_i\), where \(\nu_i\sim N(0,0.25^2)\). Only \(X_i\) can be missing.

Let \(M_i=0\) when \(X_i\) is observed. We consider three mechanisms:
\[
\begin{aligned}
\text{A:}\quad P(M_i=0\mid D_i,\widetilde X_i)&=0.875-0.30D_i,\\
\text{B:}\quad P(M_i=0\mid D_i,\widetilde X_i)&=\kappa\,\mathbf{1}\{\widetilde X_i\geq-1.5\},\\
\text{C:}\quad P(M_i=0\mid D_i,\widetilde X_i)&=\begin{cases}
p_0,&D_i=0,\\
\kappa\,\mathbf{1}\{\widetilde X_i\geq-1.5\},&D_i=1.
\end{cases}
\end{aligned}
\]
The constants \(\kappa\simeq0.9000\) and \(p_0\simeq0.6946\) give a 25\% missingness rate under each mechanism. All three are MAR because observation depends only on the fully observed \(D_i\) and \(\widetilde X_i\). As in Section~\ref{sec:motivating-simulation}, we report complete-case regression and treatment IPW; we also report linear and correctly specified multiple imputation (MI)\footnote{Linear MI uses a multinomial-logit model for \(X_i\) with linear predictors \(D_i\), \(Y_i\), and \(\widetilde X_i\). Correctly specified MI adds \(\log\cosh\{(Y_i-D_i-x)/0.5^2+(\widetilde X_i-x)/0.25^2\}\) for \(x\in\{-1,0,1\}\). Both use 20 imputations.}.

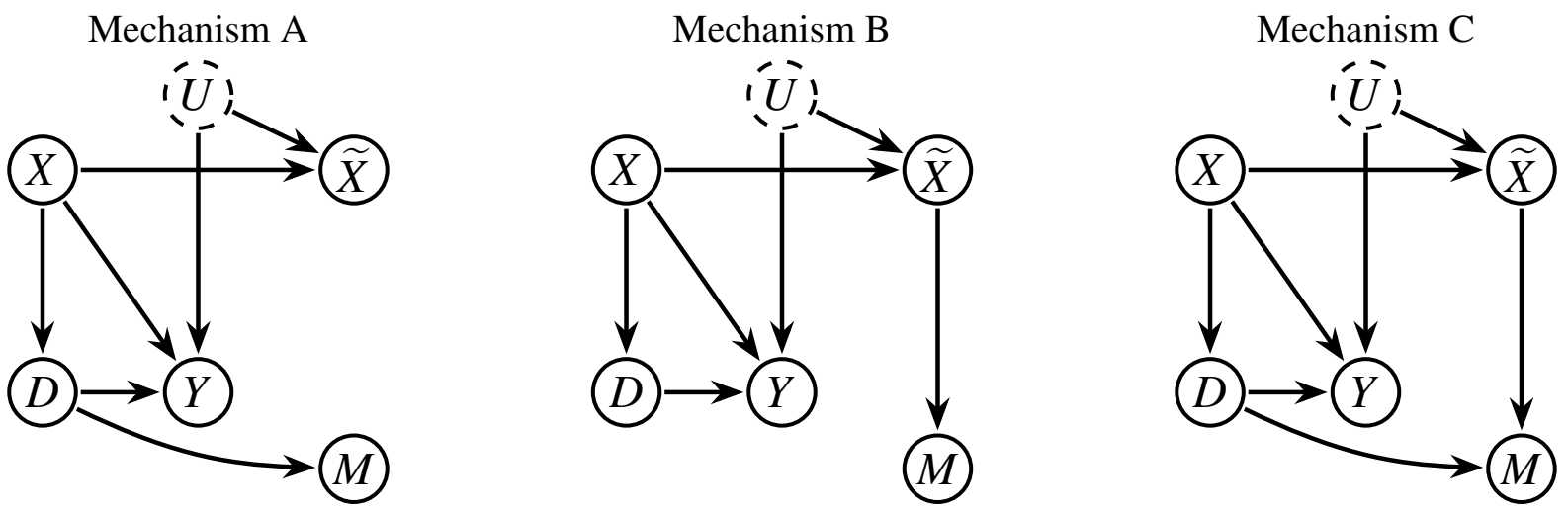
\begin{figure}[H]
\centering
\begin{tikzpicture}[manuscript dag,scale=0.62,transform shape]
\foreach \mechanism/\offset in {A/0,B/6.3,C/12.6} {
  \begin{scope}[xshift=\offset cm]
    \node at (0,1.9) {Mechanism \mechanism};
    \node[manuscript dag node] (X) at (-1.12,0.88) {\(X\)};
    \node[manuscript dag node,dashed] (U) at (0,1.42) {\(U\)};
    \node[manuscript dag node] (D) at (-1.12,-0.72) {\(D\)};
    \node[manuscript dag node] (Y) at (0,-0.72) {\(Y\)};
    \node[manuscript dag node] (proxy) at (1.12,0.88) {\(\widetilde X\)};
    \node[manuscript dag node] (M) at (1.12,-1.27) {\(M\)};
    \path[manuscript dag edge]
      (X) edge (D) (X) edge (Y) (X) edge (proxy)
      (U) edge (Y) (U) edge (proxy) (D) edge (Y);
    \if B\mechanism\else
      \path[manuscript dag edge] (D) edge[bend right=12] (M);
    \fi
    \if A\mechanism\else
      \path[manuscript dag edge] (proxy) edge (M);
    \fi
  \end{scope}
}
\end{tikzpicture}
\caption{Causal Diagrams for the MAR Missing Mechanisms}
\label{fig:motivating-mar-dags}
\vspace{0.4em}
\begin{minipage}{0.92\linewidth}
\footnotesize\emph{Note:} The dashed circle marks unobserved $U_i$. Only $X_i$ can be missing; $M_i=0$ denotes a complete case.
\end{minipage}
\end{figure}

\begin{figure}[H]
\centering
\includegraphics[width=0.92\textwidth]{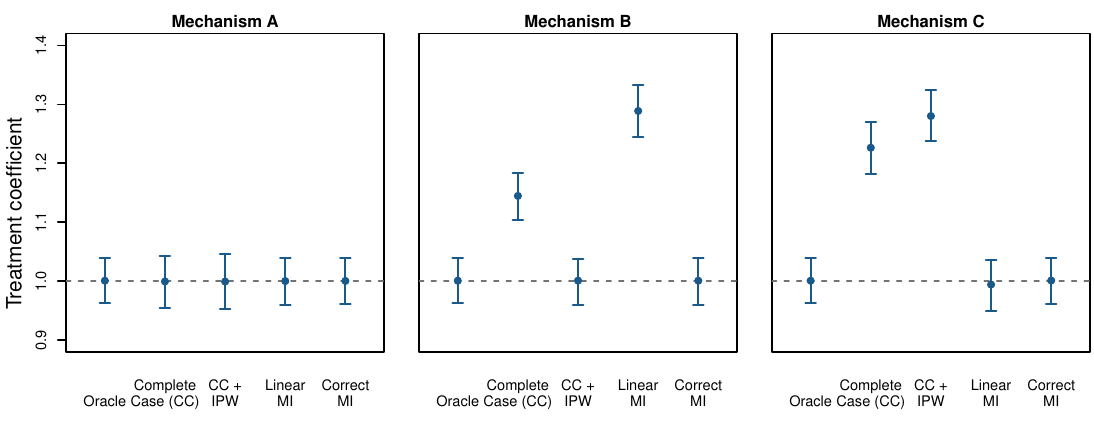}
\caption{Estimated Treatment Coefficients under MAR Missing Mechanisms}
\label{fig:motivating-mar-simulation}
\vspace{0.4em}
\begin{minipage}{0.92\linewidth}
\footnotesize\emph{Note:} Points show mean treatment-coefficient estimates across \MARreps{} replications. Bars show the 2.5th and 97.5th percentiles. The dashed line marks \(\tau=1\). CC + IPW uses inverse treatment-propensity weights.
\end{minipage}
\end{figure}

Figure~\ref{fig:motivating-mar-simulation} shows the same progression as the NMAR simulation. Moving from A to B illustrates the need to pool valid within-\(X_i\) comparisons appropriately; moving from B to C shows how the adjustment criterion can fail. Correctly specified MI remains centered near the target under all three MAR mechanisms, although linear MI misses the target under B.

\FloatBarrier
\section{Technical Definitions, Assumptions, and Proofs}
\label{app:adjustment-criterion}

\subsection{Adjustment criterion}

\paragraph{Adjustment criterion in the original paper} Definition~5 of \citet{shpitser2010validity} defines the adjustment criterion as follows, using \(\mathbf{D}\) for their treatment set \(X\) to distinguish it from our covariates \(X_i\). Let \(G\) be a causal DAG, and let \(\mathbf{D}\), \(\mathbf{Y}\), and \(\mathcal Z\) be pairwise disjoint sets of treatment, outcome, and observed adjustment variables. Let \(G_{\overline{\mathbf{D}}}\) denote the graph obtained by deleting all arrows into nodes in \(\mathbf{D}\). A directed path from \(\mathbf{D}\) to \(\mathbf{Y}\) is \emph{proper} if its only treatment node is its starting node. The set \(\mathcal Z\) satisfies the adjustment criterion relative to \((\mathbf{D},\mathbf{Y})\) in \(G\) if:
\begin{enumerate}
    \item In \(G_{\overline{\mathbf{D}}}\), no member of \(\mathcal Z\) may descend from a node \(v\notin\mathbf{D}\) that lies on a proper causal path from \(\mathbf{D}\) to \(\mathbf{Y}\) in \(G\).
    \item \(\mathcal Z\) blocks every noncausal path from \(\mathbf{D}\) to \(\mathbf{Y}\) in \(G\).
\end{enumerate}
A node counts as its own descendant. A noncausal path is one that is not directed from its treatment endpoint to its outcome endpoint.

\paragraph{Condition 1 with a single treatment} Our analysis has one treatment \(D\) and one outcome \(Y\), so every directed path from \(D\) to \(Y\) is proper. For any node \(v\neq D\) on such a path, no directed path from \(v\) to a descendant can pass through \(D\), because that would create a cycle. Deleting arrows into \(D\) therefore leaves the descendants of \(v\) unchanged:
\[
\operatorname{De}_{G_{\bar D}}(v)=\operatorname{De}_{G}(v).
\]
Condition 1 thus excludes nodes other than \(D\) on directed paths from \(D\) to \(Y\), together with their descendants in the original DAG.

\paragraph{Condition 2 with a single treatment} Condition 2 is unchanged: \(\mathcal Z\) must block every noncausal path from \(D\) to \(Y\) in the original graph \(G\), as stated in Section~\ref{sec:causal-conditions}.

\subsection{SUTVA, Oracle Samples, Missing Mechanism, and Proof of Proposition~\ref{prop:adjustment-ignorability}}

\begin{assumption}[Stable unit treatment value assumption (SUTVA)]
\label{ass:sutva}
Each treatment level has a single outcome-relevant version. Let \(\mathbf d=(d_1,\ldots,d_n)\) be a treatment assignment vector. For every unit \(i\) and every \(\mathbf d\), the unit's potential outcome depends only on its own assignment: \(Y_i(\mathbf d)=Y_i(d_i)\). For the realized assignment, \(Y_i=Y_i(D_i)\).
\end{assumption}

\begin{definition}[Oracle potential-outcome sample and oracle sample]
\label{def:oracle-samples} Let \(Y_i(d)\) be unit \(i\)'s potential outcome under treatment \(d\in\{0,1\}\). The oracle potential-outcome sample is
\[\mathcal S_{\mathrm{po}} =\{(Y_i(0),Y_i(1),D_i,X_i,\mathbf M_i):i=1,\ldots,n\}, \qquad (Y_i(0),Y_i(1),D_i,X_i,\mathbf M_i) \overset{\mathrm{i.i.d.}}{\sim}P_{\mathrm{po}}, \]
where \(P_{\mathrm{po}}\) is the joint population distribution of the respective unit-level vector.

Under SUTVA, let \(Y_i=Y_i(D_i)\) denote unit \(i\)'s realized outcome. The oracle sample is
\[\mathcal S_{\mathrm{or}}=\{(Y_i,D_i,X_i,\mathbf M_i):i=1,\ldots,n\},\qquad(Y_i,D_i,X_i,\mathbf M_i)\overset{\mathrm{i.i.d.}}{\sim}P_{\mathrm{or}},\]
where \(P_{\mathrm{or}}\) is the joint distribution induced by
\(P_{\mathrm{po}}\) and \(Y_i=Y_i(D_i)\). The sample contains every unit's
realized values, including those missing in the observed data, and retains
each unit's missing pattern.
\end{definition}

\paragraph{Missing mechanism}
\label{app:missing-mechanism}

\begin{definition}[Missing mechanism]
The missing mechanism specifies the conditional probability of
each missing pattern given the full realized data:
\[
\pi(\mathbf m\mid y,d,x)
=P(\mathbf M_i=\mathbf m\mid Y_i=y,D_i=d,X_i=x).
\]
\end{definition}

\begin{assumption}[Complete-case treatment overlap]
\label{ass:cc-treatment-overlap}
Let \(e_0(x)=P(D_i=1\mid X_i=x,M_i=0)\). For every \(x\) in the complete-case support, \(0<e_0(x)<1\).
\end{assumption}

\begin{proof}[Proof of Proposition~\ref{prop:adjustment-ignorability}(i)]
Apply \citet[Theorem~4]{shpitser2010validity} with their treatment \(X\) equal to \(D_i\) and their adjustment set \(Z\) equal to \(\{X_i,M_i\}\). Their intervention-defined counterfactual \(Y_x\) is our potential outcome \(Y_i(d)\) when \(x=d\).
\end{proof}

\begin{proof}[Proof of Proposition~\ref{prop:adjustment-ignorability}(ii)]
Fix a supported complete-case stratum \(X_i=x,M_i=0\) satisfying treatment overlap. The first equality follows from consistency, \(Y_i=Y_i(D_i)\), which means that a unit assigned \(d\) has realized outcome \(Y_i(d)\), while the second follows from Proposition~\ref{prop:adjustment-ignorability}(i):
\[
\begin{aligned}
\mathbb E[Y_i\mid D_i=d,X_i=x,M_i=0]
&=\mathbb E[Y_i(d)\mid D_i=d,X_i=x,M_i=0]\\
&=\mathbb E[Y_i(d)\mid X_i=x,M_i=0].
\end{aligned}
\]
Subtracting the expressions for \(d=1\) and \(d=0\) gives
\[
\begin{aligned}
&\mathbb E[Y_i\mid D_i=1,X_i=x,M_i=0]
-\mathbb E[Y_i\mid D_i=0,X_i=x,M_i=0]\\
&\qquad=\mathbb E[Y_i(1)-Y_i(0)\mid X_i=x,M_i=0]=\delta_{\mathrm{cc}}(x).
\end{aligned}
\]
Because \(D_i\) is binary, the population coefficient in a regression on an intercept and \(D_i\) within this stratum equals this difference in means. Under a constant unit-level treatment effect, \(\delta_{\mathrm{cc}}(x)=\tau\).
\end{proof}

\subsection{Proof of Proposition~\ref{prop:cc-ols-preservation}}
\label{app:proof-cc-ols-preservation}

\begin{assumption}[Complete-case OLS regularity]
\label{ass:cc-ols-regularity}
Let \(Z_i=(g(X_i)^\top,D_i h(X_i)^\top)^\top\), where \(g\) and \(h\) are fixed finite-dimensional vectors of chosen terms, each including an intercept. For the additive specification, take \(g(X_i)=b(X_i)\) and \(h(X_i)=1\). Suppose \(P(M_i=0)>0\),
\[
\mathbb E[Y_i^2+\|Z_i\|^2+\|h(X_i)\|^2\mid M_i=0]<\infty,
\]
and \(G_0=\mathbb E[Z_iZ_i^\top\mid M_i=0]\) is positive definite.
\end{assumption}

\begin{proof}[Proof of Proposition~\ref{prop:cc-ols-preservation}]
Correct specification of the complete-case conditional mean gives
\[
\mathbb E[Y_i\mid D_i,X_i,M_i=0]
=g(X_i)^\top\alpha_{\mathrm{cc}}+D_i h(X_i)^\top\beta_{\mathrm{cc}}.
\]
Under Assumption~\ref{ass:cc-treatment-overlap} (complete-case treatment overlap), both treatment arms have positive probability at each supported \(X_i=x\). Subtracting their conditional means cancels the baseline term \(g(x)^\top\alpha_{\mathrm{cc}}\), giving
\[
\mathbb E[Y_i\mid D_i=1,X_i=x,M_i=0]
-\mathbb E[Y_i\mid D_i=0,X_i=x,M_i=0]
=h(x)^\top\beta_{\mathrm{cc}}.
\]
By SUTVA (Assumption~\ref{ass:sutva}), a unit assigned \(d\) has realized outcome \(Y_i(d)\). The adjustment criterion then makes the comparison causal, as established in Proposition~\ref{prop:adjustment-ignorability}(ii). Hence
\[
\delta_{\mathrm{cc}}(x)=h(x)^\top\beta_{\mathrm{cc}}.
\]
Averaging over \(X_i\mid M_i=0\) and applying the law of iterated expectations gives
\[
\mathbb E[Y_i(1)-Y_i(0)\mid M_i=0]
=\mathbb E[h(X_i)^\top\beta_{\mathrm{cc}}\mid M_i=0]
=\tau_{\mathrm{cc}}.
\]
The integrability needed for this averaging follows from Assumption~\ref{ass:cc-population-regularity}.

For consistency, let \(\vartheta_{\mathrm{cc}}=(\alpha_{\mathrm{cc}}^\top,\beta_{\mathrm{cc}}^\top)^\top\). Multiplying the zero conditional residual mean by \(Z_i\) and averaging gives the complete-case population OLS normal equations:
\[
\mathbb E[Z_i(Y_i-Z_i^\top\vartheta_{\mathrm{cc}})\mid M_i=0]=0.
\]
Their solution is unique under Assumption~\ref{ass:cc-ols-regularity}. Define
\[
\widehat G_0=\frac{1}{n_{\mathrm{cc}}}\sum_{i:M_i=0}Z_iZ_i^\top,
\qquad
\widehat q_0=\frac{1}{n_{\mathrm{cc}}}\sum_{i:M_i=0}Z_iY_i.
\]
The i.i.d. sampling specified in Subsection~\ref{subsec:oracle-regression}, positive complete-case probability, and finite moments in Assumption~\ref{ass:cc-ols-regularity} imply, by the law of large numbers applied to selected sums and the complete-case fraction,
\[
\widehat G_0\xrightarrow{p}G_0,
\qquad
\widehat q_0\xrightarrow{p}G_0\vartheta_{\mathrm{cc}}.
\]
Positive definiteness of \(G_0\) and the continuous mapping theorem give
\[
\widehat\vartheta_{\mathrm{cc}}=\widehat G_0^{-1}\widehat q_0
\xrightarrow{p}\vartheta_{\mathrm{cc}}.
\]
In particular, \(\widehat\beta_{\mathrm{cc}}\xrightarrow{p}\beta_{\mathrm{cc}}\). The finite moments of \(h(X_i)\) also give
\[
\frac{1}{n_{\mathrm{cc}}}\sum_{i:M_i=0}h(X_i)
\xrightarrow{p}\mathbb E[h(X_i)\mid M_i=0].
\]
Taking the inner product of this sample average with \(\widehat\beta_{\mathrm{cc}}\) therefore yields
\[
\widehat\tau_{\mathrm{cc}}
=\left\{\frac{1}{n_{\mathrm{cc}}}\sum_{i:M_i=0}h(X_i)\right\}^\top\widehat\beta_{\mathrm{cc}}
\xrightarrow{p}\mathbb E[h(X_i)\mid M_i=0]^\top\beta_{\mathrm{cc}}
=\tau_{\mathrm{cc}}.
\]
\end{proof}

\paragraph{Treatment-induced attrition} The complete-case ATE, \(\tau_{\mathrm{cc}}=\mathbb E[Y_i(1)-Y_i(0)\mid M_i=0]\), averages effects among units observed under the realized assignment. If treatment affects attrition, this group may differ under another assignment. In this case, Propositions~\ref{prop:adjustment-ignorability}--\ref{prop:graphical-cc-ipw} still hold. Under Corollary~\ref{cor:cc-effect-transport}'s transport and support conditions, averaging complete-case conditional effects over the population covariate distribution recovers the population ATE.

\subsection{Oracle specification and the indicator--error independence condition of \cite{arel2018can}}
\label{app:oracle-mean-preservation}

For the additive special case \(g(X_i)=b(X_i)\) and \(h(X_i)=1\), the indicator--error independence condition of \citet{arel2018can}, written in our notation as \(M_i\perp\varepsilon_{\mathrm{or},i}\mid D_i,X_i\), together with correct oracle specification, implies the zero conditional mean requirement in Proposition~\ref{prop:cc-ols-preservation}. Proposition~\ref{prop:cc-ols-preservation} relaxes these specification and independence restrictions. Its additional graphical condition establishes the causal interpretation as the complete-case ATE, which is not addressed by their coefficient-preservation result.

\paragraph{Sufficient conditions}
Suppose the oracle conditional mean is correctly specified:
\[
Y_i=\beta_{\mathrm{or}}D_i+b(X_i)^\top\gamma_{\mathrm{or}}+\varepsilon_{\mathrm{or},i},
\qquad
\mathbb E[\varepsilon_{\mathrm{or},i}\mid D_i,X_i]=0.
\]
Suppose also that
\[
M_i\perp\varepsilon_{\mathrm{or},i}\mid D_i,X_i,
\]
and Assumption~\ref{ass:cc-ols-regularity} (complete-case OLS regularity) holds. Then the same regression among complete cases satisfies
\[
\mathbb E[\varepsilon_{\mathrm{cc},i}\mid D_i,X_i,M_i=0]=0.
\]
If \(\{X_i,M_i\}\) also satisfies the adjustment criterion and Assumptions~\ref{ass:sutva} (SUTVA), \ref{ass:cc-treatment-overlap} (complete-case treatment overlap), and~\ref{ass:cc-population-regularity} (complete-case population conditions) hold, then
\[
\tau_{\mathrm{cc}}=\mathbb E[Y_i(1)-Y_i(0)\mid M_i=0],
\qquad
\widehat\tau_{\mathrm{cc}}\xrightarrow{p}\tau_{\mathrm{cc}}.
\]

\begin{proof}
Conditional independence and the correctly specified oracle conditional mean imply, on the complete-case support,
\[
\mathbb E[\varepsilon_{\mathrm{or},i}\mid D_i,X_i,M_i=0]
=\mathbb E[\varepsilon_{\mathrm{or},i}\mid D_i,X_i]=0.
\]
Consequently,
\[
\mathbb E[Y_i\mid D_i,X_i,M_i=0]
=\beta_{\mathrm{or}}D_i+b(X_i)^\top\gamma_{\mathrm{or}}.
\]
These coefficients satisfy the complete-case population OLS normal equations. Assumption~\ref{ass:cc-ols-regularity} ensures their uniqueness, giving \(\beta_{\mathrm{cc}}=\beta_{\mathrm{or}}\) and \(\gamma_{\mathrm{cc}}=\gamma_{\mathrm{or}}\). Thus the complete-case residual \(\varepsilon_{\mathrm{cc},i}\) equals \(\varepsilon_{\mathrm{or},i}\), and
\[
\mathbb E[\varepsilon_{\mathrm{cc},i}\mid D_i,X_i,M_i=0]=0.
\]
Under the adjustment criterion and Assumptions~\ref{ass:sutva}, \ref{ass:cc-treatment-overlap}, and~\ref{ass:cc-population-regularity}, Proposition~\ref{prop:cc-ols-preservation} therefore gives
\[
\tau_{\mathrm{cc}}=\mathbb E[Y_i(1)-Y_i(0)\mid M_i=0],
\]
and consistency of \(\widehat\tau_{\mathrm{cc}}\) under the maintained i.i.d. sampling in Subsection~\ref{subsec:oracle-regression}.
\end{proof}

These assumptions are stronger than the complete-case conditional mean condition in Proposition~\ref{prop:cc-ols-preservation}: the proof needs only preservation of the zero conditional residual mean, whereas conditional independence preserves the entire conditional outcome distribution given the regressors.

\paragraph{Violation of conditional independence in Mechanism B} Mechanism B violates this independence condition. Its oracle residual is \(U_i+\xi_i\), with \(\mathbb E[U_i+\xi_i\mid D_i,X_i]=0\). Because selection depends on \(U_i\), however,
\[
\mathbb E[U_i+\xi_i\mid D_i,X_i,M_i=0]=X_i^2/2.
\]
Thus the oracle residual is not conditionally independent of \(M_i\) given \(D_i,X_i\), and the same intercept-and-linear-\(X_i\) specification no longer captures the complete-case mean. Nevertheless, the graphical adjustment condition still gives valid within-\(X_i\) treatment comparisons. Saturated outcome adjustment or treatment IPW can recover their common effect, as shown in the motivating simulation.

MAR does not generally imply the indicator--error independence condition: it permits selection on observed outcome values \citep{seaman2013what}. If only \(X_i\) can be missing while \(Y_i,D_i\) are observed, \(P(M_i=0\mid Y_i,D_i,X_i)=P(M_i=0\mid Y_i,D_i)\) is MAR, yet selection through \(Y_i\) may depend on the oracle residual. If only \(Y_i\) can be missing and \(D_i,X_i\) are fully observed, MAR gives the indicator--error independence condition. With a correctly specified oracle conditional mean, it then preserves the zero conditional residual mean among complete cases.

\subsection{Proof of Proposition~\ref{prop:cc-ipw}}
\label{app:proof-cc-ipw}

\begin{assumption}[Complete-case population conditions]
\label{ass:cc-population-regularity}
\(P(M_i=0)>0\) and \(\mathbb E[|Y_i(d)|\mid M_i=0]<\infty\) for \(d\in\{0,1\}\).
\end{assumption}

\begin{assumption}[Estimated inverse-weight consistency]
\label{ass:cc-ipw-weight-consistency}
With probability tending to one, \(0<\widehat e_0(x)<1\) throughout the complete-case support, and the estimated inverse weights converge uniformly over that support:
\[
\sup_x\left|\frac{1}{\widehat e_0(x)}-\frac{1}{e_0(x)}\right|=o_p(1),
\qquad
\sup_x\left|\frac{1}{1-\widehat e_0(x)}-\frac{1}{1-e_0(x)}\right|=o_p(1).
\]
\end{assumption}

\begin{propositionA}[Complete-case treatment IPW]
\label{prop:cc-ipw}
Let \(e_0(x)=P(D_i=1\mid X_i=x,M_i=0)\). Suppose Assumption~\ref{ass:sutva} (SUTVA), Assumption~\ref{ass:cc-treatment-overlap} (complete-case treatment overlap), and Assumption~\ref{ass:cc-population-regularity} (complete-case population conditions) hold, and \(Y_i(d)\perp D_i\mid X_i,M_i=0\) for \(d\in\{0,1\}\). Then
\[
\mathbb E\!\left[\left.\frac{D_iY_i}{e_0(X_i)}
-\frac{(1-D_i)Y_i}{1-e_0(X_i)}\,\right|M_i=0\right]
=\mathbb E[Y_i(1)-Y_i(0)\mid M_i=0].
\]

Under Assumption~\ref{ass:cc-ipw-weight-consistency}, the sample analogue
\[
\widehat\tau_{\mathrm{IPW}}
=\frac{1}{n_{\mathrm{cc}}}\sum_{i:M_i=0}
\left\{\frac{D_iY_i}{\widehat e_0(X_i)}
-\frac{(1-D_i)Y_i}{1-\widehat e_0(X_i)}\right\},
\qquad n_{\mathrm{cc}}=\sum_{i=1}^n\mathbf 1\{M_i=0\},
\]
converges in probability to \(\mathbb E[Y_i(1)-Y_i(0)\mid M_i=0]\).
\end{propositionA}

Here and below, \(o_p(1)\) denotes convergence to zero in probability, and \(O_p(1)\) denotes boundedness in probability.

\paragraph{Proof of Proposition \ref{prop:cc-ipw}} For treated units, Assumption~\ref{ass:sutva} (SUTVA) gives \(D_iY_i=D_iY_i(1)\). The condition \(Y_i(1)\perp D_i\mid X_i,M_i=0\) in Proposition~\ref{prop:cc-ipw} and the definition of \(e_0(x)\) then imply, for each supported \(x\),
\[
\begin{aligned}
\mathbb E\!\left[\left.\frac{D_iY_i}{e_0(X_i)}\,\right|X_i=x,M_i=0\right]
&=\frac{\mathbb E[D_i\mid X_i=x,M_i=0]}{e_0(x)}
\mathbb E[Y_i(1)\mid X_i=x,M_i=0]\\
&=\mathbb E[Y_i(1)\mid X_i=x,M_i=0].
\end{aligned}
\]
The same argument for untreated units gives
\[
\mathbb E\!\left[\left.\frac{(1-D_i)Y_i}{1-e_0(X_i)}\,\right|X_i=x,M_i=0\right]
=\mathbb E[Y_i(0)\mid X_i=x,M_i=0].
\]
Subtracting and averaging over \(X_i\mid M_i=0\) proves the stated identity.

Specifically, define the sample analogue using the true complete-case propensity score as
\[
\widetilde\tau_{\mathrm{IPW}}
=\frac{1}{n_{\mathrm{cc}}}\sum_{i:M_i=0}
\left\{\frac{D_iY_i}{e_0(X_i)}
-\frac{(1-D_i)Y_i}{1-e_0(X_i)}\right\}.
\]
Let \(\psi_i=D_iY_i/e_0(X_i)-(1-D_i)Y_i/[1-e_0(X_i)]\) when \(M_i=0\), and set \(\psi_i=0\) otherwise. The weighted terms have finite absolute means: Assumption~\ref{ass:sutva}, the conditional independence in Proposition~\ref{prop:cc-ipw}, and Assumption~\ref{ass:cc-population-regularity} give
\[
\begin{aligned}
\mathbb E\!\left[\left.\frac{D_i|Y_i|}{e_0(X_i)}\,\right|M_i=0\right]
&=\mathbb E[|Y_i(1)|\mid M_i=0]<\infty,\\
\mathbb E\!\left[\left.\frac{(1-D_i)|Y_i|}{1-e_0(X_i)}\,\right|M_i=0\right]
&=\mathbb E[|Y_i(0)|\mid M_i=0]<\infty.
\end{aligned}
\]
Because \((Y_i(0),Y_i(1),D_i,X_i,\mathbf M_i)\overset{\mathrm{i.i.d.}}{\sim}P_{\mathrm{po}}\), as specified in the oracle potential-outcome sample definition, the complete cases are draws from the conditional distribution given \(M_i=0\). Under Assumption~\ref{ass:cc-population-regularity}, the law of large numbers gives
\[
\widetilde\tau_{\mathrm{IPW}}
=\frac{n^{-1}\sum_{i=1}^n\mathbf 1\{M_i=0\}\psi_i}{n^{-1}\sum_{i=1}^n\mathbf 1\{M_i=0\}}
\xrightarrow{p}\mathbb E[\psi_i\mid M_i=0]
=\mathbb E[Y_i(1)-Y_i(0)\mid M_i=0].
\]
Moreover,
\[
|\widehat\tau_{\mathrm{IPW}}-\widetilde\tau_{\mathrm{IPW}}|
\leq
\max\!\left\{
\sup_x\left|\frac{1}{\widehat e_0(x)}-\frac{1}{e_0(x)}\right|,
\sup_x\left|\frac{1}{1-\widehat e_0(x)}-\frac{1}{1-e_0(x)}\right|
\right\}
\frac{1}{n_{\mathrm{cc}}}\sum_{i:M_i=0}|Y_i|.
\]
Assumption~\ref{ass:cc-ipw-weight-consistency} makes the maximum of the two uniform inverse-weight differences \(o_p(1)\). Under the i.i.d. sampling and finite-moment conditions in Assumption~\ref{ass:cc-population-regularity}, the complete-case sample mean \(n_{\mathrm{cc}}^{-1}\sum_{i:M_i=0}|Y_i|\) is \(O_p(1)\). Thus \(\widehat\tau_{\mathrm{IPW}}-\widetilde\tau_{\mathrm{IPW}}=o_p(1)\), and \(\widehat\tau_{\mathrm{IPW}}\) converges to the same population contrast.

\subsection{Proof of Corollary~\ref{cor:cc-ols-ipw-equivalence}}
\label{app:proof-cc-ols-ipw-equivalence}

Under the conditions of Proposition~\ref{prop:graphical-cc-ipw}, Proposition~\ref{prop:adjustment-ignorability}(i) gives \(Y_i(d)\perp D_i\mid X_i,M_i\) for \(d\in\{0,1\}\), so the IPW identity in Proposition~\ref{prop:cc-ipw} identifies the complete-case ATE. With \(e_0(x)=\bar e_0\) constant, treatment is independent of \(X_i\) among complete cases, and \(P(D_i=1\mid M_i=0)=\mathbb E[e_0(X_i)\mid M_i=0]=\bar e_0\). Thus, the distribution of \(X_i\) is the same in the treatment and control groups, and the IPW identity reduces to the difference between their average outcomes. For the treated term,
\[
\begin{aligned}
\mathbb E\!\left[\left.\frac{D_iY_i}{\bar e_0}\right|M_i=0\right]
&=\frac{P(D_i=1\mid M_i=0)}{\bar e_0}\mathbb E[Y_i\mid D_i=1,M_i=0]\\
&=\mathbb E[Y_i\mid D_i=1,M_i=0].
\end{aligned}
\]
Similarly, the untreated term equals \(\mathbb E[Y_i\mid D_i=0,M_i=0]\). Hence,
\[
\mathbb E[Y_i\mid D_i=1,M_i=0]-\mathbb E[Y_i\mid D_i=0,M_i=0]=\tau_{\mathrm{cc}}.
\]

By the Frisch--Waugh--Lovell theorem, \(\beta_{\mathrm{cc}}\) is the coefficient from regressing \(Y_i\) on the residual from projecting \(D_i\) on \(g(X_i)\) among complete cases. Because \(\mathbb E[D_i\mid X_i,M_i=0]=\bar e_0\) and \(g(X_i)\) includes an intercept, that residual is \(D_i-\bar e_0\). Let \(\mu_d=\mathbb E[Y_i\mid D_i=d,M_i=0]\) for \(d\in\{0,1\}\). Since \(D_i\) is binary,
\[
\begin{aligned}
\operatorname{Cov}(D_i,Y_i\mid M_i=0)
&=\bar e_0\mu_1-\bar e_0\{\bar e_0\mu_1+(1-\bar e_0)\mu_0\}\\
&=\bar e_0(1-\bar e_0)(\mu_1-\mu_0),\\
\operatorname{Var}(D_i\mid M_i=0)&=\bar e_0(1-\bar e_0).
\end{aligned}
\]
Subtracting \(\bar e_0\) from \(D_i\) changes neither its covariance with \(Y_i\) nor its variance. Thus,
\[
\begin{aligned}
\beta_{\mathrm{cc}}
&=\frac{\operatorname{Cov}(D_i-\bar e_0,Y_i\mid M_i=0)}{\operatorname{Var}(D_i-\bar e_0\mid M_i=0)}\\
&=\mathbb E[Y_i\mid D_i=1,M_i=0]-\mathbb E[Y_i\mid D_i=0,M_i=0]
=\tau_{\mathrm{cc}}.
\end{aligned}
\]
Consistency follows as in Proposition~\ref{prop:cc-ols-preservation}. \(\blacksquare\)

\subsection{Proof of Corollary~\ref{cor:graphical-cc-ipw-constant-effect}}

Under a constant unit-level treatment effect, \(\mathbb E[Y_i(1)-Y_i(0)\mid M_i=m]=\tau\). Thus, by the law of total expectation,
\[
\mathbb E[Y_i(1)-Y_i(0)]
=\int \mathbb E[Y_i(1)-Y_i(0)\mid M_i=m]\,dF_M(m)
=\int \tau\,dF_M(m)=\tau.
\]

In particular, \(\mathbb E[Y_i(1)-Y_i(0)\mid M_i=0]=\tau\). Substituting this equality into Proposition~\ref{prop:cc-ols-preservation} gives \(\tau_{\mathrm{cc}}=\tau\) and consistency of \(\widehat\tau_{\mathrm{cc}}\) under that proposition's conditions. Proposition~\ref{prop:graphical-cc-ipw} gives \(\beta^*=\tau\) under its own conditions and consistency of \(\widehat\beta\) under Assumption~\ref{ass:cc-ipw-weight-consistency} (estimated inverse-weight consistency). Corollary~\ref{cor:cc-ols-ipw-equivalence} gives \(\beta_{\mathrm{cc}}=\tau_{\mathrm{cc}}=\tau\) and consistency of \(\widehat\beta_{\mathrm{cc}}\) under its own conditions.

\subsection{Proof of Corollary~\ref{cor:cc-effect-transport}}
\label{app:proof-cc-effect-transport}

The complete-case support condition in Corollary~\ref{cor:cc-effect-transport}, \(P(M_i=0\mid W_i=w)>0\) for \(F_{\mathrm{po}}^W\)-almost every \(w\), ensures that \(\tau_{\mathrm{cc}}(w)\) is defined throughout the population support of \(W_i\) up to a set of probability zero. By mean exchangeability given \(W_i\) and the law of iterated expectations,
\[
\begin{aligned}
\int\tau_{\mathrm{cc}}(w)\,dF_{\mathrm{po}}^W(w)
&=\int\mathbb E[Y_i(1)-Y_i(0)\mid W_i=w] \,dF_{\mathrm{po}}^W(w)\\
&=\mathbb E[Y_i(1)-Y_i(0)]=\tau.
\end{aligned}
\]
For the observed-outcome representation, Assumption~\ref{ass:sutva} (SUTVA), Proposition~\ref{prop:adjustment-ignorability}(i) (conditional ignorability given the full adjustment vector), and Assumption~\ref{ass:cc-treatment-overlap} (complete-case treatment overlap) give, for \(d\in\{0,1\}\),
\[
\begin{aligned}
\mathbb E[Y_i\mid D_i=d,X_i=x,M_i=0]
&=\mathbb E[Y_i(d)\mid D_i=d,X_i=x,M_i=0]\\
&=\mathbb E[Y_i(d)\mid X_i=x,M_i=0].
\end{aligned}
\]
Subtracting the expression for \(d=0\) from that for \(d=1\) gives \(\delta_{\mathrm{cc}}(x)\). Since \(W_i\) is a subvector of \(X_i\), iterated expectations within complete cases give
\[
\begin{aligned}
\int\delta_{\mathrm{cc}}(x)\,dF_{\mathrm{cc}}(x\mid W_i=w)
&=\mathbb E[Y_i(1)-Y_i(0)\mid W_i=w,M_i=0]\\
&=\tau_{\mathrm{cc}}(w).
\end{aligned}
\]
This inner averaging uses only the covariate support among complete cases within each \(W_i=w\) group. The transport and support assumptions at the \(W_i\) level then justify the outer population average. When \(W_i=X_i\), the inner distribution is concentrated at \(x=w\), recovering the full-covariate result.

\subsection{Complete-case IPW regression}
\label{app:proof-cc-ipw-wls}

The following result applies \citet[Section~3 and Supporting Information~S1]{gabriel2024inverse} to complete cases.

\begin{assumption}[Complete-case regression moments]
\label{ass:cc-wls-moments}
\(\mathbb E[Y_i(d)^2\mid M_i=0]<\infty\) for \(d\in\{0,1\}\), \(\mathbb E[\|X_i\|^2\mid M_i=0]<\infty\), and
\(\mathbb E[(1,X_i^\top)^\top(1,X_i^\top)\mid M_i=0]\) is positive definite.
\end{assumption}

\begin{propositionA}[Complete-case IPW regression]
\label{prop:cc-ipw-wls}
Let \(\omega_i=D_i/e_0(X_i)+(1-D_i)/\{1-e_0(X_i)\}\), and let \(\widehat\omega_i\) be its analogue with \(\widehat e_0\) in place of \(e_0\). Let \((\widehat\alpha,\widehat\beta,\widehat\gamma)\) minimize
\[
\sum_{i:M_i=0}\widehat\omega_i\left(Y_i-\alpha-\beta D_i-X_i^\top\gamma\right)^2 .
\]
Suppose the conditions of Proposition~\ref{prop:cc-ipw} and Assumption~\ref{ass:cc-wls-moments} (complete-case regression moments) hold. Then the population weighted least-squares coefficient on \(D_i\) among complete cases equals \(\mathbb E[Y_i(1)-Y_i(0)\mid M_i=0]\), whether or not \(\mathbb E[Y_i\mid D_i,X_i,M_i=0]\) is linear in \(D_i\) and \(X_i\). Under Assumption~\ref{ass:cc-ipw-weight-consistency}, \(\widehat\beta\) converges in probability to \(\mathbb E[Y_i(1)-Y_i(0)\mid M_i=0]\).
\end{propositionA}

\paragraph{Proof of Proposition~\ref{prop:cc-ipw-wls}}

Let \(Z_i=(1,D_i,X_i^\top)^\top\) be the regressor vector and \(\theta=(\alpha,\beta,\gamma^\top)^\top\) be the coefficient vector.

\emph{Step 1: The population coefficients exist and are unique.}
Let \(\phi(X_i)\) be any function with \(\mathbb E[|\phi(X_i)|\mid M_i=0]<\infty\). Because \(\mathbb E[D_i\mid X_i,M_i=0]=e_0(X_i)\), treatment overlap gives
\[
\mathbb E[\omega_iD_i\phi(X_i)\mid X_i,M_i=0]
=\frac{\phi(X_i)}{e_0(X_i)}\mathbb E[D_i\mid X_i,M_i=0]
=\phi(X_i).
\]
The same calculation with \(1-D_i\) gives \(\mathbb E[\omega_i(1-D_i)\phi(X_i)\mid X_i,M_i=0]=\phi(X_i)\). Taking expectations over \(X_i\mid M_i=0\) yields
\[
\mathbb E[\omega_iD_i\phi(X_i)\mid M_i=0]
=\mathbb E[\omega_i(1-D_i)\phi(X_i)\mid M_i=0]
=\mathbb E[\phi(X_i)\mid M_i=0].
\tag{\thesection.1}\label{eq:cc-ipw-covariate-moments}
\]
For a treated complete case, \(\omega_i\) is the inverse treatment probability \(1/e_0(X_i)\), while for an untreated complete case, it is \(1/\{1-e_0(X_i)\}\). Equation \eqref{eq:cc-ipw-covariate-moments} shows that weighting either treatment arm this way reproduces the covariate moments of the complete-case population.

Hence \(Q=\mathbb E[\omega_iZ_iZ_i^\top\mid M_i=0]\) is finite under Assumption~\ref{ass:cc-wls-moments}. 

For any \(a=(a_0,a_1,a_2^\top)^\top\), \eqref{eq:cc-ipw-covariate-moments} gives
\[
\begin{aligned}
a^\top Qa
&=\mathbb E\!\left[\omega_i(a^\top Z_i)^2\mid M_i=0\right]\\
&=\mathbb E\!\left[
\frac{D_i}{e_0(X_i)}(a_0+a_1+a_2^\top X_i)^2
+\frac{1-D_i}{1-e_0(X_i)}(a_0+a_2^\top X_i)^2
\,\middle|\,M_i=0\right]\\
&=\mathbb E\!\left[(a_0+a_1+a_2^\top X_i)^2\mid M_i=0\right]
+\mathbb E\!\left[(a_0+a_2^\top X_i)^2\mid M_i=0\right].
\end{aligned}
\]
The first equality follows from \(Q=\mathbb E[\omega_iZ_iZ_i^\top\mid M_i=0]\) and \(a^\top Z_iZ_i^\top a=(a^\top Z_i)^2\). The second substitutes the definitions of \(Z_i\) and \(\omega_i\): when \(D_i=1\), \(a^\top Z_i=a_0+a_1+a_2^\top X_i\) and \(\omega_i=1/e_0(X_i)\); when \(D_i=0\), \(a^\top Z_i=a_0+a_2^\top X_i\) and \(\omega_i=1/\{1-e_0(X_i)\}\).

The expression for \(a^\top Qa\) is the conditional expectation of two squared quantities. Each square is nonnegative. If their sum has expectation zero, both must be zero almost surely among complete cases. Thus
\[
a_0+a_1+a_2^\top X_i=0
\quad\text{and}\quad
a_0+a_2^\top X_i=0
\qquad\text{almost surely given }M_i=0.
\]
Subtracting the second equation from the first gives \(a_1=0\).

By the positive-definiteness condition in Assumption~\ref{ass:cc-wls-moments}, \(a_0+a_2^\top X_i=0\) almost surely implies \(a_0=0\) and \(a_2=0\). Hence \(a=0\), so \(Q\) is positive definite.

Let \(s_i=(1+D_i^2+X_i^\top X_i)^{1/2}\), the length of the regressor vector \(Z_i\). Similarly, we have
\[
\begin{aligned}
\mathbb E[\omega_i s_i\,|Y_i|\mid M_i=0]
&=\sum_{d\in\{0,1\}}\mathbb E\!\left[(1+d^2+X_i^\top X_i)^{1/2}\,|Y_i(d)|\,\middle|\,M_i=0\right]\\
&\le\sum_{d\in\{0,1\}}
\left\{\mathbb E[1+d^2+X_i^\top X_i\mid M_i=0]\right\}^{1/2}
\left\{\mathbb E[Y_i(d)^2\mid M_i=0]\right\}^{1/2}
<\infty,
\end{aligned}
\]

For the equality, SUTVA gives \(Y_i=Y_i(d)\) when \(D_i=d\), and then \(s_i=(1+d^2+X_i^\top X_i)^{1/2}\). Conditional independence of \(Y_i(d)\) and \(D_i\) given \(X_i,M_i=0\) lets the treatment probability \(e_0(X_i)\) for \(d=1\), or \(1-e_0(X_i)\) for \(d=0\), cancel its inverse in \(\omega_i\). The inequality applies the Cauchy--Schwarz inequality to \((1+d^2+X_i^\top X_i)^{1/2}\) and \(|Y_i(d)|\) for each \(d\). The final bound is finite by Assumption~\ref{ass:cc-wls-moments}. Because each component of \(Z_i\) satisfies \(|Z_{ij}|\le s_i\), this bound makes every component of \(\mathbb E[\omega_iZ_iY_i\mid M_i=0]\) finite. Thus \(q=\mathbb E[\omega_iZ_iY_i\mid M_i=0]\) is finite. 

The population weighted least-squares coefficient vector minimizes the expected weighted squared error
\[
\mathcal L(\theta)=\mathbb E[\omega_i(Y_i-Z_i^\top\theta)^2\mid M_i=0].
\]
Expanding the square and using the definitions of \(Q\) and \(q\) gives
\[
\begin{aligned}
\mathcal L(\theta)
&=\mathbb E[\omega_iY_i^2\mid M_i=0]
-2\theta^\top\mathbb E[\omega_iZ_iY_i\mid M_i=0]
+\theta^\top\mathbb E[\omega_iZ_iZ_i^\top\mid M_i=0]\theta\\
&=\mathbb E[\omega_iY_i^2\mid M_i=0]-2\theta^\top q+\theta^\top Q\theta.
\end{aligned}
\]
The first term is finite because the same weighting argument gives \(\mathbb E[\omega_iY_i^2\mid M_i=0]=\sum_{d\in\{0,1\}}\mathbb E[Y_i(d)^2\mid M_i=0]<\infty\). Since \(Q\) and \(q\) are also finite, \(\mathcal L(\theta)\) is finite for every \(\theta\).

The first-order condition gives \(Q\theta=q\). Because \(Q\) is positive definite, the objective is strictly convex and its unique minimizer is \(\theta^*=(\alpha^*,\beta^*,\gamma^{*\top})^\top=Q^{-1}q\). Substituting the definitions of \(Q\) and \(q\) into \(Q\theta^*=q\) gives
\[
\mathbb E[\omega_iZ_i(Y_i-Z_i^\top\theta^*)\mid M_i=0]=0 .
\tag{\thesection.2}\label{eq:cc-ipw-wls-normal-equations}
\]

\emph{Step 2: The coefficient on \(D_i\) is the complete-case ATE.}

Let \(r_i=Y_i-\alpha^*-\beta^*D_i-X_i^\top\gamma^*\), where \(\theta^*=(\alpha^*,\beta^*,\gamma^{*\top})^\top=Q^{-1}q\) is the population weighted least-squares coefficient vector.

Since \(Z_i=(1,D_i,X_i^\top)^\top\), equation \eqref{eq:cc-ipw-wls-normal-equations} can be written as
\[
\mathbb E\!\left[
\omega_i
\begin{pmatrix}
r_i\\
D_i r_i\\
X_i r_i
\end{pmatrix}
\middle| M_i=0\right]=0.
\]
Its first two components give \(\mathbb E[\omega_ir_i\mid M_i=0]=0\) and \(\mathbb E[\omega_iD_ir_i\mid M_i=0]=0\). Hence \(\mathbb E[\omega_i(1-D_i)r_i\mid M_i=0]=0\). 

As in the proof of Proposition~\ref{prop:cc-ipw}, applying IPW to the regression residuals, together with SUTVA and \(Y_i(d)\perp D_i\mid X_i,M_i=0\), gives, for each supported \(x\), 
\[
\begin{aligned}
\mathbb E[\omega_iD_ir_i\mid X_i=x,M_i=0]
&=\mathbb E[Y_i(1)\mid X_i=x,M_i=0]-\alpha^*-\beta^*-x^\top\gamma^*,\\
\mathbb E[\omega_i(1-D_i)r_i\mid X_i=x,M_i=0]
&=\mathbb E[Y_i(0)\mid X_i=x,M_i=0]-\alpha^*-x^\top\gamma^*.
\end{aligned}
\]
For the first equality, untreated units contribute zero because \(D_i=0\). Among treated units at \(X_i=x\), the residual is \(Y_i(1)-\alpha^*-\beta^*-x^\top\gamma^*\) and the weight is \(1/e_0(x)\). Since the treated fraction is \(e_0(x)\), the fraction and its inverse cancel. Conditional ignorability then replaces the treated mean of \(Y_i(1)\) with its mean among all complete cases at \(x\). Thus, at fixed \(x\), the weighted treated residual mean equals the mean treated potential outcome minus the treated fitted value \(\alpha^*+\beta^*+x^\top\gamma^*\). The same reasoning applies to the untreated term. 

Averaging over \(X_i\mid M_i=0\) and setting both expressions to zero gives
\[
\begin{aligned}
\mathbb E[Y_i(1)\mid M_i=0]&=\alpha^*+\beta^*+\mathbb E[X_i\mid M_i=0]^\top\gamma^*,\\
\mathbb E[Y_i(0)\mid M_i=0]&=\alpha^*+\mathbb E[X_i\mid M_i=0]^\top\gamma^*.
\end{aligned}
\]
Subtracting yields \(\beta^*=\mathbb E[Y_i(1)-Y_i(0)\mid M_i=0]\). 

We note here that the argument does not assume a linear model for \(\mathbb E[Y_i\mid D_i,X_i,M_i=0]\). To identify \(\beta^*\), it uses only the intercept and \(D_i\) components of \eqref{eq:cc-ipw-wls-normal-equations}. Further, the terms involving \(\gamma^*\) cancel when the treated and untreated equations are subtracted.

\emph{Step 3: Consistency.}
Let \(\widetilde Q\) and \(\widetilde q\) be the complete-case sample averages of \(\omega_iZ_iZ_i^\top\) and \(\omega_iZ_iY_i\), and let \(\widehat Q\) and \(\widehat q\) be the same averages with \(\widehat\omega_i\) in place of \(\omega_i\). Whenever \(\widehat Q\) is invertible, \(\widehat\theta=\widehat Q^{-1}\widehat q\). By Step 1 these weighted terms have finite means. As in the proof of Proposition~\ref{prop:cc-ipw}, the law of large numbers gives \(\widetilde Q\xrightarrow{p}Q\) and \(\widetilde q\xrightarrow{p}q\). Let
\[
r_n=\max\!\left\{\sup_x\left|\frac{1}{\widehat e_0(x)}-\frac{1}{e_0(x)}\right|,\ \sup_x\left|\frac{1}{1-\widehat e_0(x)}-\frac{1}{1-e_0(x)}\right|\right\},
\]
which is \(o_p(1)\) by Assumption~\ref{ass:cc-ipw-weight-consistency}. Since \(|\widehat\omega_i-\omega_i|\le r_n\),
\[
\|\widehat Q-\widetilde Q\|\le r_n\frac{1}{n_{\mathrm{cc}}}\sum_{i:M_i=0}s_i^2,
\qquad
\|\widehat q-\widetilde q\|\le r_n\frac{1}{n_{\mathrm{cc}}}\sum_{i:M_i=0}s_i\,|Y_i|.
\]
Both averages are \(O_p(1)\) by Assumption~\ref{ass:cc-wls-moments} and the Cauchy--Schwarz inequality, so both differences are \(o_p(1)\). Hence \(\widehat Q\xrightarrow{p}Q\) and \(\widehat q\xrightarrow{p}q\). Because \(Q\) is invertible, \(\widehat Q\) is invertible with probability tending to one. The continuous mapping theorem then gives \(\widehat\theta\xrightarrow{p}\theta^*\), and in particular \(\widehat\beta\xrightarrow{p}\mathbb E[Y_i(1)-Y_i(0)\mid M_i=0]\).

\FloatBarrier
\section{Simulation of the Collider--Mediator Dilemma}
\label{app:collider-mediator-simulation}

We assume the following data generation process, where the total causal effect of $D$ on $Y$ is $0.8\times1.5+0.8=2$, while the controlled direct effect \citep{imai2011unpacking} of $D$ on $Y$ is $0.8$.
The outcome $Y$ is missing according to the latent index $L_i=1.5D_i+2A_i+\eta_i$.
The threshold $c$ in the missingness determination equation is selected such that the expected probability of missingness is $30\%$, i.e.
$\mathbb{E}\left[M_{i}\right]=\mathbb{P}\left(L_{i}\geq c\right)=0.3$.

\begin{align*}
D_{i}&\sim Bernoulli\left(0.5\right)\\
A_{i}&=1+0.8D_{i}+u_{i},\qquad u_{i}\sim N\left(0,1\right)\\
Y_{i}&=1+0.8D_{i}+1.5A_{i}+\zeta_{i},\qquad\zeta_{i}\sim N\left(0,1\right)\\
M_{i}&=I\left\{ 1.5D_{i}+2A_{i}+\eta_{i}\geq c\right\} ,\qquad\eta_{i}\sim N\left(0,1\right)
\end{align*}

We draw 10,000 samples of 5,000 observations. As in the main-text illustration, we estimate the oracle regression, complete-case regression with and without \(A_i\), and multiple imputation. Here, the synthetic models use the variables in the DGP above, without additional baseline controls. For MI, we impute \(Y\) using \((D,A)\), or \(D\) using \((Y,A)\), and then regress \(Y\) on \(D\) in the imputed datasets \citep{little2019statistical}.

\paragraph{Simulation Results}

\begin{figure}[htbp]
\centering
\begin{minipage}[t]{0.49\textwidth}
\centering
\includegraphics[width=\linewidth]{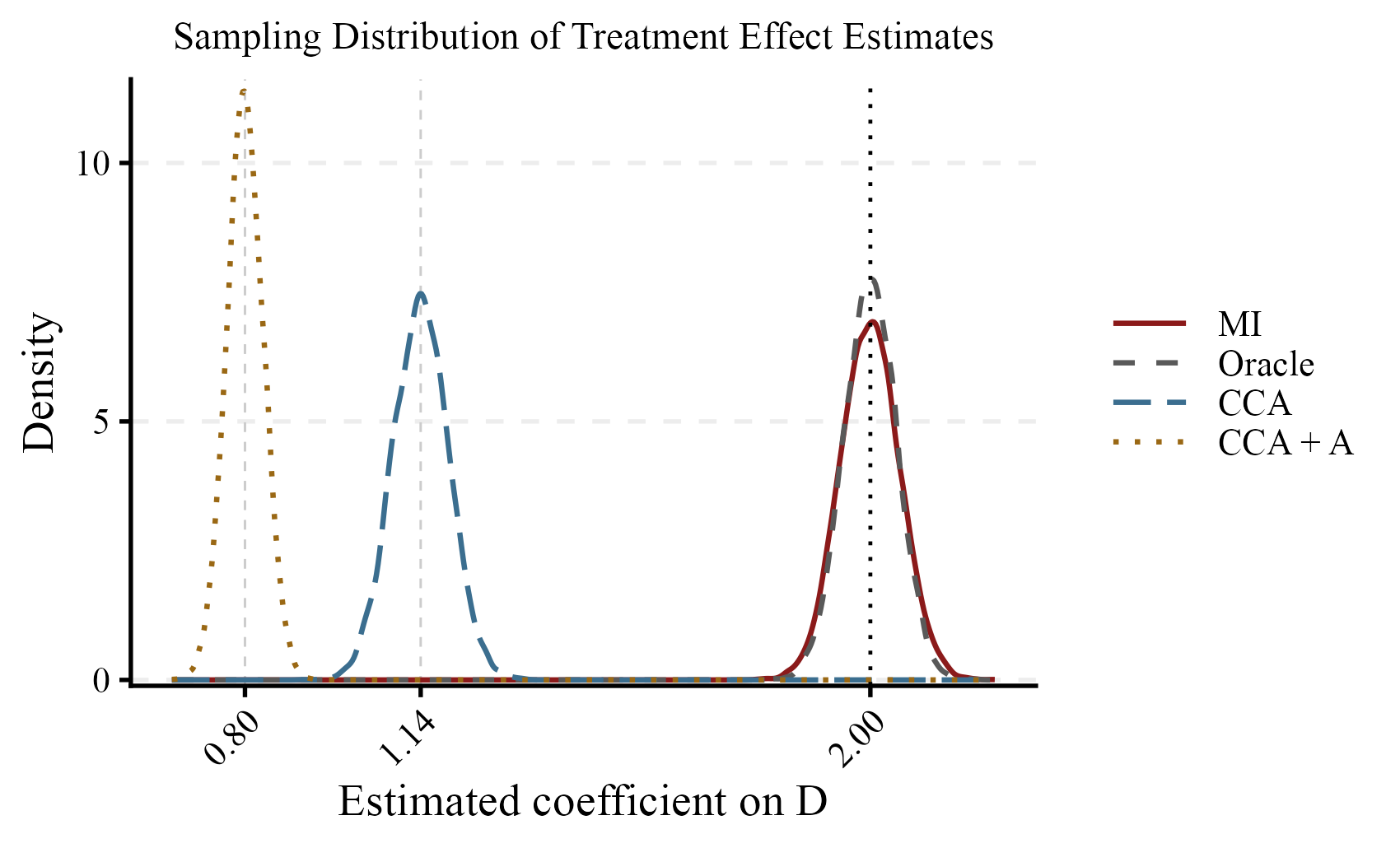}
\caption{Sampling Distribution of the Estimated Total Effect of $D$ on $Y$: Missing $Y$}
\label{fig:CM-Dilemma-1}
\end{minipage}\hfill
\begin{minipage}[t]{0.49\textwidth}
\centering
\includegraphics[width=\linewidth]{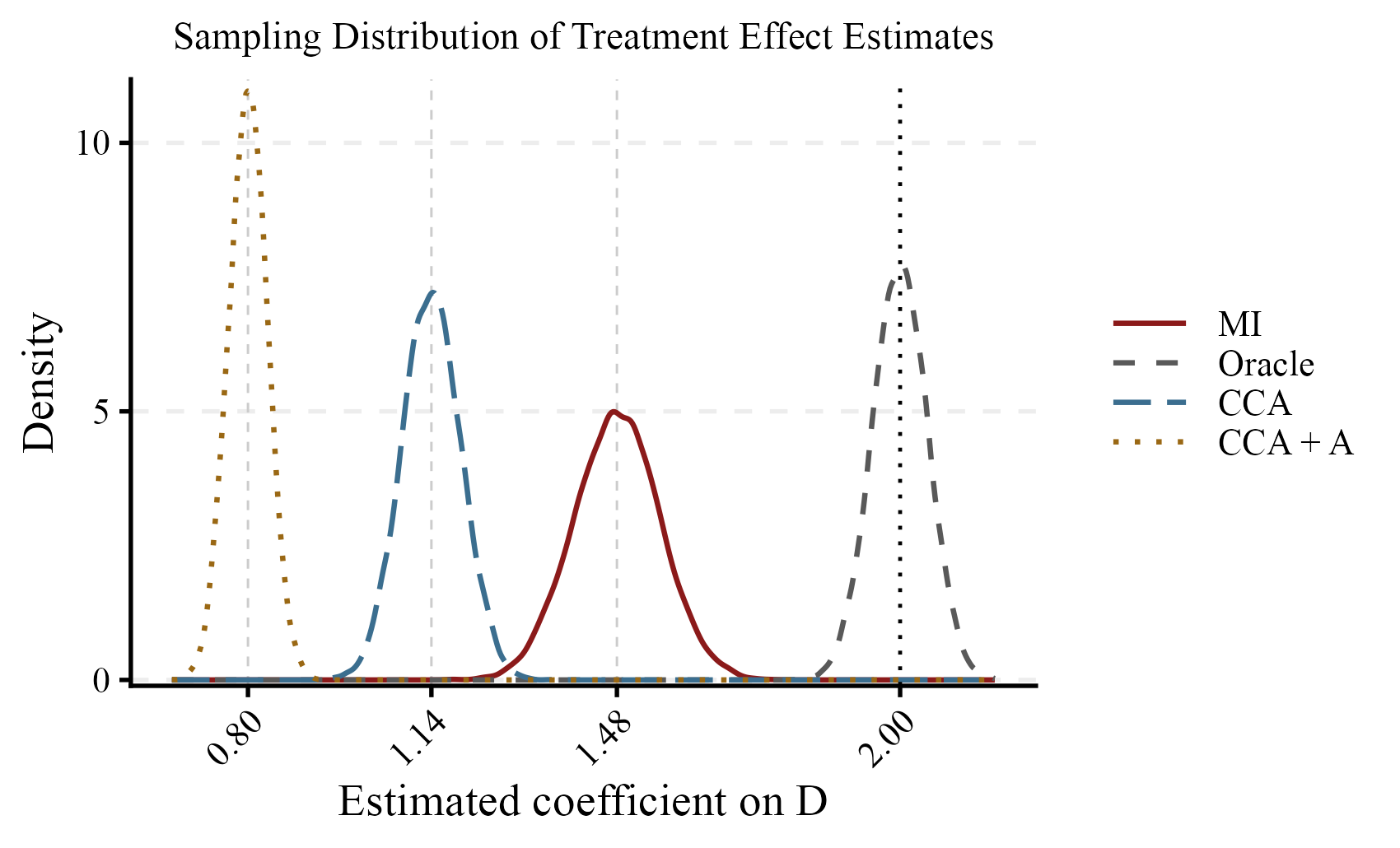}
\caption{Sampling Distribution of the Estimated Total Effect of $D$ on $Y$: Missing $D$}
\label{fig:CM-Dilemma-D}
\end{minipage}
\end{figure}

When \(Y\) is missing (Figure~\ref{fig:CM-Dilemma-1}), neither complete-case regression estimates the total effect consistently: without \(A\), selection opens a noncausal path; controlling for \(A\) blocks that path but also blocks the mediated effect. CCA+\(A\) instead estimates the controlled direct effect, while MI recovers the total effect under the specified MAR mechanism. When \(D\) is missing under the same index (Figure~\ref{fig:CM-Dilemma-D}), neither complete-case regression nor MI recovers the total effect; missingness in \(D\) depends on its own value.

\FloatBarrier
\section{Reweighting for Generalized Linear Models}
\label{app:glm-reweighting}
\setcounter{table}{0}
\renewcommand{\thetable}{\thesection\arabic{table}}
\renewcommand{\theHtable}{\thesection\arabic{table}}

The distinction between conditional comparisons and their aggregation also applies to generalized linear models. Suppose the oracle conditional mean is
\[
m_{\mathrm{po}}(d,x)
:=\mathbb E[Y_i\mid D_i=d,X_i=x]
=\ell^{-1}\!\left(\alpha+\beta_Dd+x^\top\gamma\right).
\]
For continuous \(D_i\), the marginal effect is
\[
a_{\mathrm{po}}(d,x)
:=\frac{\partial m_{\mathrm{po}}(d,x)}{\partial d}
=\beta_D\left(\ell^{-1}\right)'
\!\left(\alpha+\beta_Dd+x^\top\gamma\right).
\]
Although \(\beta_D\) is constant, the outcome-scale marginal effect generally varies with \(d\) and \(x\). The oracle response function has a causal interpretation if \(X_i\) satisfies the adjustment criterion and SUTVA and treatment overlap hold.

Let \(m_{\mathrm{cc}}(d,x):=\mathbb E[Y_i\mid D_i=d,X_i=x,M_i=0]\) and \(a_{\mathrm{cc}}(d,x):=\partial m_{\mathrm{cc}}(d,x)/\partial d\). With \(H_{\mathrm{cc}}\) and \(H_{\mathrm{po}}\) denoting the complete-case and population distributions of \((D_i,X_i)\), the corresponding average marginal effects are
\[
\theta_{\mathrm{cc}}=\int a_{\mathrm{cc}}(d,x)\,dH_{\mathrm{cc}}(d,x),
\qquad
\theta_{\mathrm{po}}=\int a_{\mathrm{po}}(d,x)\,dH_{\mathrm{po}}(d,x).
\]
As in Subsection~\ref{subsec:reweighting-hte}, recovering the population average requires both transport of the conditional effects and averaging over the population distribution. A sufficient condition for transporting the marginal-effect function is
\[
\mathbb E[Y_i\mid D_i=d,X_i=x,M_i=0]
=\mathbb E[Y_i\mid D_i=d,X_i=x],
\]
together with complete-case support; equality of the response functions implies equality of their derivatives where these exist. We can then fit a correctly specified GLM among complete cases and average its fitted marginal effects over \(H_{\mathrm{po}}\). If regressors are missing, their target distribution must be recovered, as discussed after Corollary~\ref{cor:cc-effect-transport}. Consistent estimation of the GLM coefficients alone does not ensure consistent estimation of the population AME.

For binary treatment, use the probability contrast \(\delta_{\mathrm{cc}}(x)=m_{\mathrm{cc}}(1,x)-m_{\mathrm{cc}}(0,x)\) in place of a derivative. Mean exchangeability of selection and treatment effect heterogeneity can transport this contrast without requiring equality of both response levels. Its population average then follows the same averaging argument as in Subsection~\ref{subsec:reweighting-hte}.

\FloatBarrier
\section{Estimates for the Reanalysis of Bodea and Hicks (2015)}
\label{app:bodea-hicks}
\setcounter{table}{0}
\renewcommand{\thetable}{\thesection\arabic{table}}
\renewcommand{\theHtable}{\thesection\arabic{table}}

\begin{table}[H]
\centering
\caption{Reanalysis of \citet{bodeaHicks2015}: Supporting Estimates}
\label{tab:bodea-hicks-appendix}
\footnotesize
\setlength{\tabcolsep}{1.5pt}
\renewcommand{\arraystretch}{1.0}
\begin{tabular}{@{}>{\raggedright\arraybackslash}m{0.128\linewidth}*{7}{>{\centering\arraybackslash}m{0.110\linewidth}}@{}}
\toprule
\multicolumn{8}{@{}l}{\textbf{Panel A: CBI effect on inflation without a moderator}} \\
Version & \shortstack{Countries in\\dataset} & \shortstack{CC\\countries} & \shortstack{CC\\observations} & \multicolumn{2}{c}{\shortstack{Coefficient per\\1.0 CBI (SE)}} & \multicolumn{2}{c}{\shortstack{Difference,\\expanded minus original}} \\
\shortstack[l]{Original\\data} & 79 & 78 & 1,769 & \multicolumn{2}{c}{$-$0.465 (0.143)} & \multicolumn{2}{c}{\multirow{2}{*}{\shortstack[c]{$-$0.081\\[-2.2pt]{\fontsize{7.2}{8.4}\selectfont [$-$0.182, 0.002]}\\[-2.2pt]{\fontsize{7.2}{8.4}\selectfont $p=0.054$}}}} \\
\addlinespace[3pt]
\shortstack[l]{Expanded data\\(cbi144)} & 132 & 115 & 2,198 & \multicolumn{2}{c}{$-$0.546 (0.154)} & \multicolumn{2}{c}{} \\
\midrule
\multicolumn{8}{@{}l}{\textbf{Panel B: Average effect by moderator, expanded data, per 0.1 CBI}} \\
\multicolumn{8}{@{}c@{}}{\begingroup\setlength{\tabcolsep}{0.5pt}\begin{tabular}{@{}>{\raggedright\arraybackslash}m{0.112\linewidth}>{\centering\arraybackslash}m{0.113\linewidth}*{3}{>{\centering\arraybackslash}m{0.112\linewidth}}>{\centering\arraybackslash}m{0.058\linewidth}*{2}{>{\centering\arraybackslash}m{0.112\linewidth}}>{\centering\arraybackslash}m{0.058\linewidth}>{\centering\arraybackslash}m{0.064\linewidth}@{}}
 &  &  & \multicolumn{3}{c}{Updated-CBI countries} & \multicolumn{3}{c}{All countries} &  \\
\cmidrule(lr){4-6}\cmidrule(lr){7-9}
Moderator & \shortstack{Observations\\(countries)} & CC & Weighted & Difference & \shortstack{\scriptsize $p$-value\\[-2pt]\scriptsize Raw\\[-2pt]\scriptsize Holm} & Weighted & Difference & \shortstack{\scriptsize $p$-value\\[-2pt]\scriptsize Raw\\[-2pt]\scriptsize Holm} & \shortstack{\scriptsize Thinnest\\\scriptsize support} \\
\addlinespace[2pt]
Executive constraints & 2,174 (115) & \shortstack[c]{$-$0.031\\{\fontsize{7.2}{8.4}\selectfont [$-$0.074, $-$0.001]}} & \shortstack[c]{$-$0.007\\{\fontsize{7.2}{8.4}\selectfont [$-$0.059, 0.034]}} & \shortstack[c]{+0.024\\{\fontsize{7.2}{8.4}\selectfont [0.006, 0.046]}} & \shortstack[c]{0.007\\{\fontsize{7.2}{8.4}\selectfont 0.028}} & \shortstack[c]{0.005\\{\fontsize{7.2}{8.4}\selectfont [$-$0.053, 0.051]}} & \shortstack[c]{+0.036\\{\fontsize{7.2}{8.4}\selectfont [0.010, 0.065]}} & \shortstack[c]{0.005\\{\fontsize{7.2}{8.4}\selectfont 0.020}} & 37.0\% \\
\addlinespace[2pt]
Political constraints & 2,273 (116) & \shortstack[c]{$-$0.042\\{\fontsize{7.2}{8.4}\selectfont [$-$0.075, $-$0.010]}} & \shortstack[c]{$-$0.017\\{\fontsize{7.2}{8.4}\selectfont [$-$0.055, 0.035]}} & \shortstack[c]{+0.025\\{\fontsize{7.2}{8.4}\selectfont [0.009, 0.053]}} & \shortstack[c]{0.003\\{\fontsize{7.2}{8.4}\selectfont 0.015}} & \shortstack[c]{$-$0.006\\{\fontsize{7.2}{8.4}\selectfont [$-$0.047, 0.057]}} & \shortstack[c]{+0.036\\{\fontsize{7.2}{8.4}\selectfont [0.014, 0.076]}} & \shortstack[c]{0.003\\{\fontsize{7.2}{8.4}\selectfont 0.015}} & 32.8\% \\
\addlinespace[2pt]
Freedom House & 2,231 (116) & \shortstack[c]{$-$0.048\\{\fontsize{7.2}{8.4}\selectfont [$-$0.086, $-$0.018]}} & \shortstack[c]{$-$0.024\\{\fontsize{7.2}{8.4}\selectfont [$-$0.065, 0.012]}} & \shortstack[c]{+0.024\\{\fontsize{7.2}{8.4}\selectfont [0.009, 0.045]}} & \shortstack[c]{0.002\\{\fontsize{7.2}{8.4}\selectfont 0.012}} & \shortstack[c]{$-$0.005\\{\fontsize{7.2}{8.4}\selectfont [$-$0.051, 0.037]}} & \shortstack[c]{+0.043\\{\fontsize{7.2}{8.4}\selectfont [0.018, 0.074]}} & \shortstack[c]{0.002\\{\fontsize{7.2}{8.4}\selectfont 0.012}} & 33.3\% \\
\addlinespace[2pt]
Polity & 2,198 (115) & \shortstack[c]{$-$0.030\\{\fontsize{7.2}{8.4}\selectfont [$-$0.079, 0.006]}} & \shortstack[c]{0.004\\{\fontsize{7.2}{8.4}\selectfont [$-$0.061, 0.050]}} & \shortstack[c]{+0.034\\{\fontsize{7.2}{8.4}\selectfont [0.007, 0.060]}} & \shortstack[c]{0.020\\{\fontsize{7.2}{8.4}\selectfont 0.060}} & \shortstack[c]{0.018\\{\fontsize{7.2}{8.4}\selectfont [$-$0.056, 0.071]}} & \shortstack[c]{+0.048\\{\fontsize{7.2}{8.4}\selectfont [0.010, 0.082]}} & \shortstack[c]{0.019\\{\fontsize{7.2}{8.4}\selectfont 0.057}} & 24.6\% \\
\addlinespace[2pt]
Press freedom & 1,933 (116) & \shortstack[c]{$-$0.017\\{\fontsize{7.2}{8.4}\selectfont [$-$0.063, 0.016]}} & \shortstack[c]{$-$0.007\\{\fontsize{7.2}{8.4}\selectfont [$-$0.056, 0.029]}} & \shortstack[c]{+0.011\\{\fontsize{7.2}{8.4}\selectfont [0.000, 0.022]}} & \shortstack[c]{0.050\\{\fontsize{7.2}{8.4}\selectfont 0.100}} & \shortstack[c]{0.006\\{\fontsize{7.2}{8.4}\selectfont [$-$0.048, 0.043]}} & \shortstack[c]{+0.023\\{\fontsize{7.2}{8.4}\selectfont [0.003, 0.040]}} & \shortstack[c]{0.029\\{\fontsize{7.2}{8.4}\selectfont 0.058}} & 29.0\% \\
\addlinespace[2pt]
Checks and balances & 2,073 (116) & \shortstack[c]{$-$0.072\\{\fontsize{7.2}{8.4}\selectfont [$-$0.115, $-$0.030]}} & \shortstack[c]{$-$0.077\\{\fontsize{7.2}{8.4}\selectfont [$-$0.124, $-$0.028]}} & \shortstack[c]{$-$0.005\\{\fontsize{7.2}{8.4}\selectfont [$-$0.018, 0.011]}} & \shortstack[c]{0.591\\{\fontsize{7.2}{8.4}\selectfont 0.591}} & \shortstack[c]{$-$0.078\\{\fontsize{7.2}{8.4}\selectfont [$-$0.128, $-$0.027]}} & \shortstack[c]{$-$0.007\\{\fontsize{7.2}{8.4}\selectfont [$-$0.025, 0.014]}} & \shortstack[c]{0.578\\{\fontsize{7.2}{8.4}\selectfont 0.578}} & 38.2\% \\
\end{tabular}\endgroup} \\
\midrule
\multicolumn{8}{@{}l}{\textbf{Panel C: Polity categories, expanded data}} \\
Category & CC observations & Share, CC & Share, cbi144 & Share, worldwide & Slope per 0.1 CBI & Countries (with CBI variation) &  \\
Autocracy ($-10$ to $-6$) & 282 & 12.8\% & 29.7\% & 36.1\% & +0.140 & 38 (11) &  \\
Anocracy ($-5$ to $5$) & 439 & 20.0\% & 21.9\% & 23.7\% & $-$0.033 & 57 (14) &  \\
Democracy (6 to 10) & 1,477 & 67.2\% & 48.4\% & 40.2\% & $-$0.062 & 83 (53) &  \\
\bottomrule
\end{tabular}
\end{table}
\begingroup\footnotesize\setstretch{1}\medskip
\textit{Notes:} CC means complete-case. The outcome is the signed log of inflation. All models include country fixed effects and the controls of \citet{bodeaHicks2015}: lagged log inflation, lagged money growth, lagged log GDP, lagged GDP growth, lagged trade openness, a de jure exchange-rate regime indicator, lagged fiscal balance, election indicators, and lagged world inflation. Panel A also includes Polity and reports coefficients and their difference per unit of CBI. The original-data estimate reproduces their Table 1, Model 5. The expanded data add country-years from countries covered by the updated CBI data, with controls rebuilt from public sources. The updated-CBI country target is labeled cbi144; the updated CBI data cover 143 countries and the European Central Bank. CC have every control observed. Panel B fits a separate regression for each moderator category and averages the category slopes using the CC shares, the shares among the updated-CBI countries (cbi144), or the shares across all countries Polity5 covers (worldwide), 1973--2008. Differences are the reweighted minus the CC average. Intervals are 95\% percentile intervals from 2,000 country-cluster bootstrap draws, all draws succeeded. Target shares are held fixed in the bootstrap; CC shares are recomputed in each draw. Panel B reports raw and Holm-adjusted two-sided $p$-values (raw/adjusted), with adjustment across the six moderators separately for each target; Panel A reports its unadjusted $p$-value. Thinnest support is the smallest share, across a moderator\textquotesingle s categories, of countries whose CBI changes within the category. The moderator categories are defined as follows: Polity, $-10$ to $-6$, $-5$ to $5$, and $6$ to $10$; executive constraints, 1--3, 4--6, and 7; political constraints, at most 0.2, above 0.2 through 0.4, and above 0.4; checks and balances, 1--2, 3, and 4 or more; Freedom House average rating, at most 2.5, above 2.5 through 5, and above 5; and press freedom, not free, partly free, and free.
\par\endgroup

\FloatBarrier
\section{Treatment-IPW Comparison for the Canvassing Experiment}
\label{app:kb-probit-ipw}
\setcounter{table}{0}
\renewcommand{\thetable}{\thesection\arabic{table}}
\renewcommand{\theHtable}{\thesection\arabic{table}}
\setcounter{figure}{0}
\renewcommand{\thefigure}{\thesection\arabic{figure}}
\renewcommand{\theHfigure}{\thesection\arabic{figure}}

In another hypothetical example, had the researcher used a randomization scheme in which treatment probabilities differed by participants' place of residence, assignment would depend on a component of \(X\), creating the arrow \(X\rightarrow D\). Suppose further that follow-up response depended on both \(X\) and an unobserved determinant of the outcome, \(U\), as in Figure~\ref{fig:kb-conditional-randomization}. Adjusting for \(X\) would still preserve valid within-\(X\) treatment comparisons, satisfying Box 1. However, selection could make the baseline outcome mean nonlinear in \(X\), as in mechanism B of the motivating simulation. A regression controlling only linearly for \(X\) could then pool those comparisons incorrectly, so Box 2 would require an additional specification or weighting check.

\begin{figure}[H]
\centering
\begin{tikzpicture}[manuscript dag,scale=0.62,transform shape]
\node[manuscript dag node] (D) at (0,0) {D};
\node[manuscript dag node] (Y) at (3,0) {Y};
\node[manuscript dag node] (X) at (0,1.5) {X};
\node[manuscript dag node,dashed] (U) at (3,1.5) {U};
\node[manuscript dag node] (M) at (1.5,2.7) {M};
\node[manuscript dag node] (W) at (1.5,1.5) {W};
\path[manuscript dag edge]
 (D) edge (Y) (X) edge (D) (X) edge (Y)
 (U) edge (Y) (X) edge (M) (U) edge (M) (W) edge (Y) (W) edge (M);
\end{tikzpicture}
\par\medskip
\begin{minipage}{\linewidth}
\footnotesize
\textit{Note:} In this hypothetical design, \(W\) is baseline college education, \(X\) includes residence and the other baseline controls, and dashed \(U\) is unobserved. Treatment probabilities vary with \(X\). Conditioning on \(M=0\) opens a noncausal path, which adjustment for \(X\) and \(W\) blocks. With treatment overlap, the adjustment criterion is satisfied among complete cases.
\end{minipage}
\caption{A Hypothetical Conditionally Randomized Design with Selection on an Unobserved Outcome Determinant}
\label{fig:kb-conditional-randomization}
\end{figure}

\begin{table}[H]
\centering
\caption{Complete-Case and Probit-IPW Estimates in the Two-Arm Canvassing Sample}
\label{tab:kb-probit-ipw}
\small
\setlength{\tabcolsep}{4pt}
\renewcommand{\arraystretch}{1.15}
\begin{tabular}{@{}l r c c c@{}}
\toprule
Group & CC observations & CC regression & Probit-IPW & Difference (IPW minus CC)\\
\midrule
All complete cases & 1,172 & \shortstack{0.0925 (0.0226)\\{[0.0483, 0.1357]}} & \shortstack{0.0932 (0.0231)\\{[0.0474, 0.1375]}} & \shortstack{0.0007 (0.0062)\\{[-0.0112, 0.0134]}} \\
\bottomrule
\end{tabular}
\par\medskip
\begin{minipage}{\linewidth}
\footnotesize
\textit{Notes:} CC means complete-case. The sample includes only the full-intervention and placebo arms of Study 1 in \citet{kallaBroockman2023}. The outcome is the first-follow-up omnibus index of immigrant attitudes and policy preferences. Estimates are index units; bootstrap standard errors are in parentheses and 95\% percentile intervals are in brackets. The CC regression includes the authors' baseline controls except college education. A probit of treatment assignment on education and these controls estimates the complete-case treatment propensity $\widehat e_0(W,X)$. IPW estimates come from a regression of the outcome on an intercept and treatment, weighted by $D/\widehat e_0(W,X)+(1-D)/[1-\widehat e_0(W,X)]$. No weights are trimmed. In 2,000 household-bootstrap draws, households are resampled from the target sample, and both regressions and the probit model are refitted. All draws succeeded. Differences use the same paired draws. These weights address treatment assignment among complete cases, rather than response selection or population education composition. The use of estimated propensities is an illustrative choice; the actual design assigns equal probabilities within matched household blocks.
\end{minipage}
\end{table}

\begin{table}[H]
\centering
\caption{Complete-Case Estimation and Reweighting by Education}
\label{tab:kb-education-reweight}
\small
\setlength{\tabcolsep}{4pt}
\renewcommand{\arraystretch}{1.0}
\begin{tabular}{@{}lccc@{}}
\toprule
 & Pooled CC & No college & College \\
\midrule
\multicolumn{4}{@{}l}{\textbf{Complete-case treatment effects}} \\
Treatment effect & 0.0925 & 0.0497 & 0.1107 \\
Bootstrap SE & (0.0226) & (0.0410) & (0.0372) \\
Complete-case observations & 1,172 & 548 & 624 \\
\midrule
\multicolumn{4}{@{}l}{\textbf{Education shares and averages}} \\
Complete-case share & --- & 46.8\% & 53.2\% \\
Target-sample share & --- & 52.5\% & 47.5\% \\
\midrule
Average using complete-case shares & \multicolumn{3}{c}{0.0822} \\
Average using target-sample shares & \multicolumn{3}{c}{0.0787} \\
Difference: reweighted minus pooled CC & \multicolumn{3}{c}{$-0.0138$} \\
95\% bootstrap interval for difference & \multicolumn{3}{c}{$[-0.0488, 0.0174]$} \\
\bottomrule
\end{tabular}
\par\medskip
\begin{minipage}{\linewidth}
\footnotesize
\textit{Notes:} CC means complete-case. The sample comprises 1,900 participants in the full-intervention and placebo arms who confirmed their identity at the door and were eligible for follow-up surveys, of whom 1,172 have the first-follow-up outcome. Education-specific effects come from separate treatment-IPW regressions within each education group. Within each group, a separate complete-case probit models treatment assignment using the remaining baseline controls. The pooled CC regression omits education and includes the same baseline controls. The difference compares the target-share average of education-specific effects with the pooled CC regression. The complete-case-share average and the pooled CC estimate can differ in a finite sample because they use different specifications. Standard errors and percentile intervals use 2,000 paired household-bootstrap draws from the target sample, refitting the propensity model and regressions and recomputing the education shares. All draws succeeded. No weights are trimmed. Calculations use unrounded values.
\end{minipage}
\end{table}

\end{document}